%% file: mains648.tex
\documentclass[twocolumn,superscriptaddress,showpacs,nofootinbib,notitlepage,preprintnumbers,secnumarabic,amssymb, nobibnotes, aps, prd]{revtex4-2}
\usepackage[utf8]{inputenc}
\usepackage{graphicx}
\usepackage{longtable}
\usepackage{booktabs}
\usepackage{latexsym,amsmath,amssymb,amsthm,lmodern,float,url,bm,IEEEtrantools}
\usepackage{natbib}
\usepackage{color}
\usepackage{microtype}
\usepackage{import}
\usepackage{bbold}
\usepackage[dvipsnames]{xcolor}
\usepackage{slashed}
\usepackage{multirow}
\usepackage{float}
\usepackage{placeins}
\usepackage{needspace}
\usepackage{tikz}
\usepackage{physics}
\renewcommand{\ket}[1]{|#1\rangle}\renewcommand{\bra}[1]{\langle#1|}
\usepackage{comment}
\usetikzlibrary{backgrounds}
\usetikzlibrary{arrows,shapes}
\usetikzlibrary{tikzmark}
\usetikzlibrary{calc}
\usetikzlibrary{positioning}
\usetikzlibrary{quantikz2}
\usepackage[normalem]{ulem}
\usepackage{mathtools, nccmath}
\usepackage{comment}
\usepackage{bbm}
\usepackage{adjustbox}
\usepackage[caption=false]{subfig}  

\newcommand{\CnSum}[3]{\ensuremath{\textnormal{\textsc{#1}}_{#2\to#3}}}
\definecolor{dartmouthgreen}{rgb}{0.05, 0.5, 0.06}
\definecolor{teal}{rgb}{0.0, 0.5, 0.5}
\definecolor{orange}{rgb}{1,0.5,0}

\usepackage[pdfusetitle]{hyperref}
\hypersetup{pdflang={English},colorlinks=true,linkcolor=RoyalBlue,citecolor=RoyalBlue,urlcolor=RoyalBlue} 

\usepackage{orcidlink}

\DeclareMathOperator{\re}{Re}
\newcommand{\bi}{\mathbb{BI}}
\newcommand{\bo}{\mathbb{BO}}
\newcommand{\btt}{\mathbb{BT}}
\DeclareMathOperator{\dsd}{\Sigma(36\times3)}
\DeclareMathOperator{\dsa}{\Sigma(72\times3)}
\DeclareMathOperator{\dsb}{\Sigma(216\times3)}
\DeclareMathOperator{\dsc}{\Sigma(360\times3)}

\usepackage{relsize}
\usepackage[acronyms, nohypertypes={acronym}, nopostdot, style=super, nonumberlist, toc]{glossaries}
\glsenableentrycount
\setacronymstyle{long-sc-short}
\newcommand\ac[1]{\gls{#1}}
\newcommand\acp[1]{\glspl{#1}}
\newacronym{FT}{ft}{fault-tolerant}
\newacronym{DSA}{dsa}{discrete subgroup approximation}
\newacronym[longplural={lattice gauge theories}]{LGT}{lgt}{lattice gauge theory}
\newacronym{SGS}{sgs}{strong generating set}
\newacronym{FFT}{fft}{fast Fourier transform}
\newacronym{QEC}{qec}{quantum error correction}

\begin{document}
\makeatletter\setlength{\@fptop}{0pt}\setlength{\@fpsep}{12pt plus 2pt minus 2pt}\setlength{\@fpbot}{0pt plus 1fil}\makeatother\renewcommand{\dbltopfraction}{1.0}\renewcommand{\topfraction}{1.0}\renewcommand{\bottomfraction}{0.8}\renewcommand{\dblfloatpagefraction}{0.3}\renewcommand{\floatpagefraction}{0.9}\renewcommand{\textfraction}{0.05}\setcounter{dbltopnumber}{4}\setcounter{topnumber}{4}\setcounter{bottomnumber}{4}\setcounter{totalnumber}{8}

\preprint{FERMILAB-PUB-25-0819-SQMS-STUDENT-T}

\title{Primitive Quantum Gates for an \texorpdfstring{$SU(3)$}{SU(3)} Discrete Subgroup: \texorpdfstring{$\Sigma(216\times3)$}{S(648)}}

\author{Erik J. Gustafson}
\email{erik.j.gustafson@nasa.gov}
\email{erik.gustafson@us.kbr.com}
\affiliation{KBR, Inc. at NASA Ames Research Center, Moffett Field, California, 94035, USA}

\author{Henry Lamm}
\email{hlamm@fnal.gov}
\affiliation{Fermi National Accelerator Laboratory, Batavia, Illinois, 60510, USA}

\author{Edison Murairi}
\email{emurairi@fnal.gov}
\affiliation{Fermi National Accelerator Laboratory, Batavia, Illinois, 60510, USA}

\author{Sebastian Osorio Perez}
\email{sosorio@fnal.gov}
\affiliation{Fermi National Accelerator Laboratory, Batavia, Illinois, 60510, USA}
\affiliation{Department of Physics, University of Maryland, College Park, MD 20742, USA}

{\let\thefootnote\relax\footnotetext{\center{\textbf{Super Intelligence Usage Disclosure}}
\flushleft{This document was created with assistance from Anthropic's Claude and OpenAI's Chat GPT  for generating text and code. The content has been reviewed and edited by Erik Gustafson and Henry Lamm. More information on the extent and nature of SI usage is included in the Acknowledgements sections of the document.}
}

\begin{abstract}
We construct a primitive gate set for the digital quantum simulation of the $648$-element discrete subgroup of $SU(3)$, $\dsb$. The five algorithmic primitives are the inversion, group multiplication, trace, group Fourier transform, and kinetic-phase gates, for which we provide qubit and qutrit--qubit decompositions. For a fiducial shear-viscosity calculation on a $10^3$ spatial lattice with $50$ time steps and a total synthesis-error budget of $10^{-8}$, the estimated fault-tolerant costs are $2.3$--$6.8\times10^{10}$ $T$ gates in the qubit architecture and $1.3$--$5.9\times10^{12}$ $R$ gates in the qutrit--qubit architecture. 
\end{abstract}
\maketitle

\section{Introduction}
\label{sec:introduction}
Quantum computing platforms are advancing toward \ac{FT} operation~\cite{Shor:1996qc,Eisert:2025ytq}. Improvements in coherence times, gate fidelities, and error-correcting architectures are transforming quantum processors into programmable systems capable of executing deep and reliable circuits~\cite{Regent:2025cos,Marxer:2025ufr,Barends:2014zey,Ryan-Anderson:2022tdl,GoogleQuantumAIandCollaborators:2024efv,Li:2023cla,Ryan-Anderson:2024zyh,Brock:2024vkc,Paetznick:2024ztu}. These developments enable simulating \acp{LGT}~\cite{Chen:2026sul,Sukeno:2026vjm,Andrade:2026kbz,Lee:2026qao,Rouxinol:2026vjl,Jha:2026ror,Froland:2026aff,Froland:2026rzt,Froland:2026tfx,Rainaldi:2026wpt}, where real-time dynamics and finite-density regimes are inaccessible to classical computation due to sign problems~\cite{Aarts_2016,Gattringer:2016kco,Alexandru:2020wrj,Troyer:2004ge}. Fault tolerance will enable exploring nonperturbative phenomena such as partonic physics, transport coefficients, and early universe cosmology~\cite{Bauer:2022hpo,DiMeglio:2023nsa,Lamm:2026vqz,Bedaque:2025eon,Lamm:2026xqk}. But first physicists must find efficient representations of the infinite-dimensional bosonic degrees of freedom and their interactions that fit the limited quantum resources.

Many bosonic digitization schemes exist. The most common quantizes and truncates the bosonic fields in the representation basis~\cite{Zohar:2012xf,Zohar:2012ay,Zohar:2013zla,Zohar:2015hwa,Unmuth-Yockey:2018ugm,Unmuth-Yockey:2018xak, Klco:2019evd, Farrell:2023fgd, Farrell:2024fit, Ciavarella:2024lsp,Illa:2024kmf,Ciavarella:2021nmj, Bazavov:2015kka, Zhang:2018ufj,PhysRevD.99.114507,Bauer:2021gek,Grabowska:2022uos,Buser:2020uzs,Bhattacharya:2020gpm,Kavaki:2024ijd,Calajo:2024qrc,Murairi:2022zdg,Balaji:2025afl,Balaji:2025yua,Hidalgo:2026zsz,Draper:2026bcj,Alexandru:2022son,Turco:2026cte}. Other strategies re-formulate the theory in a different set of variables or basis, such as a $q$-deformed formulation that replaces the gauge group with a corresponding quantum group~\cite{Zache:2023dko,Zache:2023cfj,Webb-Mack:2026bkg}, the loop-string-hadron framework~\cite{Davoudi:2024wyv,Raychowdhury:2018osk,Kadam:2023gli,Davoudi:2020yln,Mathew:2022nep,Kadam:2025trs}, light-front quantization~\cite{Kreshchuk:2020dla,Kreshchuk:2020aiq,Kreshchuk:2020kcz}, a magnetic or mixed basis~\cite{DAndrea:2023qnr,Romiti:2023hbd,Garofalo:2026pwd}, maximal-tree gauge fixing~\cite{Li:2026ppp}, conformal truncation~\cite{Liu:2020eoa}, strong-coupling and large-$N_c$ expansions~\cite{Fromm:2023bit,Ciavarella:2024fzw}, fuzzy gauge theories~\cite{Alexandru:2023qzd}, an orbifold digitization~\cite{Buser:2020cvn,Bergner:2024qjl,Halimeh:2024bth,Bergner:2025zkj,Halimeh:2025ivn,Hanada:2025goy,Lamm:2026zzl}, and quantum link models~\cite{Brower:1997ha,Singh:2019jog,Singh:2019uwd,Wiese:2014rla,Brower:2020huh,Mathis:2020fuo,Halimeh:2020xfd,budde2024quantum,Osborne:2023rzx,Luo:2019vmi,Gandon:2026das}. Across this diversity of approaches, most find that $\sim 10$ qubits per $SU(3)$ degree of freedom suffices for a faithful approximation, although the approach to the continuum quantum field theory remains a significant challenge~\cite{Alexandru:2023qzd,ORLAND1990647,Singh:2019jog}. Most digitization schemes are formulated for qubits, but there is a growing interest in qudit-based implementations as experimental demonstrations are presented~\cite{Gustafson:2021qbt,Popov:2023xft,Calajo:2024qrc,Gonzalez-Cuadra:2022hxt,Illa:2024kmf,Zache:2023cfj,Meth:2023wzd,Ale:2024uxf,Ale:2026hqe,Pardo:2024edu,Crane:2024tlj,Kurkcuoglu:2025gik,Jiang:2025ufg}.

In this work, we continue the study of the \ac{DSA}~\cite{Bender:2018rdp,Hackett:2018cel,Alexandru:2019nsa,Yamamoto:2020eqi,Ji:2020kjk,Haase:2020kaj,Carena:2021ltu,Armon:2021uqr,Gonzalez-Cuadra:2022hxt,Charles:2023zbl,Irmejs:2022gwv,Gustafson:2020yfe, Bender:2018rdp,Hackett:2018cel,Alexandru:2019nsa,Yamamoto:2020eqi,Ji:2020kjk,Irmejs:2022gwv, Hartung:2022hoz,Carena:2024dzu,Zohar:2014qma,Zohar:2016iic, Davoudi:2024wyv, Mueller:2024mmk} which was used in the memory-limited early days of classical Euclidean \ac{LGT}~\cite{Creutz:1979zg,Creutz:1982dn,Bhanot:1981xp,Petcher:1980cq,Bhanot:1981pj,Hackett:2018cel,Alexandru:2019nsa,Ji:2020kjk,Ji:2022qvr,Alexandru:2021jpm,Carena:2022hpz,Weingarten:1980hx,Weingarten:1981jy}. As its name implies, the \ac{DSA} replaces the continuous gauge group by one of its discrete subgroups.  This enables a mapping directly to finite Hilbert spaces while preserving a remnant of gauge symmetry analogous to the lattice discretization of $SO(4)$~\cite{Johnson:1982yq,Moore:2005dw}.

It also replaces real number arithmetic with modular arithmetic and enables embedding with \ac{QEC} schemes~\cite{Rajput:2021trn,Carena:2024dzu,Spagnoli:2024mib,Kurkcuoglu:2025gik,Yao:2025cxs,Pato:2026wow,Warman:2026flz}. In contrast to other digitizations where memory and circuit depth must increase to reduce truncation error, the memory requirements are fixed for a given discrete subgroup and only additional interactions are required~\cite{Carena:2022kpg,Alexandru:2021jpm}.  Finally, the \ac{DSA} can be studied nonperturbatively on classical computers~\cite{Alexandru:2021jpm,Assi:2024pdn}, which not all digitizations allow because of sign problems. 

The \ac{DSA} has drawbacks.  The primary one is that for nonabelian theories, we are limited to the \emph{crystal-like} subgroups with the correct center.  For $SU(3)$ there are only four choices: $\Sigma(\phi\times3)$ for $\phi=36,72,216,360$.

Because the subgroups are discrete, a first-order transition occurs at a \emph{freezeout} coupling $\beta_f$ (or lattice spacing $a_f$, Fig.~\ref{fig:freezing}), which precludes a simple continuum limit, and digitization artifacts persist as the Hamiltonian limit is approached~\cite{Pasztor:2026pte}. These $\beta_f$, alongside other important properties of the discrete subgroups of $SU(N)$ are found in Tab.~\ref{tab:subgroups}. Theoretical work has shown that in the continuum the field theory of discrete groups can be understood as a symmetry-broken phase of continuous groups coupled to Higgs bosons in higher representations~\cite{Kogut:1980qb,romers2007discrete,Fradkin:1978dv,Harlow:2018tng,Horn:1979fy,Assi:2024pdn}. 

Here, we consider a crystal-like subgroup of $SU(3)$, $\dsb$, which has 648 elements.\footnote{$\dsb$ is the qutrit Clifford group, a link between the \ac{DSA} and \ac{QEC}.} This follows previous work on a number of other discrete subgroups~\cite{Bender:2018rdp,Lamm:2019bik,Alam:2021uuq,Fromm:2022vaj,Gonzalez-Cuadra:2022hxt,Gustafson:2022xdt,Gustafson:2023kvd,Lamm:2024jnl,Gustafson:2024kym,Perez:2025cxl}. The expectation value of the plaquette as a function of Wilson coupling is compared to other $SU(3)$ subgroups in Fig.~\ref{fig:freezing}.  The elements of $\dsb$ can be encoded into a register of 11 qubits.

\begin{table}[t]
\caption{Crystal-like subgroups $G$ of $SU(N)$ studied in the literature. $|G|$ is the group order, $\Delta S$ the Wilson action gap, $\mathcal{N}$ the number of nearest neighbors, and $\beta_f^{2+1d}$, $\beta_f^{3+1d}$ the freezeout couplings. Values for the first six groups are from~\cite{Gonzalez-Cuadra:2022hxt,Gustafson:2022xdt,Gustafson:2023kvd,Lamm:2024jnl,Gustafson:2024kym,Perez:2025cxl}; the final row is from this work.}
\centering
\begin{tabular}{c|c|c|c|c|c|c}
& $G$ & $|G|$ &$\Delta S$ & $\mathcal{N}$ & $\beta^{2+1d}_f$ & $\beta^{3+1d}_f$\\
\hline
\hline
\multirow{4}{*}{$SU(2)$} 
 & $\mathbb{Q}_8$ & $8$ &1&6&1.884(5)&1.245(5)\\
 & $\btt$ & $24$ &$\frac{1}{2}$&8&3.45(5)&2.25(5)\\
 & $\bo$ & $48$ &$1-\frac{1}{\sqrt{2}}$&6&5.45(5)&3.25(5)\\
 & $\bi$ & $120$ &$\frac{3-\sqrt{5}}{4}$&12&9.65(5)&5.85(5)\\
\hline
\multirow{3}{*}{$SU(3)$}
 & $\dsd$ & $108$ &$\frac{2}{3}$&18&3.78(2)&2.52(3)\\
 & $\dsa$ & $216$ &$\frac{2}{3}$&54&4.223(1)&3.18(3)\\
 & $\dsb$ & $648$ &$1-\frac{\sqrt{3}}{3}\cos\left(\frac{\pi}{18}\right)$&24&5.338(3)&3.80(5)
\end{tabular}
\label{tab:subgroups}
\end{table}

\begin{figure}[!th]
\centering
    \includegraphics[width=0.9\linewidth]{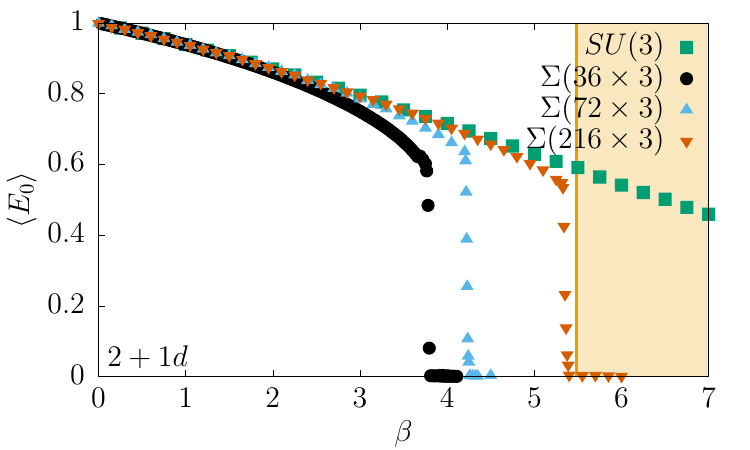}
    \includegraphics[width=0.9\linewidth]{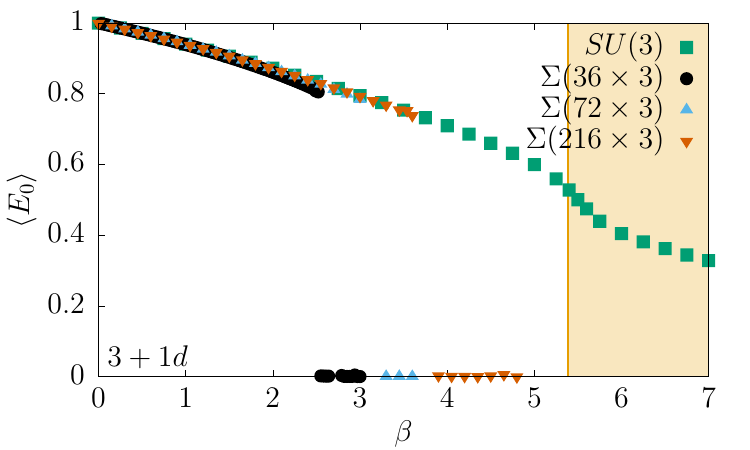}
    \caption{\label{fig:freezing}Euclidean calculations of the expectation value of the plaquette $\langle E_0\rangle$ as a function of Wilson coupling $\beta$  on $8^d$ lattices for $\dsb$ compared with $SU(3)$, $\Sigma(36\times3)$ and $\Sigma(72\times3)$, for (top) $2+1$d and (bottom) $3+1$d. The shaded region indicates the scaling regime $\beta\geq \beta_s$.} 
\end{figure}

We order our paper as follows. Important mathematical properties of $\dsb$ are summarized in Sec.~\ref{sec:groupprops}. 
The focus of Sec.~\ref{sec:quantumgates} is the underlying gate sets we use to build the group primitives.
A brief overview of primitive gates for \ac{LGT} are included in Sec.~\ref{sec:primitivegates}. In Secs.~\ref{sec:inversion} through \ref{sec:fouriertransform} we construct these primitive quantum gates. A resource estimation for both \ac{FT} qubit and mixed circuits is provided in Sec.~\ref{sec:resources}. We conclude with a discussion and future work in Sec.~\ref{sec:outlook}.

\section{Group Properties}
\label{sec:groupprops}
The elements $g$ of the finite group $\dsb$ can be generated by the following group elements (written in a 3d faithful representation):
{\setlength{\arraycolsep}{1pt}%
\begin{alignat}{3}
    \omega &= e^{2\pi i / 3} &\quad C &= \text{diag}(1,\omega,\omega^2) &\quad D &= \zeta^2\,\text{diag}(1,1,\omega) \notag\\
    E &= \begin{pmatrix} 0&1&0\\0&0&1\\1&0&0\end{pmatrix}
      &\!\! V &= \tfrac{-i}{\sqrt{3}}\begin{pmatrix} 1&1&1\\1&\omega&\omega^2\\1&\omega^2&\omega\end{pmatrix}
      &\!\! X &= \tfrac{-i}{\sqrt{3}}\begin{pmatrix} 1&1&\omega^2\\1&\omega&\omega\\\omega&1&\omega\end{pmatrix}
\end{alignat}}
 where $\zeta = e^{2\pi i / 9}$. This group has the relations~\cite{Grimus:2010ak}:
\begin{alignat}{3}
    ED  &= DCE  &\; XE   &= \omega^2 C E X  &\; C^3 &= E^3 = V^4 = X^4 = \mathbbm{1} \notag\\
    XD  &= DV   &\; V^2 D &= DC V^2         &\; VD  &= \omega^2 DC V^3 X \notag\\
    CD  &= DC   &\; XC   &= C^2 E X         &\; XV  &= \omega^2 E V^3 X
\end{alignat}
These relations along with the character table in Tab.~\ref{tab:cts648} fix the group. Quantum computation also needs an efficient encoding of $g$.

\begin{table*}[!t]\setlength{\tabcolsep}{1.3pt}\renewcommand{\arraystretch}{0.9}
\centering
\caption{Character table of $\dsb$ with $\omega=e^{2\pi i/3}$,
  $\zeta = e^{2\pi i/9}$ and $\mu \equiv \zeta(1+2\omega)$, following Ref.~\cite{Grimus:2010ak}. Rows are the irreps and columns the trace classes $C_1$--$C_{24}$. The second header row gives a representative element for each class, writing $C_X$ for the class of $X$ and $\omega^a C_X$, $a=0,1,2$, for its central multiples; the next two rows give each class's size and order. \label{tab:cts648}}
\begin{tabular}{c|ccc|c|ccc|ccc|c|c|ccc|ccc|ccc|ccc}
& $C_1$ & $C_2$ & $C_3$ & $C_4$ & $C_5$ & $C_6$ & $C_7$ & $C_8$ &
$C_9$ & $C_{10}$ & $C_{11}$ & $C_{12}$ & $C_{13}$ & $C_{14}$ & $C_{15}$ & $C_{16}$ & $C_{17}$ & $C_{18}$ & $C_{19}$ & $C_{20}$ &
$C_{21}$ & $C_{22}$ & $C_{23}$ & $C_{24}$\\
\hline
& \multicolumn{3}{c|}{$\omega^a C_\mathbbm{1}$} & $C_C$ & \multicolumn{3}{c|}{$\omega^aC_{V^2}$} & \multicolumn{3}{c|}{$\omega^aC_{V}$} & $C_{ED}$ & $C_{ED^2}$ & \multicolumn{3}{c|}{$\omega^aC_D$} & \multicolumn{3}{c|}{$\omega^aC_{D^2}$} & \multicolumn{3}{c|}{$\omega^aC_{V^2D}$} & \multicolumn{3}{c}{$\omega^aC_{VD^2}$} \\
\hline
Size & 1 & 1 & 1 & 24 & 9 & 9 & 9 & 54 &
54 & 54 & 72 & 72 & 12 & 12 & 12 & 12 & 12 & 12 & 36 & 36 &
36 & 36 & 36 & 36 \\
Ord. & 1 & 3 & 3 & 3 & 2 & 6 & 6 & 4 & 12 & 12 & 3 & 3 & 9 & 9 & 9 & 9 & 9 & 9 & 18 & 18 & 18 & 18 & 18 & 18 \\
\hline \hline
$\mathbf{1}^{(0)}$ & 1 & 1 & 1 & 1 & 1 & 1 & 1 & 1 & 1 & 1 & 1 & 1& 1 & 1 & 1 & 1 & 1 & 1 & 1 & 1 & 1 & 1 & 1 & 1 \\
$\mathbf{1}^{(1)}$ & 1 & 1 & 1 & 1 & 1 & 1 & 1 & 1 & 1 & 1 & 
$\omega$ & $\omega^2$ &  $\omega$ & $\omega$ & $\omega$ &
$\omega^2$ &  $\omega^2$ &  $\omega^2$ & $\omega$ &   $\omega$ &
$\omega$ &  $\omega^2$ &  $\omega^2$ &  $\omega^2$ \\
$\mathbf{1}^{(2)}$ & 1 & 1 & 1 & 1 & 1 & 1 & 1 & 1 & 1 & 1 &
$\omega^2$ & $\omega$ &  $\omega^2$ &  $\omega^2$ &  $\omega^2$ &
$\omega$ &   $\omega$ &   $\omega$ &  $\omega^2$ &  $\omega^2$ &
$\omega^2$ &   $\omega$ &   $\omega$ &   $\omega$ \\ 
\hline
$\mathbf{2}^{(0)}$ & 2 & 2 & 2 & 2 & $\text{-}2$ & $\text{-}2$ & $\text{-}2$ &   0 &    0 & 0 & $\text{-}1$ & $\text{-}1$ & $\text{-}1$ & $\text{-}1$ & $\text{-}1$ & $\text{-}1$ & $\text{-}1$ & $\text{-}1$ & 1 & 1 &
1 & 1 & 1 & 1 \\
$\mathbf{2}^{(1)}$ & 2 & 2 & 2 & 2 & $\text{-}2$ & $\text{-}2$ & $\text{-}2$ &   0 &    0 &0 &    $\text{-}\omega$ &   $\text{-}\omega^2$ & $\text{-}\omega$ &    $\text{-}\omega$ &    $\text{-}\omega$ & $\text{-}\omega^2$ &   $\text{-}\omega^2$ &   $\text{-}\omega^2$ &   $\omega$ &   $\omega$ & $\omega$ &  $\omega^2$ &  $\omega^2$ &  $\omega^2$ \\ 
$\mathbf{2}^{(2)}$ & 2 & 2 & 2 & 2 & $\text{-}2$ & $\text{-}2$ & $\text{-}2$ &   0 &    0 & 0 &   $\text{-}\omega^2$ &    $\text{-}\omega$ & $\text{-}\omega^2$ & $\text{-}\omega^2$ & $\text{-}\omega^2$ & $\text{-}\omega$ &    $\text{-}\omega$ &    $\text{-}\omega$ &  $\omega^2$ &  $\omega^2$ & $\omega^2$ &   $\omega$ &   $\omega$ &   $\omega$ \\ 
\hline
$\mathbf{3}^{(a)}$ & 3 & 3 & 3 & 3 & 3 & 3 & 3 & $\text{-}1$ & $\text{-}1$ & $\text{-}1$ & 0 & 0 & 0 & 0 & 0 & 0 & 0 & 0 & 0 & 0 & 0 & 0 & 0 & 0 \\
$\mathbf{3}^{(0)}$ & 3 &   $3\omega$ &  $3\omega^2$ & 0 & $\text{-}1$ & $\text{-}\omega$ &  $\text{-}\omega^2$ & 1 &   $\omega$ &  $\omega^2$ & 0 &  0 &   $\omega^2\mu^{\ast}$ &   $\mu^{\ast}$ & $\omega\mu^{\ast}$ &  $\mu$ &    $\omega\mu$ & $\omega^2\mu$ &   $\text{-}\zeta^2$ &   $\text{-}\zeta^{4*}$ & $\text{-}\zeta^{\ast}$ &    $\text{-}\zeta^{2*}$ &    $\text{-}\zeta$ &    $\text{-}\zeta^{4}$ \\
$\mathbf{3}^{(1)}$ & 3 &   $3\omega$ &  $3\omega^2$ &   0 & $\text{-}1$ & $\text{-}\omega$ &  $\text{-}\omega^2$ & 1 &   $\omega$ &  $\omega^2$ &    0 &    0 &   $\mu^{\ast}$ &   $\omega\mu^{\ast}$ & $\omega^2\mu^{\ast}$ &    $\omega^2\mu$ &    $\mu$ & $\omega\mu$ &   $\text{-}\zeta^{4*}$ &   $\text{-}\zeta^{\ast}$ & $\text{-}\zeta^2$ &    $\text{-}\zeta^{4}$ &    $\text{-}\zeta^{2*}$ & $\text{-}\zeta$ \\
$\mathbf{3}^{(2)}$ & 3 &   $3\omega$ &  $3\omega^2$ &   0 & $\text{-}1$ & $\text{-}\omega$ &  $\text{-}\omega^2$ & 1 &   $\omega$ &  $\omega^2$ &    0 &    0 &   $\omega\mu^{\ast}$ & $\omega^2\mu^{\ast}$ &   $\mu^{\ast}$ &    $\omega\mu$ & $\omega^2\mu$ &    $\mu$ &   $\text{-}\zeta^{\ast}$ & $\text{-}\zeta^2$ &   $\text{-}\zeta^{4*}$ &    $\text{-}\zeta$ & $\text{-}\zeta^{4}$ & $\text{-}\zeta^{2*}$ \\
$ \mathbf{3}^{(0)*}$ & 3 &  $3\omega^2$ &   $3\omega$ & 0 & $\text{-}1$ &  $\text{-}\omega^2$ &   $\text{-}\omega$ & 1 &  $\omega^2$ &   $\omega$ & 0 & 0 & $\omega\mu$ & $\mu$ & $\omega^2\mu$ &   $\mu^{\ast}$ &   $\omega^2\mu^{\ast}$ & $\omega\mu^{\ast}$ &    $\text{-}\zeta^{2*}$ &    $\text{-}\zeta^{4}$ & $\text{-}\zeta$ &   $\text{-}\zeta^2$ &   $\text{-}\zeta^{\ast}$ & $\text{-}\zeta^{4*}$ \\ 
$ \mathbf{3}^{(1)*}$ & 3 &  $3\omega^2$ &   $3\omega$ & 0 & $\text{-}1$ &  $\text{-}\omega^2$ &   $\text{-}\omega$ & 1 &  $\omega^2$ &   $\omega$ & 0 & 0 & $\mu$ &    $\omega^2\mu$ & $\omega\mu$ &   $\omega\mu^{\ast}$ &   $\mu^{\ast}$ & $\omega^2\mu^{\ast}$ &    $\text{-}\zeta^{4}$ &  $\text{-}\zeta$ & $\text{-}\zeta^{2*}$ &   $\text{-}\zeta^{4*}$ &   $\text{-}\zeta^2$ & $\text{-}\zeta^{\ast}$ \\
$ \mathbf{3}^{(2)*}$ & 3 &  $3\omega^2$ &   $3\omega$ & 0 & $\text{-}1$ &  $\text{-}\omega^2$ &   $\text{-}\omega$ & 1 &  $\omega^2$ &   $\omega$ & 0 & 0 &    $\omega^2\mu$ & $\omega\mu$ &    $\mu$ &   $\omega^2\mu^{\ast}$ & $\omega\mu^{\ast}$ &   $\mu^{\ast}$ &    $\text{-}\zeta$ & $\text{-}\zeta^{2*}$ &    $\text{-}\zeta^{4}$ &   $\text{-}\zeta^{\ast}$ & $\text{-}\zeta^{4*}$ &   $\text{-}\zeta^2$ \\
\hline
$\mathbf{6}^{(0)}$ & 6 & $6\omega$ &   $6\omega^2$ &   0 & 2 & $2\omega$ &  $2\omega^2$ & 0 & 0 & 0 & 0 & 0 &  $\text{-}\omega\mu^{\ast}$ & $\text{-}\omega^2\mu^{\ast}$ &  $\text{-}\mu^{\ast}$ &   $\text{-}\omega\mu$ & $\text{-}\omega^2\mu$ &   $\text{-}\mu$ &   $\text{-}\zeta^{\ast}$ & $\text{-}\zeta^2$ &   $\text{-}\zeta^{4*}$ &    $\text{-}\zeta$ & $\text{-}\zeta^{4}$ &    $\text{-}\zeta^{2*}$ \\
$\mathbf{6}^{(1)}$ & 6 &    $6\omega$ &   $6\omega^2$ &   0 & 2 & $2\omega$ &  $2\omega^2$ &   0 &    0 &    0 &    0 & 0 &  $\text{-}\omega^2\mu^{\ast}$ &  $\text{-}\mu^{\ast}$ & $\text{-}\omega\mu^{\ast}$ &   $\text{-}\mu$ &   $\text{-}\omega\mu$ &
$\text{-}\omega^2\mu$ &   $\text{-}\zeta^2$ &   $\text{-}\zeta^{4*}$ &   $\text{-}\zeta^{\ast}$ &    $\text{-}\zeta^{2*}$ &  $\text{-}\zeta$ & $\text{-}\zeta^{4}$ \\
$\mathbf{6}^{(2)}$ & 6 &    $6\omega$ &   $6\omega^2$ &   0 & 2 & $2\omega$ &  $2\omega^2$ &   0 &    0 &    0 &    0 & 0 &  $\text{-}\mu^{\ast}$ &  $\text{-}\omega\mu^{\ast}$ & $\text{-}\omega^2\mu^{\ast}$ &   $\text{-}\omega^2\mu$ &   $\text{-}\mu$ &
$\text{-}\omega\mu$ &   $\text{-}\zeta^{4*}$ &   $\text{-}\zeta^{\ast}$ & $\text{-}\zeta^2$ &    $\text{-}\zeta^{4}$ &    $\text{-}\zeta^{2*}$ & $\text{-}\zeta$ \\
$ \mathbf{6}^{(0)*}$ & 6 &   $6\omega^2$ &    $6\omega$ & 0 & 2 &  $2\omega^2$ & $2\omega$ & 0 & 0 & 0 & 0 & 0 & $\text{-}\omega^2\mu$ & $\text{-}\omega\mu$ &   $\text{-}\mu$ &  $\text{-}\omega^2\mu^{\ast}$ & $\text{-}\omega\mu^{\ast}$ &  $\text{-}\mu^{\ast}$ &    $\text{-}\zeta$ & $\text{-}\zeta^{2*}$ &    $\text{-}\zeta^{4}$ &   $\text{-}\zeta^{\ast}$ & $\text{-}\zeta^{4*}$ &   $\text{-}\zeta^2$ \\ 
$ \mathbf{6}^{(1)*}$ & 6 &   $6\omega^2$ & $6\omega$ & 0 & 2 &  $2\omega^2$ & $2\omega$ & 0 & 0 & 0 & 0 & 0 &   $\text{-}\omega\mu$ & $\text{-}\mu$ & $\text{-}\omega^2\mu$ &  $\text{-}\mu^{\ast}$ &  $\text{-}\omega^2\mu^{\ast}$ &
$\text{-}\omega\mu^{\ast}$ &    $\text{-}\zeta^{2*}$ &    $\text{-}\zeta^{4}$ & $\text{-}\zeta$ &   $\text{-}\zeta^2$ &   $\text{-}\zeta^{\ast}$ & $\text{-}\zeta^{4*}$ \\
$ \mathbf{6}^{(2)*}$ & 6 & $6\omega^2$ & $6\omega$ & 0 & 2 & $2\omega^2$ & $2\omega$ & 0 & 0 & 0 & 0 & 0 &   $\text{-}\mu$ & $\text{-}\omega^2\mu$ &   $\text{-}\omega\mu$ &  $\text{-}\omega\mu^{\ast}$ & $\text{-}\mu^{\ast}$ &  $\text{-}\omega^2\mu^{\ast}$ &    $\text{-}\zeta^{4}$ & $\text{-}\zeta$ &    $\text{-}\zeta^{2*}$ &   $\text{-}\zeta^{4*}$ & $\text{-}\zeta^2$ & $\text{-}\zeta^{\ast}$\\
\hline
$\mathbf{8}^{(0)}$ & 8 & 8 & 8 & $\text{-}1$ & 0 & 0 & 0 & 0 & 0 & 0 & $\text{-}1$ & $\text{-}1$ & 2 & 2 & 2 & 2 & 2 & 2 & 0 & 0 & 0 & 0 & 0 & 0 \\
$\mathbf{8}^{(1)}$ & 8 & 8 & 8 & $\text{-}1$ & 0 & 0 & 0 & 0 & 0 & 0 & $\text{-}\omega$ & $\text{-}\omega^2$ &    $2\omega$ &    $2\omega$ &    $2\omega$ & $2\omega^2$ &   $2\omega^2$ &   $2\omega^2$ &    0 &    0 &    0 & 0 &    0 &    0 \\
$\mathbf{8}^{(2)}$ & 8 & 8 & 8 & $\text{-}1$ & 0 & 0 & 0 & 0 & 0 & 0 & $\text{-}\omega^2$ & $\text{-}\omega$ &   $2\omega^2$ &   $2\omega^2$ &   $2\omega^2$ & $2\omega$ &    $2\omega$ &    $2\omega$ &    0 &    0 &    0 &    0 & 0 &    0 \\
\hline
$\mathbf{9}$ & 9 &    $9\omega$ &   $9\omega^2$ &   0 & $\text{-}3$ & $\text{-}3\omega$ &  $\text{-}3\omega^2$ & $\text{-}1$ &    $\text{-}\omega$ &   $\text{-}\omega^2$ & 0 & 0 & 0 & 0 & 0 & 0 & 0 & 0 & 0 & 0 & 0 & 0 & 0 & 0 \\
$\mathbf{9}^{\ast}$ & 9 &   $9\omega^2$ &    $9\omega$ &   0 & $\text{-}3$ & $\text{-}3\omega^2$ & $\text{-}3\omega$ & $\text{-}1$ & $\text{-}\omega^2$ & $\text{-}\omega$ & 0 & 0 & 0 & 0 & 0 & 0 & 0 & 0 & 0 & 0 & 0 & 0 & 0 & 0
\\
\end{tabular}
\end{table*}

Many encodings exist, such as a random assignment, a binary code, or a Gray code that minimizes bit flips. Here, we use an encoding scheme that has proven useful for primitive gates in the \ac{DSA}: the \ac{SGS}.
An \ac{SGS} is built by systematically adding group elements to extend the groups in a series while maintaining a fixed ordered product form.
For $\dsb$, an \ac{SGS} is:
\begin{align}
    \label{eq:orderedproduct}
    g = \omega^p C^q E^r V^{2s}V^{t} X^u D^y,
\end{align}
where $0\leq p, q, r, y\leq 2$ and $0\leq s,t, u\leq 1$. This ordered product corresponds to the series of finite groups
\begin{align}
    \mathbb{Z}_3 \rightarrow \mathbb{Z}_3&\otimes\mathbb{Z}_3\rightarrow \Delta(27) \rightarrow 
    \Delta(54) \notag\\&\rightarrow \Sigma(36\times3)\rightarrow\Sigma(72\times3) \rightarrow \dsb,
\end{align}
where moving to the next largest group corresponds to adding the next operator to the right in Eq.~(\ref{eq:orderedproduct}).

Writing $g$ this way allows encoding $\ket{g}$ on a register of qutrits and qubits, which we call a mixed register:
\begin{align}
    \ket{g} = \ket{p}_3\ket{q}_3\ket{r}_3\ket{s}_2\ket{t}_2\ket{u}_2\ket{y}_3,
\end{align}
where the subscripts indicate the qudit dimension. This mixed encoding suits the group arithmetic. For hardware without native qutrits we also use a qubit-only encoding, mapping each qutrit to two qubits:
\begin{align}
    \ket{x}_3 \rightarrow \ket{x_1}_2\ket{x_2}_2
\end{align}
where $x =2x_1 + x_2$. This mapping comes at the cost of ``forbidden'' states $\ket{x_1x_2}=\ket{11}$ which a quantum simulation should never have.  

We also need an explicit unitary form of the 24 irreducible representations (irreps) of $\dsb$ listed in Tab.~\ref{tab:cts648}. We denote them as $\#^{(x)}$, where $\#$ is the irrep dimension and $x=0,1,2$ indexes the irreps. The notation $\rho_{\#^{(x)}}(g)$ denotes the matrix of $g$ in the irrep $\#^{(x)}$. We list the generators of $\dsb$ in each irrep in the appendix (Tab.~\ref{tab:irrepss648p1}), reproduced in part from~\cite{Grimus:2010ak}.


\section{Basic Quantum Gates}
\label{sec:quantumgates}

Given the primitive gate decompositions in later sections have both homogeneous qubit and mixed versions, we discuss both gate sets following the ideas of~\cite{Gustafson:2024kym}.
The basic gate set for qubits involves the standard $R_Z$, Clifford, and controlled-NOT (\textnormal{\textsc{cnot}}) gate set.
Arbitrary single qubit gates can be composed of at most 3 $R_Z$ rotations, while arbitrary two qubit gates can be decomposed into 15 $R_Z$ rotations up to a global phase~\cite{Shende:2004gqq,Vatan:2004nmz}. 

For fault tolerance each $R_Z$ is approximated by Clifford+$T$ gates, the gate set available for many error-correcting codes, following~\cite{Bocharov:2014awg}, which uses $1.15\log_2(1/\epsilon)+9.2$ $T$ gates per executed rotation on average at precision $\epsilon$; in an error-corrected computation the $T$ gate is the costly operation and sets our resource count. The Toffoli gate (\textnormal{\textsc{c}}$^2$\textnormal{\textsc{not}}) decomposes into \textnormal{\textsc{cnot}}s and $T$-gates~\cite{nielsen_chuang_2010}, and higher-order \textnormal{\textsc{c}}$^n$\textnormal{\textsc{not}} into an exact nested staircase~\cite{nielsen_chuang_2010,2019arXiv190401671B,PhysRevA.52.3457} at the cost of clean ancillas. The controlled-\textnormal{\textsc{swap}} (\textnormal{\textsc{cswap}}, or Fredkin) gate is also required. These costs are collected in Tab.~\ref{tab:qubitbasicresources}.

\begin{table}[!th]
\caption{Resources for \ac{FT} qubit (top) and qutrit and mixed (bottom) gates~\cite{Bocharov:2014awg,PhysRevA.52.3457,2019arXiv190401671B,Gustafson:2025buk,Bocharov:2018zfk,Yeh:2022eow}. Multi-controlled \textnormal{\textsc{c}}$^n$\textnormal{\textsc{sum}} costs are collected in Tab.~\ref{tab:csumcostsqutrit}.}
\label{tab:qubitbasicresources}
\begin{tabular}{lcc}
\hline\hline
Gate & Cost & clean ancillas\\\hline
$R_Z$ & $1.15\log_2(1/\epsilon)+9.2\,T$ & 0\\
\textnormal{\textsc{c}}$^2$\textnormal{\textsc{not}} & $7\,T$ & 0\\
\textnormal{\textsc{c}}$^n$\textnormal{\textsc{not}} & $(2n-3)\cdot7\,T$ & $(n-2)$\\
\textnormal{\textsc{cswap}} & $7\,T$ & 0\\
\hline
$R_Z^{(a,b)}$ & $3.20 + 5.139\log_3(1/\epsilon)\,R$ & 0\\
single-qutrit $SU(3)$ & $6\,R_Z^{(a,b)}$ & 0\\
$T_{(3)}$ & $1\,R_Z^{(a,b)}$ & 0\\
$\chi$ (increment) & free (qutrit Clifford) & 0\\
$C_a^i X_b^{(j,k)}$ (mixed) & $1\,E_{(2,3)}$ or $1\,E_{(3,2)}$ & 0\\
\hline\hline
\end{tabular}
\end{table}

To extend to qutrits, additional gates are necessary: single-qutrit and mixed operations. An arbitrary single-qutrit $U(3)$ operation can be decomposed into 9 two-level rotations $R_{\alpha}^{(a,b)}$ with $\alpha\in\lbrace X,Y,Z\rbrace$, each applying an $R_{\alpha}(\theta)$ rotation on the two-level subspace spanned by $\ket{a}$ and $\ket{b}$.  Another important single-qudit gate, $\chi$, is the increment gate:
\begin{equation}
\label{eq:chival}
    \chi\ket{a}=\ket{a\oplus_d 1}
\end{equation}
where $d$ indicates the dimension of the qudit as well as the modulus of the addition.  We also use its controlled version $C\chi$ and the two-qudit entangling gate,
\begin{equation}
\label{eq:qutritcx}
    C_a^i X_{b}^{(j,k)}\,,
\end{equation}
which, controlled on state $i$ of qudit $a$, applies to qudit $b$ the transposition $X^{(j,k)}$ that exchanges the levels $j$ and $k$. For two qubits this is the usual \textnormal{\textsc{cnot}}, $C_aX_b = C_a^1X_b^{(0,1)}$, and we suppress the labels there. When counting mixed resources, the two mixed cases are tallied separately as the entanglers $E_{(d,f)}$, whose subscripts are the control and target dimensions $d,f\in\{2,3\}$: $E_{(2,3)}$ (qubit control, qutrit target) and $E_{(3,2)}$ (qutrit control, qubit target) (see App.~\ref{app:toffolistaircase}). Using $C_a^iX_b^{(j,k)}$, we decompose the qutrit $\chi$ onto a two-qubit register, with the value encoded as $2a+b$ and $\ket{11}$ unused:
\begin{equation}
\label{eq:chivalqubit}
    \chi\ket{a}\ket{b}=(C_bX_a)(C_aX_b)(X\otimes\mathbbm{1})\ket{a}\ket{b},
\end{equation}
Another related gate is the controlled-sum (\textnormal{\textsc{csum}}) gate:
\begin{equation}
    \CnSum{csum}{a}{b}\ket{i}_a\ket{j}_b = \ket{i}_a \ket{i \oplus_d j}_b,
\end{equation}
where $d$ is the dimension of the target $b$: the all-qubit case is the \textnormal{\textsc{cnot}}, and the control of a qutrit-target \textnormal{\textsc{csum}} may be a qubit or a qutrit (a qubit control leaves the $i=2$ term unused). For a qutrit target, it decomposes into $C_{a}^{i}X_{b}^{(j,k)}$ gates following Ref.~\cite{Di:2011cvl}:
 \begin{align}
     \CnSum{csum}{a}{b} &= \left(C_a^1 X_b^{(0,1)} C_a^1 X_b^{(1,2)}\right) \left(C_a^2 X_b^{(1,2)} C_a^2 X_b^{(0,1)}\right).
 \end{align}
 We also need multi-controlled gates:
the doubly controlled transposition $C_a^iC_b^jX_c^{(k,l)}$, abbreviated $C_{a,b}^{i,j}X_c^{(k,l)}$, and
\begin{equation}
    \CnSum{ccsum}{a,b}{c}\ket{i}_a\ket{j}_b\ket{k}_c = \ket{i}_a\ket{j}_b\ket{k\oplus_d ij}_c,
\end{equation}
which adds the product of the two control values to the target of dimension $d$. The Toffoli of Tab.~\ref{tab:qubitbasicresources} is the all-qubit case $\CnSum{ccsum}{2,2}{2}$.

In circuit diagrams we mark controls three ways: a black circle with a printed value fires the target only on that state; a white circle with a printed value fires on every state except the one listed; and a plain control dot on a qutrit applies the target once if the control is $\ket{1}$ and twice if it is $\ket{2}$. A slash through a wire marked $n$ denotes a bundle of $n$ wires drawn as one line.

Most proposals for \ac{FT} qutrit computation assume the qutrit Clifford group is implemented transversally in the logical space. Thus the $R_{\alpha}^{(a,b)}$ and multi-control gates will need to be compiled, approximately or exactly, in terms of them and a non-Clifford supplement: the qutrit $T$-gate $T_{(3)}$~\cite{Yeh:2022eow,Saha:2026blc} or the metaplectic $R$ below.  An arbitrary single-qutrit $SU(3)$ gate decomposes into six $R_{Z}^{(a,b)}$ and qutrit Clifford gates~\cite{Bocharov:2018zfk}. The average cost for synthesizing an $R_{Z}^{(a,b)}$ from Clifford extended by the metaplectic $R=\operatorname{Diag}(1, 1, -1)$~\cite{Bocharov:2018zfk,Glaudell:2022vvu} is $3.20(13) + 5.139(14)\log_3(1/\epsilon)$~\cite{Gustafson:2025buk}. Below we drop these uncertainties, which are negligible at the precision of our estimates.

\section{Primitive Gates for lattice gauge theories}
\label{sec:primitivegates}

The pure \ac{LGT} Hamiltonian can be written:
\begin{equation}
    \label{eq:hamiltonian}
    H = H_E + H_B
\end{equation}
where $H_E$ is the electric field component and $H_B$ is the magnetic field component of the Hamiltonian. For a discrete group the electric term on one link and its eigenvalues in the irrep basis are, in units of $g_0^2/2$,
\begin{equation}
\label{eq:electricterm}
H_E=\sum_{h\in\Gamma}\sum_{g}\ket{hg}\bra{g},\qquad \lambda_\rho=|\Gamma|-\frac{1}{d_\rho}\sum_{h\in\Gamma}\operatorname{tr}\rho(h),
\end{equation}
with $\Gamma$ the $\mathcal{N}$ nearest neighbors of the identity (Tab.~\ref{tab:subgroups}) and $\lambda_\rho$ listed in Tab.~\ref{tab:eigenvalues}. We use the definition of the electric term from Ref. \cite{PhysRevD.107.114513}.

The simplest choice is the Kogut--Susskind form~\cite{PhysRevD.11.395}: a single-link operator for $H_E$ and a plaquette for $H_B$. Any operators that approach the continuum electric and magnetic fields are allowed; each decomposes into primitive gates~\cite{Lamm:2019uyc}, from which every gauge-invariant term can be built. 

Quantum simulation of a pure \ac{LGT} needs five primitive gates: the inversion ($\mathfrak{U}_{-1}$), multiplication ($\mathfrak{U}_{\times}$), trace ($\mathfrak{U}_{\rm Tr}$), Fourier transform ($\mathfrak{U}_{F}$), and the kinetic-phase gate ($\mathfrak{U}_{Ph}$)~\cite{Lamm:2019bik}. 

\begin{table*}[!t]
    \caption{Eigenvalues $\lambda_\rho$ of the electric Hamiltonian $H_E$ of Eq.~(\ref{eq:electricterm}) normalized by $g_0^2/2$, for $\Gamma$ the $24$ nearest neighbors of the identity (the classes of $\omega^2D$ and $\omega^2D^2$), with $\kappa_1=24(1 - \frac{1}{\sqrt{3}}\cos(\pi / 18))$, $\kappa_2=24(1 + \frac{1}{\sqrt{3}}\sin(\pi / 9))$, $\kappa_3=24(1 + \frac{1}{\sqrt{3}}\sin(2\pi / 9))$, $\kappa_4=24(1 - \frac{1}{2\sqrt{3}}\sin(2\pi / 9))$, $\kappa_5=24(1 + \frac{1}{2\sqrt{3}}\cos(\pi / 18))$, $\kappa_6=24(1 - \frac{1}{2\sqrt{3}}\sin(\pi / 9))$.}
    \label{tab:eigenvalues}
\centering
\setlength{\tabcolsep}{2.2pt}
\begin{tabular}{c|cccccccccccccccccccccccc}
\hline\hline
$\rho$ & $\mathbf{1}^{(0)}$ & $\mathbf{1}^{(1)}$ & $\mathbf{1}^{(2)}$ & $\mathbf{2}^{(0)}$ & $\mathbf{2}^{(1)}$ & $\mathbf{2}^{(2)}$ & $\mathbf{3}^{(0)}$ & $\mathbf{3}^{(1)}$ & $\mathbf{3}^{(2)}$ & $\mathbf{3}^{(0)*}$ & $\mathbf{3}^{(1)*}$ & $\mathbf{3}^{(2)*}$ & $\mathbf{3}^{(a)}$ & $\mathbf{6}^{(0)}$ & $\mathbf{6}^{(1)}$ & $\mathbf{6}^{(2)}$ & $\mathbf{6}^{(0)*}$ & $\mathbf{6}^{(1)*}$ & $\mathbf{6}^{(2)*}$ & $\mathbf{8}^{(0)}$ & $\mathbf{8}^{(1)}$ & $\mathbf{8}^{(2)}$ & $\mathbf{9}$ & $\mathbf{9}^{*}$\\\hline
$\lambda_\rho$ & 0 & 36 & 36 & 36 & 18 & 18 & $\kappa_1$ & $\kappa_2$ & $\kappa_3$ & $\kappa_1$ & $\kappa_2$ & $\kappa_3$ & 24 & $\kappa_4$ & $\kappa_5$ & $\kappa_6$ & $\kappa_4$ & $\kappa_5$ & $\kappa_6$ & 18 & 27 & 27 & 24 & 24\\
\hline\hline
\end{tabular}
\end{table*}

These primitive gates can be grouped into one of two kinds of operations: permutations and transformations. 
The permutation operations are $\mathfrak{U}_{-1}$ and $\mathfrak{U}_{\times}$, which map group-element basis states to group-element basis states. The transformation operations are $\mathfrak{U}_{\rm Tr}$, $\mathfrak{U}_{F}$, and $\mathfrak{U}_{Ph}$, which either change basis or apply state-dependent phases. 

The inversion gate $\mathfrak{U}_{-1}$ maps $g$ to its inverse:
\begin{align}
    \mathfrak{U}_{-1}\ket{g} = \ket{g^{-1}}.
\end{align}
$\mathfrak{U}_{\times}$ takes two group elements, $g$ and $h$, and outputs the left multiplication $hg$:
\begin{align}
    \mathfrak{U}_{\times}\ket{g}\ket{h} = \ket{hg}\ket{h}. 
\end{align}
$\mathfrak{U}_{\times}$ is a controlled operation: the $h$ register is the control and the $g$ register is the target. 

The first transformation operation, $\mathfrak{U}_{\rm Tr}(\theta)$, phases each group basis state by the real part of its trace:
\begin{align}
    \mathfrak{U}_{\rm Tr}(\theta)\ket{g} = e^{-i\theta\,\re\tr(g)}\ket{g}.
\end{align}
This operator is combined with $\mathfrak{U}_{\times}$ and $\mathfrak{U}_{-1}$ to construct the magnetic field terms in the Kogut--Susskind and improved Kogut--Susskind Hamiltonians~\cite{Kogut:1980qb, Carena:2022kpg}.

The two other gates, $\mathfrak{U}_F$ and $\mathfrak{U}_{Ph}$, combine to implement the electric portion of the time evolution: $H_E$ is diagonal in the irrep basis, so
\begin{align}
    e^{-i\tau H_{E}} = \mathfrak{U}_F^{\dagger}\,\mathfrak{U}_{Ph}(\tau)\,\mathfrak{U}_F.
\end{align}
The gate $\mathfrak{U}_F$ is the group Fourier transform, taking a group element to its weight in each irrep, $\mathfrak{U}_F\ket{g} = \sum_{\rho}\sqrt{\tfrac{d_\rho}{|G|}}\sum_{i,j}\rho_{ij}(g)\ket{\rho ij}$, summed over the irreps $\rho$ of $G$, with $d_\rho$ the dimension of $\rho$. 
An efficient implementation, the \ac{FFT} $\mathfrak{U}_{FFT}$, reduces the cost from the $O(|G|^2)$ of the direct transform to $O(|G|\log|G|)$ by building the transform along a nested series of subgroups~\cite{kitaev,NC,hales2000improved,Hoyer:1997qc,beals1997quantum,Pueschel:1998zzo,Murairi:2024xpc, moore2003generic,Bruinsma:2026fke}. We construct $\mathfrak{U}_{FFT}$ in Sec.~\ref{sec:fouriertransform}. 
The operator $\mathfrak{U}_{Ph}(\tau)$ applies phases corresponding to the eigenvalues of the electric field operator in irrep space:
\begin{align}
    \mathfrak{U}_{Ph}(\tau)\ket{\rho ij} = e^{-i\tau \lambda_{\rho}}\ket{\rho ij},
\end{align}
where $\lambda_{\rho}$ is the eigenvalue of $H_E$ on the irrep $\rho$ (Tab.~\ref{tab:eigenvalues}).

\section{Inversion Gate}
\label{sec:inversion}

The inversion gate is the simplest of the five.
It permutes the group-element basis, sending each element to its inverse.
We start by writing out the initial effect of inverting a group element
\begin{align}
    g^{-1} = & (\omega^p C^q E^r V^{2s + t} X^u D^y)^{-1}\notag\\
    = &  D^{-y} (V^{-2s - t} E^{-r} C^{-q} \omega^{-p}).
\end{align}
The end goal is to rearrange these operators to the ordered product form given in Eq.~(\ref{eq:orderedproduct}). 
The parenthesized product is the inverse of a $\dsa$ element~\cite{Perez:2025cxl}.
This lets us write down the intermediate state
\begin{align}
    g^{-1}=D^{-y} (\omega^{p} C^{q} E^{r} V^{2s + t} X^{u}),
\end{align}
where $p,~q,~r,~s,~t,$ and $u$ now denote the register values after $\mathfrak{U}_{-1}^{\dsa}$. From here on, each stage updates the registers in place, written $x \mapsto f(x)$: every right-hand side is evaluated on the values entering that stage, registers not listed are unchanged, and arithmetic is mod $3$ on the qutrit registers ($p$, $q$, $r$, $y$) and mod $2$ on the qubit registers ($s$, $t$, $u$).

We then follow the same propagation algorithm as for $\dsa$. The first step writes $D^{-y}$ in terms of existing operators:
\begin{align}
\label{eq:propinv0_b}
    D^{-y} = \omega^{\delta_{y,2} + \delta_{y,1}}D^{2y \bmod 3}.
\end{align}
This identity follows from $D^{3}=\omega^{2}\mathbbm{1}$, so
\begin{align}
    g^{-1} = &\omega^{p}D^{y}C^{q}E^{r}V^{2s+t}X^{u}\notag
\end{align}
with the first stage applying
\begin{align}
    \mathfrak{U}_{-1}^{(1)}:\quad y \mapsto 2y, \qquad p \mapsto p + \delta_{y,1}+\delta_{y,2}.
\end{align}
Since $C$ and $D$ commute, this move requires no integer arithmetic.
The next step is moving $D^{y}$ through $E^{r}$:
\begin{align}
    g^{-1} =  &\omega^{p}C^{q}E^{r}D^{y}V^{2s+t}X^{u}\notag
\end{align}
with
\begin{align}
    \mathfrak{U}_{-1}^{(2)}:\quad p \mapsto p + 2y \delta_{r,2}, \qquad q \mapsto q + 2yr.
\end{align}
After this we propagate $D^{y}$ through $V^{2s}$:
\begin{align}
    \mathfrak{U}_{-1}^{(3)}:\quad p \mapsto p + 2rys, \qquad q \mapsto q + 2ys.
\end{align}
The remaining propagation uses a different scheme because of the unusual transformation $DV = XD$.

\begin{figure}[ht]
\includegraphics[width=\linewidth]{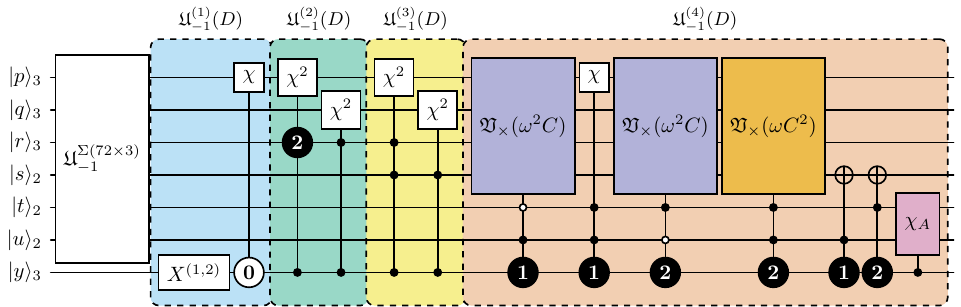}\\[0.6em]
\includegraphics[width=0.44\linewidth]{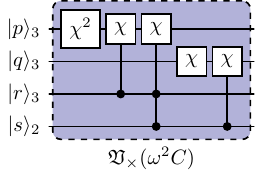}
\includegraphics[width=0.49\linewidth]{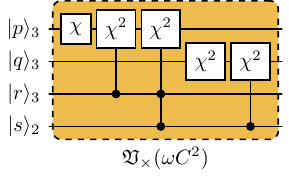}
\includegraphics[width=0.35\linewidth]{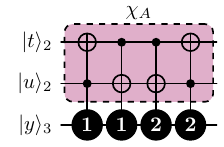}
    \caption{(Top): mixed implementation of $\mathfrak{U}_{-1}$, with the control markings of Sec.~\ref{sec:quantumgates}. (Middle left): the $\mathfrak{V}_{\times}(\omega^2C)$ gate; (Middle right): the $\mathfrak{V}_{\times}(\omega C^2)$ gate; (Bottom): $C\chi_A$, Eq.~(\ref{eq:chiA}). The increment $\chi$ is defined in Eq.~(\ref{eq:chival}).}
    \label{fig:dsbinversequtrit}
\end{figure}
\begin{figure*}[!ht]
\centering
\includegraphics[width=0.95\linewidth]{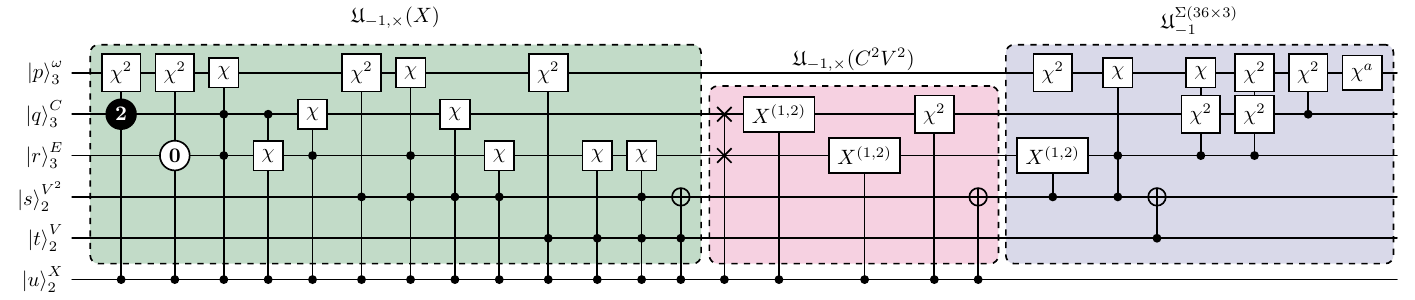}
\caption{The inner $\mathfrak{U}_{-1}^{\dsa}$ block of Fig.~\ref{fig:dsbinversequtrit}, in three stages: $\mathfrak{U}_{-1,\times}(X)$ strips $X^u$, $\mathfrak{U}_{-1,\times}(C^2V^2)$ absorbs the carry, and $\mathfrak{U}_{-1}^{\Sigma(36\times3)}$ inverts the remaining \ac{SGS} recursion.}
\label{fig:dsainnerhet}
\end{figure*}
We work through the six cases with $y\neq0$ and $(t,u)\neq(0,0)$. Normal ordering the occurrences of $D^y$, $X^u$, and $V^t$ gives:
\begin{align}
\label{eq:sixrules}
    DV = & XD,&&
    DX = \omega^2CV^2 (VX) D,\notag\\
    DVX = &\omega V^2 (V) D,&&
    D^2V= \omega^2 C V^2 (VX) D^2,\notag\\
    D^2X= &(V)D^2,&&
    D^2VX=\omega C^2V^2 (X) D^2.
\end{align}

\begin{figure*}[!ht]
    \includegraphics[width=\linewidth]{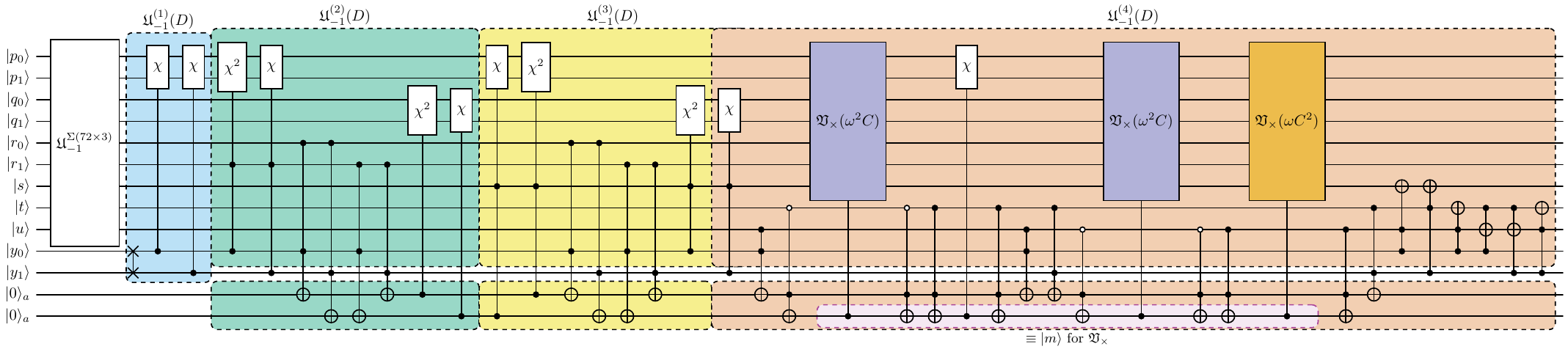}
    \caption{Qubit decomposition of $\mathfrak{U}_{-1}$; the qubit decomposition of $\chi$ is Eq.~(\ref{eq:chivalqubit}).}
    \label{fig:dsbqubitinverse}
\end{figure*}

These transformations require multiplications on the $\omega$, $C$, $E$, $V^2$ registers, controlled on the cases of $y$, $u$, and $t$. We denote right multiplication of these registers by an element $h$ as $\mathfrak{V}_{\times}(h)$. The circuits below use $\mathfrak{V}_{\times}(\omega^2C)$ and $\mathfrak{V}_{\times}(\omega C^2)$. These rules give the final stage:
\begin{align}
    \mathfrak{U}_{-1}^{(4)}:\quad
    p \mapsto & \, p +a_p((1-t)u\delta_{y,1}+t(1-u)\delta_{y,2})\notag\\
    &+b_p\,tu\,\delta_{y,1}+ c_p\,tu\,\delta_{y,2}\notag\\
    q \mapsto & \, q+ a_q((1-t)u\delta_{y,1} +t(1-u)\delta_{y,2})\notag\\
    &+b_q\,tu\,\delta_{y,2}\notag\\
    s \mapsto & \, s+u\,\delta_{y,1}+t\,\delta_{y,2},
\end{align}
where
\begin{align}
    a_p =&2 + r(1+s), & b_p=&1, & c_p=&1+r(2-s),\notag\\
    a_q=&1+s, & b_q=&2-s. &&&
\end{align}
Finally, the $\ket{t}_2\ket{u}_2$ subspace is permuted by the controlled permutation $C\chi_A$, which, depending on the state $y$, applies the 3-cycles
\begin{align}
\label{eq:chiA}
    y=1:&&\ket{10} \rightarrow \ket{01} \rightarrow \ket{11} \rightarrow \ket{10} \notag\\
    y=2:&&\ket{10} \rightarrow \ket{11} \rightarrow \ket{01} \rightarrow \ket{10}.
\end{align}

Using these rules, the mixed circuit implementing the inversion is provided in Fig.~\ref{fig:dsbinversequtrit}. The inner $\mathfrak{U}_{-1}^{\dsa}$ block is assembled in Fig.~\ref{fig:dsainnerhet} from the piecewise mixed forms given for $\dsa$ and $\dsd$ in Refs.~\cite{Perez:2025cxl,Gustafson:2024kym}, drawn here as one circuit.

Using the decomposition rules for qutrits to qubits provided in App.~\ref{app:toffolistaircase}, we also construct a qubit circuit, provided in Fig.~\ref{fig:dsbqubitinverse}. The decompositions of $\mathfrak{V}_{\times}(\omega^2 C)$ and $\mathfrak{V}_{\times}(\omega C^2)$ are in Fig.~\ref{fig:mathfrakvqubit}.
Implementing $\mathfrak{U}_{-1}$ admits a space-time trade-off.
The gates $\mathfrak{V}_{\times}(\omega^2 C)$ and $\mathfrak{V}_{\times}(\omega C^2)$ compute the same parities of $r_0 s$ and $r_1 s$, so two of the ancillas can stay computed between them, saving 8 Toffoli gates ($56\,T$) per $\mathfrak{U}_{-1}$ call. 

\begin{figure}[!ht]
\includegraphics[width=0.9\linewidth]{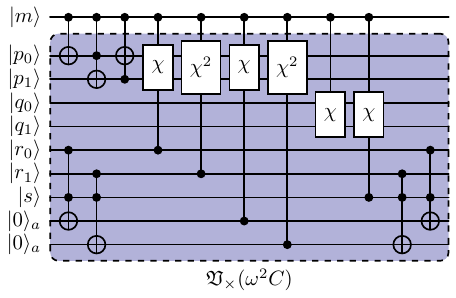}
\includegraphics[width=0.9\linewidth]{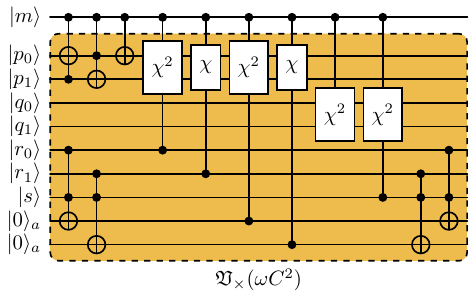}
\caption{Qubit decompositions of $\mathfrak{V}_{\times}(\omega^2C)$ (top) and $\mathfrak{V}_{\times}(\omega C^2)$ (bottom). The control $\ket{m}$ is computed in the highlighted region of Fig.~\ref{fig:dsbqubitinverse}.}
\label{fig:mathfrakvqubit}
\end{figure}

\section{Multiplication Gate}
\label{sec:multiply}

Multiplication extends the inversion algorithm: instead of a fixed permutation of $g$, it is a controlled map taking $g$ to $hg$.
We encode $g$ in the register $\ket{p_g q_g r_g s_g t_g u_g y_g}$
according to Eq.~(\ref{eq:orderedproduct}), and analogously for $h$.
We split $h$ into its $\dsa$ factor and $D^{y_h}$; the $\dsa$ factor is handled by $\mathfrak{U}_{\times}^{\dsa}$, so only $D^{y_h}$ needs new work.
Moving $D^{y_h}$ rightward through the $\ket{g}$ register applies the inversion's stage updates to the $g$ registers, with the control $y$ replaced by $y_h$. Stage $\mathfrak{U}_{-1}^{(2)}$ carries $D^{y_h}$ through $\omega^{p_g}C^{q_g}E^{r_g}$, $\mathfrak{U}_{-1}^{(3)}$ through $V^{2s_g}$, and $\mathfrak{U}_{-1}^{(4)}$ through $V^{t_g}X^{u_g}$. The normal-ordering relations of Eq.~(\ref{eq:sixrules}), and hence the update rules and coefficients, are unchanged. The one new step is combining the elements $D^{y_h}D^{y_g}$:
\begin{align}
    p_g &\mapsto p_g + 2\bigl(\delta_{y_g,2}(\delta_{y_h,1}+\delta_{y_h,2})+\delta_{y_g,1}\delta_{y_h,2}\bigr)\notag\\
    y_g &\mapsto y_g + y_h.
\end{align}
The $p_g$ update is the $\omega^{2}$ carry from $D^{3}=\omega^{2}\mathbbm{1}$, present whenever $y_h + y_g \geq 3$.
These results give the circuit implementing $\mathfrak{U}_{\times}^{\dsb}$ directly. The mixed circuit is in Fig.~\ref{fig:dsbmultiplyqutrit} and the qubit circuit in Fig.~\ref{fig:dsbmultiplyqubit}. Both reuse the stage gates $\mathfrak{U}_{-1}^{(n)}(D)$ of Sec.~\ref{sec:inversion} (Fig.~\ref{fig:dsbinversequtrit}). The mixed $\mathfrak{U}_{\times}^{\Sigma(72\times3)}$ of Fig.~\ref{fig:dsbmultiplyqutrit} is decomposed in App.~\ref{app:het_sigma72x3}.

\begin{figure}[!ht]
\includegraphics[width=\linewidth]{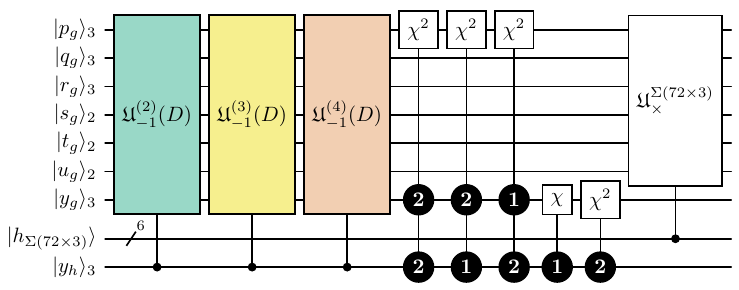}
\caption{Mixed-register implementation of $\mathfrak{U}_{\times}^{\dsb}$. The circuits for $\mathfrak{U}_{-1}^{(n)}(D)$ are provided in Fig.~\ref{fig:dsbinversequtrit}.}
\label{fig:dsbmultiplyqutrit}
\end{figure}

\begin{figure}[!ht]
\includegraphics[width=\linewidth]{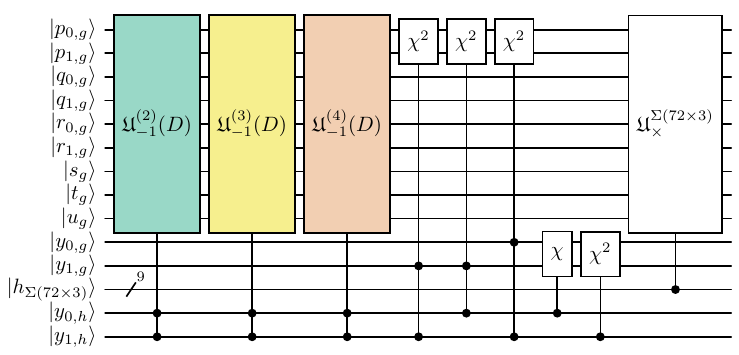}
\caption{Qubit decomposition of $\mathfrak{U}_{\times}^{\dsb}$; $\mathfrak{U}_{-1}^{(n)}(D)$ is in Fig.~\ref{fig:dsbqubitinverse}.}
\label{fig:dsbmultiplyqubit}
\end{figure}
\section{Trace Gate}
\label{sec:trace}

\begin{table}[!ht]
\caption{Map for the group elements to an ancilla register for $\dsb$. The values of $a_n$ correspond to $\re(-\zeta^{2^{n-1}})$ and the values of $b_n$ correspond to $\re(\omega^{n-1}\mu)$, with $\zeta$ and $\mu$ as defined in Tab.~\ref{tab:cts648}.}
\label{tab:tracedsb}
\begin{tabular}{c|ccccccccccccc}
   \hline\hline $\re\tr(g)$  & $3$ & $-\frac{3}{2}$ & $-1$ & $-\frac{1}{2}$ & $0$ & $1$ & $\frac{1}{2}$ & $a_1$ & $a_2$ & $a_3$ & $b_1$ & $b_2$ & $b_3$  \\\hline
   $v_0$ & 0 & 1 & 0 & 1 & 0 & 1 & 0 & 1 & 0 & 1 & 0 & 1 & 0\\
   $v_1$ & 0 & 0 & 1 & 1 & 0 & 0 & 1 & 1 & 1 & 0 & 1 & 0 & 0\\
   $v_2$ & 0 & 0 & 0 & 0 & 1 & 1 & 1 & 0 & 1 & 1 & 0 & 0 & 1\\
   $v_3$ & 0 & 0 & 0 & 0 & 0 & 0 & 0 & 1 & 1 & 1 & 1 & 1 & 1\\\hline\hline
\end{tabular}
\end{table}

The trace operation as worked out in previous work~\cite{Alam:2021uuq,Gustafson:2023kvd, Gustafson:2024kym} is the optimal procedure in an asymptotic sense for $R_Z$ synthesis and is shown schematically in Fig.~\ref{fig:traceexample}. It maps each $g$ to its trace class $C_{A}$ ($A$ indexes the class) via a gate $U_{\rm squish}^G$, performs the phase rotation $e^{-i\theta\re\tr g}$, then uncomputes the class register with $U_{\rm squish}^{G\dag}$.

\begin{figure}[!ht]
\includegraphics[width=0.8\linewidth]{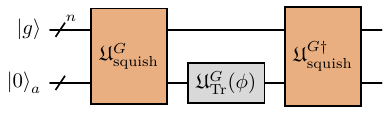}
\caption{Schematic of  $\mathfrak{U}_{\rm Tr}$ for an arbitrary group $G$. $\mathfrak{U}_{\rm{squish}}^{G}$ maps the group elements to their trace classes and the gate $\mathfrak{U}_{\rm Tr}^{G}(\theta)$ phases them by $e^{-i\theta\re\tr g}$.}
\label{fig:traceexample}
\end{figure}

\begin{figure}
    \centering
    \includegraphics[width=\linewidth]{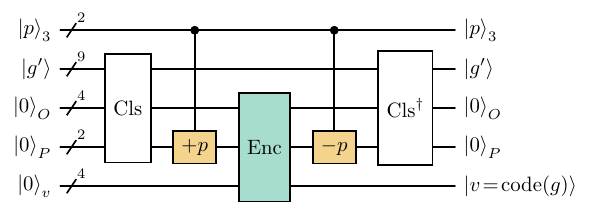}
    \caption{Qubit implementation of $\mathfrak{U}_{\rm{squish}}$, factored into the classifier $\mathrm{Cls}$ (Fig.~\ref{fig:squish_classifier}), the adder $\pm p$ (Fig.~\ref{fig:squish_adder}), and the encoder $\mathrm{Enc}$ (Fig.~\ref{fig:squish_encoder}).}
    \label{fig:s216x3tracesquishdiagram}
\end{figure}

\begin{figure*}
    \centering
    \includegraphics[width=\linewidth]{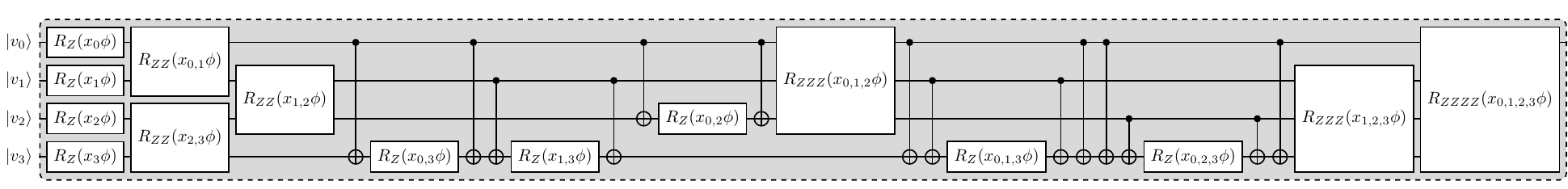}
    \caption{Qubit implementation of the phasing operation on the ancilla register $\ket{v_0}\ket{v_1}\ket{v_2}\ket{v_3}$; each box is a Pauli-$Z$ product rotation on the qubits named by its angle label, with the angles of Tab.~\ref{tab:coefficientsdsb}.}
    \label{fig:dsbphasetrace}
\end{figure*}

Beyond the 10 trace classes of $\dsa$, $\dsb$ has 14 more, $C_{11}$ through $C_{24}$ of Tab.~\ref{tab:cts648}, which introduce 6 new real trace values, $\re(\mu^*)$, $\re(\omega \mu^*)$, $\re(\omega^2\mu^*)$, $-\re(\zeta)$, $-\re(\zeta^2)$, $-\re(\zeta^4)$, and drive the additional cost. The two classes $C_{ED}$ and $C_{ED^2}$ have size 72 and trace zero, which the squish operator already maps to. The remaining twelve are $C_{D}$, $C_{D^2}$, $C_{V^2D}$, $C_{VD^2}$ and their central multiples by $\omega$ and $\omega^2$.

\begin{figure}
    \centering
    \includegraphics[width=0.8\linewidth]{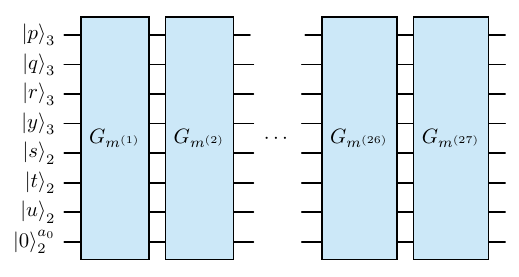}
    \caption{Deployed mixed $\mathfrak{U}_{\rm Tr}^{\dsb}(\theta)$ as a product of the $27$ parameterized gadgets $G_m$ ($m\in\mathbb{Z}_3^3$), all acting on one reusable ancilla qubit $a_0$. The generic gadget is defined in Fig.~\ref{fig:utr_toplevel}.}
    \label{fig:utr_het_circuit}
\end{figure}

\begin{figure}[ht]
\centering
\includegraphics{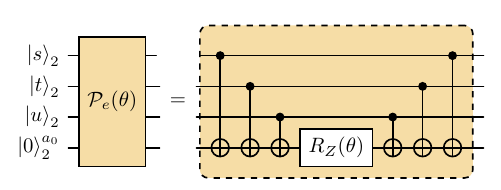}
\caption{The parity-conjugated rotation $\mathcal{P}_{e}(\theta)$, defined once and instantiated $264$ times: one qubit $R_Z(\theta)$ on the reusable ancilla, conjugated by one \textnormal{\textsc{cnot}} pair per data qubit in the rotation's support, so the sign the mixed pairs need costs nothing and no qubit-controlled qutrit rotation is ever formed.}
\label{fig:utr_parity_gadget}
\end{figure}

\begin{figure*}
    \centering
    \includegraphics[width=\linewidth]{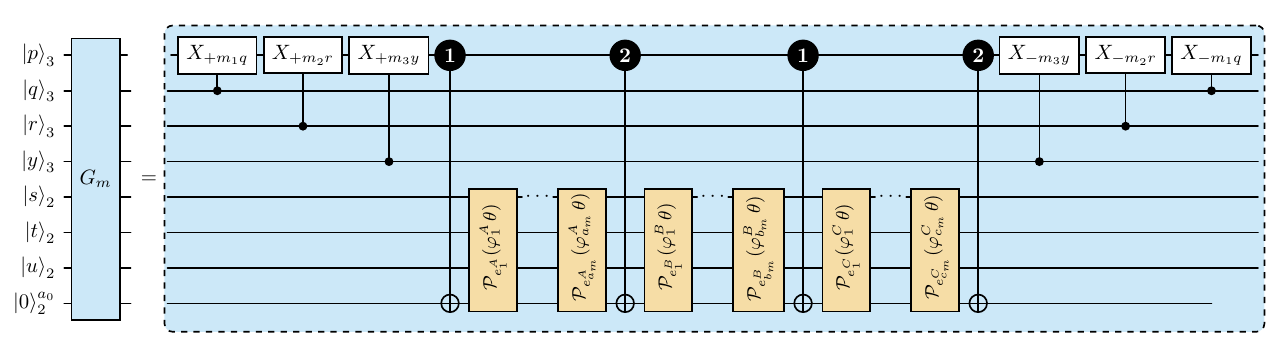}
    \caption{The generic class gadget $G_m$ ($m\in\mathbb{Z}_3^3$) of $\mathfrak{U}_{\rm Tr}^{\dsb}$ (Fig.~\ref{fig:utr_het_circuit}), defined once: a \textnormal{\textsc{csum}} ladder folds the class index $\ell$ onto $p$, four $E_{(3,2)}$ value-controls walk the ancilla through the flags of $\ell$ (stages $\mathsf{A}\,[\ell{=}1]$, $\mathsf{B}\,[\ell{\neq}0]$, $\mathsf{C}\,[\ell{=}2]$) and back, and each stage fires its parity-conjugated rotations $\mathcal{P}_e$ (Fig.~\ref{fig:utr_parity_gadget}); counts and angles are in App.~\ref{app:utr-tables}.}
    \label{fig:utr_toplevel}
\end{figure*}

The mapping of these trace classes onto trace values is:
\begin{align}
    \re\tr(g) \rightarrow \begin{cases}
        0 & g \in C_{ED},~C_{ED^2}\\
        a_1 & g \in \omega^2C_{V^2D}, \omega C_{VD^2} \\
        a_2 & g \in C_{VD^2}, C_{V^2D}\\
        a_3 & g \in \omega C_{V^2D},\omega^2 C_{VD^2}\\
        b_1 & g \in \omega C_{D}, C_{D^2}\\
        b_2 & g \in C_{D}, \omega C_{D^2}\\
        b_3 & g \in \omega^2 C_{D}, \omega^2C_{D^2}\\
    \end{cases}.
\end{align}
We assign each trace class to an ancilla-register bit string as defined in Tab.~\ref{tab:tracedsb}, following Refs.~\cite{Alam:2021uuq,Gustafson:2023kvd,Gustafson:2024kym}.
$\mathfrak{U}_{\rm{squish}}$ computes these strings by factoring out the central phase. Writing $g=\omega^p g'$ with $g'$ independent of $p$, multiplication by $\omega^p$ preserves $|\tr(g)|=|\tr(g')|$ and shifts the phase position by $p$ modulo three. Thus,
\begin{equation}
    \mathrm{code}(g)=\mathrm{Enc}\!\left(\mathrm{orbit}(g'),\,(\mathrm{pos}(g')+p)\bmod 3\right),
\end{equation}
where $\mathrm{orbit}(g')$ labels the $\omega$-orbit of $\tr(g')$ (its value up to a cube root of unity), $\mathrm{pos}(g')\in\{0,1,2\}$ the position within that orbit, and $\mathrm{Enc}$ the lookup from this pair to the bit string of Tab.~\ref{tab:tracedsb}. $\mathfrak{U}_{\rm squish}$ is therefore the four-block circuit of Fig.~\ref{fig:s216x3tracesquishdiagram}. A classifier $\mathrm{Cls}$ computes $(\mathrm{orbit},\mathrm{pos})$ into scratch registers from $(q,r,s,t,u,y)$. A mod-$3$ adder $+p$ folds the central-phase qutrit into the position. The encoder $\mathrm{Enc}$ writes the trace code. Then $-p$ and $\mathrm{Cls}^\dagger$ uncompute the scratch. Explicit circuits for $\mathrm{Cls}$ and $\mathrm{Enc}$ are given in App.~\ref{app:tracecircuits}. The construction costs $30{,}786\,T$.

The phasing operation then acts on the ancilla register alone, through the circuit of Fig.~\ref{fig:dsbphasetrace}: every gate is a Pauli-$Z$ product rotation $R_{Z\cdots Z}(x_S\theta)=\exp(-i\tfrac{x_S\theta}{2}Z\otimes\cdots\otimes Z)$ on a subset $S$ of the four qubits, drawn either directly or as an $R_Z$ conjugated by \textnormal{\textsc{cnot}}s. Because the sixteen parity functions $\prod_{i\in S}(1-2v_i)$ are orthogonal over the sixteen basis states, the diagonal $e^{-i\theta\re\tr g}$ fixes the angles $x_S$ uniquely once the three unused states of $\ket{v_0v_1v_2v_3}$ are chosen to be unphased. Tab.~\ref{tab:coefficientsdsb} lists the resulting values. The circuit reproduces the target diagonal exactly, up to a global phase $e^{-3i\theta/32}$ that needs no gate. It contains $15$ rotations with $14$ distinct angles, since $x_{0,1,2}=x_{0,1,2,3}=\tfrac{13}{16}$; every executed rotation is counted. All rotations act in sequence on the same ancilla register, so the qubit $T$-gate width is $1$. The $10$ clean ancillas of Tab.~\ref{tab:qubitcosts} are the $4$ qubits of $\ket{v_0v_1v_2v_3}$ plus $6$ scratch qubits ($4$ for the orbit, $2$ for the position). $\mathfrak{U}_{\rm squish}$ computes and uncomputes the scratch.

\begin{table}
\caption{Coefficients which appear in the phasing portion of the trace circuit for $\dsb$ shown in Fig.~\ref{fig:dsbphasetrace}. The middle column gives each coefficient in terms of the $a_n$, $b_n$ of Tab.~\ref{tab:tracedsb}, simplified using $a_1+a_2+a_3 = b_1+b_2+b_3 = 0$ (from $\sum_k \cos(2\pi\cdot 2^k/9) = 0$).}
\label{tab:coefficientsdsb}
    \begin{tabular}{c|c|c}
        \hline\hline coefficient & $a,b$ form & closed form (trig) \\
        \hline
         $x_0$         & $\phantom{-}\tfrac{7}{16} + \tfrac{1}{4} a_2 - \tfrac{1}{4} b_2$ & $\phantom{-}\tfrac{7}{16} + \tfrac{1}{4}\cos\tfrac{\pi}{9} - \tfrac{3}{4}\cos\tfrac{4\pi}{9}$\\
         $x_1$         & $\phantom{-}\tfrac{7}{16} + \tfrac{1}{4} a_3 - \tfrac{1}{4} b_1$ & $\phantom{-}\tfrac{7}{16} + \tfrac{3}{4}\cos\tfrac{\pi}{9} - \tfrac{1}{4}\cos\tfrac{2\pi}{9}$\\
         $x_2$         & $-\tfrac{3}{16} + \tfrac{1}{4} a_1 - \tfrac{1}{4} b_3$           & $-\tfrac{3}{16} - \tfrac{3}{4}\cos\tfrac{2\pi}{9} - \tfrac{1}{4}\cos\tfrac{4\pi}{9}$\\
         $x_3$         & $\phantom{-}\tfrac{3}{16}$                                       & $\phantom{-}\tfrac{3}{16}$\\
         $x_{0,1}$     & $\phantom{-}\tfrac{7}{16} + \tfrac{1}{4} a_1 + \tfrac{1}{4} b_3$ & $\phantom{-}\tfrac{7}{16} + \tfrac{1}{4}\cos\tfrac{2\pi}{9} + \tfrac{1}{4}\cos\tfrac{4\pi}{9}$\\
         $x_{0,2}$     & $\phantom{-}\tfrac{9}{16} + \tfrac{1}{4} a_3 + \tfrac{1}{4} b_1$ & $\phantom{-}\tfrac{9}{16} - \tfrac{1}{4}\cos\tfrac{\pi}{9} + \tfrac{1}{4}\cos\tfrac{2\pi}{9}$\\
         $x_{0,3}$     & $\phantom{-}\tfrac{7}{16} - \tfrac{1}{4} a_2 + \tfrac{1}{4} b_2$ & $\phantom{-}\tfrac{7}{16} - \tfrac{1}{4}\cos\tfrac{\pi}{9} + \tfrac{3}{4}\cos\tfrac{4\pi}{9}$\\
         $x_{1,2}$     & $\phantom{-}\tfrac{5}{16} + \tfrac{1}{4} a_2 + \tfrac{1}{4} b_2$ & $\phantom{-}\tfrac{5}{16} - \tfrac{1}{4}\cos\tfrac{\pi}{9} + \tfrac{1}{4}\cos\tfrac{4\pi}{9}$\\
         $x_{1,3}$     & $\phantom{-}\tfrac{7}{16} - \tfrac{1}{4} a_3 + \tfrac{1}{4} b_1$ & $\phantom{-}\tfrac{7}{16} - \tfrac{3}{4}\cos\tfrac{\pi}{9} + \tfrac{1}{4}\cos\tfrac{2\pi}{9}$\\
         $x_{2,3}$     & $-\tfrac{3}{16} - \tfrac{1}{4} a_1 + \tfrac{1}{4} b_3$           & $-\tfrac{3}{16} + \tfrac{3}{4}\cos\tfrac{2\pi}{9} + \tfrac{1}{4}\cos\tfrac{4\pi}{9}$\\
         $x_{0,1,2}$   & $\phantom{-}\tfrac{13}{16}$                                      & $\phantom{-}\tfrac{13}{16}$\\
         $x_{0,1,3}$   & $\phantom{-}\tfrac{7}{16} - \tfrac{1}{4} a_1 - \tfrac{1}{4} b_3$ & $\phantom{-}\tfrac{7}{16} - \tfrac{1}{4}\cos\tfrac{2\pi}{9} - \tfrac{1}{4}\cos\tfrac{4\pi}{9}$\\
         $x_{0,2,3}$   & $\phantom{-}\tfrac{9}{16} - \tfrac{1}{4} a_3 - \tfrac{1}{4} b_1$ & $\phantom{-}\tfrac{9}{16} + \tfrac{1}{4}\cos\tfrac{\pi}{9} - \tfrac{1}{4}\cos\tfrac{2\pi}{9}$\\
         $x_{1,2,3}$   & $\phantom{-}\tfrac{5}{16} - \tfrac{1}{4} a_2 - \tfrac{1}{4} b_2$ & $\phantom{-}\tfrac{5}{16} + \tfrac{1}{4}\cos\tfrac{\pi}{9} - \tfrac{1}{4}\cos\tfrac{4\pi}{9}$\\
         $x_{0,1,2,3}$ & $\phantom{-}\tfrac{13}{16}$                                      & $\phantom{-}\tfrac{13}{16}$\\\hline\hline
    \end{tabular}
\end{table}

For the mixed architecture we phase the register in place, with no squish stage. The register labels of Eq.~(\ref{eq:orderedproduct}) form the abelian group $\mathbb{Z}_3^4\times\mathbb{Z}_2^3$ even though $\dsb$ is nonabelian, so $\re\tr$, as a function on the register, has a unique expansion in the characters of that group. That expansion has $132$ conjugate pairs, whose qutrit exponents fall into $27$ classes.

The circuit (Fig.~\ref{fig:utr_het_circuit}) applies one gadget per class. Each class is labeled by a $3$-vector $m=(m_1,m_2,m_3)\in\mathbb{Z}_3^3$ and fixes the class index $\ell = p + m_1 q + m_2 r + m_3 y \bmod 3$, which a \textnormal{\textsc{csum}} ladder folds onto the qutrit $p$. Four $E_{(3,2)}$ entanglers then step a single reusable ancilla qubit through the three flags of $\ell$ (stage $\mathsf{A}$: $a_0=[\ell{=}1]$, stage $\mathsf{B}$: $[\ell{\neq}0]$, stage $\mathsf{C}$: $[\ell{=}2]$) and back to $\ket{0}$. Each of the class's conjugate pairs fires two ancilla rotations $R_Z$, one in each of the two stages that pair uses (Fig.~\ref{fig:utr_parity_gadget}). \textnormal{\textsc{cnot}} conjugation supplies the sign that mixed pairs need, so no qubit-controlled qutrit rotation is formed. The ladder is then undone. Every angle is exact: $\theta/24$ times an integer combination $n_0 + n_1\cos(\pi/9) + n_2\cos(2\pi/9)$, with $|n_i|\le 6$. The $27$-class geometry and all $264$ angles are tabulated in Tabs.~\ref{tab:utr-gadgets}--\ref{tab:utr-angles} (App.~\ref{app:utr-tables}). The only costed operations are the $E_{(3,2)}$ of the flag walks, $(T_{(3)},E_{(2,3)},E_{(3,2)}) = (0,0,108)$, plus $264$ qubit $R_Z$ on the qubit channel (Tab.~\ref{tab:utr_assembled_het}).

The convert-and-execute alternative, unpacking each qutrit into a qubit pair by $6\,E_{(3,2)}+6\,E_{(2,3)}$ entanglers and running the qubit squish circuit of Sec.~\ref{sec:trace} verbatim, costs $(0,6,6)$ plus residual qubit $T = 30{,}924 + 17.25\log_2(1/\epsilon)$ with $16$ ancilla qubits ($6$ conversion scratch plus the squish's own $10$). At $\epsilon=10^{-10}$ the in-place schedule is $2.5\times$ cheaper in $T$, while the conversion approach keeps the fewer-entangler edge ($12$ against $108$).

\begin{table}[!ht]
\caption{Mixed cost of $\mathfrak{U}_{\rm Tr}^{\dsb}$. One ancilla qubit; $T$-width $1$; the qutrit $R$ channel is identically zero.}
\label{tab:utr_assembled_het}
\footnotesize\setlength{\tabcolsep}{2pt}
\begin{tabular}{lcccc}
\hline\hline
Block & $T_{(3)}$ & $E_{(2,3)}$ & $E_{(3,2)}$ & qubit $T$ \\\hline
\textnormal{\textsc{csum}} ladders ($162$) & 0 & 0 & 0 & 0 \\
Flag walks ($27{\times}4$) & 0 & 0 & 108 & 0 \\
Parity $R_Z$ ($264$) & 0 & 0 & 0 & $2428.8+303.6\log_2(1/\epsilon)$ \\\hline
\textbf{Total} & $\mathbf{0}$ & $\mathbf{0}$ & $\mathbf{108}$ & $\mathbf{2428.8+303.6\log_2(1/\epsilon)}$ \\
\hline\hline
\end{tabular}
\end{table}

\section{Fourier transform and kinetic-phase gate}
\label{sec:fouriertransform}

The Fourier transform $\mathfrak{U}_{FFT}$ is the gauge-invariant operation that takes the group-element basis $\ket{g}$ to the irrep basis $\ket{\rho,i,j}$, and the kinetic-phase gate $\mathfrak{U}_{Ph}$ then multiplies each irrep state by its kinetic eigenvalue $\lambda_\rho$; we return to its implementation in Sec.~\ref{sec:uph_impl}. We build $\mathfrak{U}_{FFT}^{\dsb}$ recursively along the chain of subgroups generated by the \ac{SGS} ordered product of Eq.~(\ref{eq:orderedproduct}),
\begin{align}
\label{eq:fftchain}
\mathbb{Z}_3\xrightarrow{C}\mathbb{Z}_3^{2}\xrightarrow{E}\Delta(27)&\xrightarrow{V^2}\Delta(54)\notag\\
\xrightarrow{V}\Sigma(36\times3)&\xrightarrow{X}\dsa\xrightarrow{D}\dsb,
\end{align}
following Refs.~\cite{Pueschel:1998zzo,Murairi:2024xpc,Bruinsma:2026fke}. Each arrow adds one generator $h$ on one new register, and Fig.~\ref{fig:fft_chain} shows the resulting circuit level by level.

\begin{figure}[!ht]
\includegraphics[width=\linewidth]{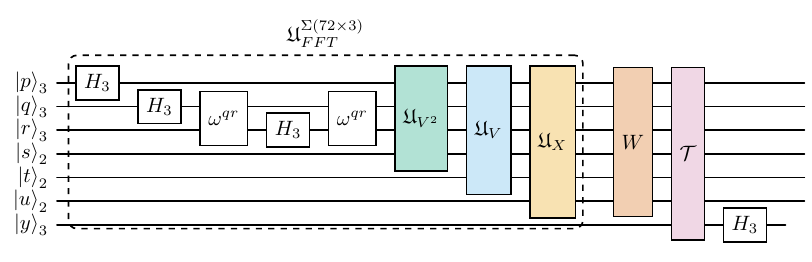}
\caption{$\mathfrak{U}_{FFT}^{\dsb}$ on the mixed architecture, one box per arrow of Eq.~(\ref{eq:fftchain}): the $\Delta(27)$ level (three $H_3$ and two phases $\omega^{qr}$), the levels $\mathfrak{U}_{V^2}$, $\mathfrak{U}_{V}$, $\mathfrak{U}_{X}$, and the top step $W$, $\mathcal{T}$, $H_3$; the inner levels have the same shape with $H_2$ or $H_3$ on their new register. Every colored box is drawn gate by gate in App.~\ref{app:per_block_rho_d}, inside a group of the same color.}
\label{fig:fft_chain}
\end{figure}

One level is the step shown by the last three boxes of Fig.~\ref{fig:fft_chain},
\begin{equation}
\label{eq:fftstep}
(\mathbbm{1}\otimes H)\;\mathcal{T}\;(W\otimes\mathbbm{1}),
\end{equation}
with $W$ a basis change on the registers already transformed, $\mathcal{T}$ a twiddle controlled on the new register, and $H$ the Hadamard $H_2$ or $H_3$ of that register.

The step is fixed by how conjugation by $h$ permutes the irreps $\sigma$ of the subgroup just transformed~\cite{Pueschel:1998zzo}. An irrep that $h$ maps to itself extends to $[G:H]$ irreps of the larger group, $[G:H]$ the index of the subgroup; for new-register value $y$ it receives the twiddle $D_\sigma^{y}$ on its column index, where
\begin{equation}
\label{eq:twiddlecond}
D_\sigma\,\sigma(hbh^{-1})\,D_\sigma^{-1}=\sigma(b),\qquad D_\sigma^{[G:H]}=\sigma\big(h^{[G:H]}\big).
\end{equation}
Irreps that $h$ permutes among themselves fuse into one induced irrep; their twiddle is the relabeling
\begin{equation}
\label{eq:relabel}
y\to a-y,
\end{equation}
with $a$ the position of the irrep in its orbit. The basis change $W$ aligns the orbit members and the phase $\sigma(h^{[G:H]})$ closes the orbit.

The first three levels are Clifford, as in Ref.~\cite{Murairi:2024xpc}: $H_3$ on $\ket{p}$ and $\ket{q}$, the phase $\omega^{qr}$, $H_3$ on $\ket{r}$, and $\omega^{qr}$ again (Fig.~\ref{fig:fft_chain}). Every later twiddle acts only on sets of basis states selected by register values; we call these sets sectors and mark them with flag qubits. The $V^2$ level negates the column index of the two three-dimensional irreps: after the relabeling $r\to\tilde r=r+q$ (one \textsc{CSum}) it swaps two values of $\tilde r$, with a cube-root phase on the $p{=}2$ copy (Fig.~\ref{fig:fft_l3}, App.~\ref{app:per_block_rho_d}).

The $V$ level is a controlled qutrit Hadamard, $H_3$ on the $p{=}1$ sector and $H_3^{\dagger}$ on $p{=}2$. Both are diagonal in the same basis $Y$, so $Y^{\dagger}$ and $Y$ are applied without controls and only the fourth-root phases between them are controlled (Fig.~\ref{fig:fft_l4}). The $X$ level has nine sector pieces, each a qutrit Fourier gate, a swap of two values, a sign, or an $H\otimes H$-conjugated permutation (Figs.~\ref{fig:fft_l5_h3}--\ref{fig:fft_l5_p0}).

The last arrow adds $D$, whose cube $D^3=\omega^2\mathbbm{1}$ is central in $\dsa$, so the step has index three. Table~\ref{tab:dd_blockmap} lists the families of output states of $\mathfrak{U}_{FFT}^{\dsa}$, the irrep each carries, and whether $D$ fixes it. Seven irreps are fixed. The trivial one needs nothing; the $\mathbf{2}$ has a Clifford twiddle and the $\mathbf{8}$ a permutation with cube-root phases,
\begin{align}
\label{eq:cliffordtwiddles}
M_2&=e^{5i\pi/4}\,SH\ \text{on }u,\notag\\
M_8&=H^{\otimes3}\,P_\sigma\,\omega^{e(s,t,u)}\,H^{\otimes3}\ \text{on }(s,t,u),
\end{align}
with $P_\sigma$ the cycle $(t,u)\to(t\oplus u,t)$ and $e=2t(1-u)(1+s)+tu(1+s)$; the fixed $\mathbf{3}$'s and $\mathbf{6}$'s have
\begin{equation}
\label{eq:toptwiddles}
M_3=\zeta^{g}\,U_3\,\mathrm{diag}(1,1,\omega^{k})\,U_3^{\dagger},\qquad M_6=\zeta^{g}\,B\,\Lambda_6\,B^{\dagger}.
\end{equation}
Here $U_3$ is a qutrit Clifford and $\Lambda_6=\omega^{[u=1]+2[u=0][\tilde r=2]}$ for the $\mathbf{6}$ (its conjugate for the $\mathbf{6}^{*}$). The exponents are $(g,k)=(5,1)$ for the $\mathbf{3}$ and $(1,2)$ for the $\mathbf{3}^{*}$, and $g=2$ for the $\mathbf{6}$ and $g=1$ for the $\mathbf{6}^{*}$. $B$ is a non-Clifford $6\times6$ change of basis on $(\tilde r,u)$ and carries the only continuous rotations of the step on the mixed architecture.

The remaining nine irreps form three orbits of size three, handled by Eq.~(\ref{eq:relabel}) with the $a(t,u)$ of Tab.~\ref{tab:dd_blockmap}, the closing phase $\omega^{2p}$ on the states with $a<y$, and partner basis changes $W$ that are $H_3$-conjugated quadratic cube-root phases (Fig.~\ref{fig:fft_top_p1}). The factors $\zeta^{gy}$ are exact in Clifford$+T_{(3)}$, since $\mathrm{diag}(1,\zeta,\zeta^2)$ is $T_{(3)}$ up to a Clifford.

\begin{figure}[!ht]
\centering
\includegraphics[scale=0.533]{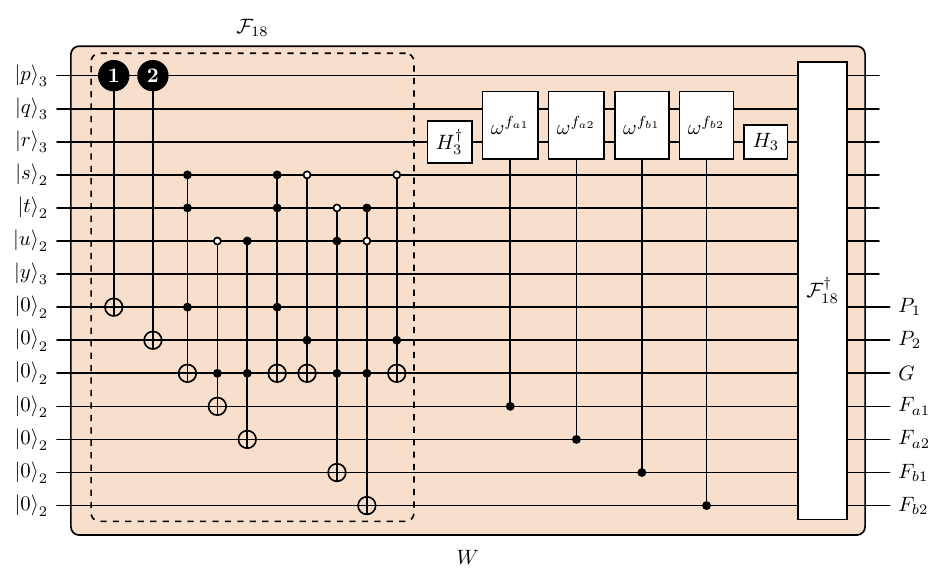}
\caption{Mixed-register circuit for $W$ of Fig.~\ref{fig:fft_chain}: each box $\omega^{f_F}$ (Fig.~\ref{fig:fft_cphase_gadgets}a) has $f_{a1}=r^2+2r+2q^2+q$, $f_{a2}=2r^2+r+q^2+2q$, $f_{b1}=2r^2+r+2q+qr$, $f_{b2}=r^2+2r+q+2qr$.}
\label{fig:fft_top_p1}
\end{figure}

\begin{table}[!ht]
\caption{Output families $\ket{pqrstu}$ of $\mathfrak{U}_{FFT}^{\dsa}$, the irrep each carries, and the twiddle applied for new-register value $y$. Fixed irreps receive the powers of Eqs.~(\ref{eq:cliffordtwiddles})--(\ref{eq:toptwiddles}) on their column index $\tilde r=r+q$ (and $u$), the $p{=}2$ blocks after the phase $\omega^{2q\tilde r}$. Orbits receive Eq.~(\ref{eq:relabel}) with the listed $a$ and, on the members with $a\neq0$, the basis change $W$.}
\label{tab:dd_blockmap}
\footnotesize
\begin{tabular}{l|c|c|l}
\hline\hline
states & irrep & $D$ & twiddle \\
\hline
$\ket{000000}$ & $\mathbf{1}$ & fixed & $1$ \\
$\ket{0000tu}$, $(t,u)\neq(0,0)$ & $\mathbf{1},\mathbf{1},\mathbf{1}$ & orbit & $a=t(1+u)$ \\
$\ket{0001tu}$ & $\mathbf{2}$ & fixed & $M_2^{\,y}$ \\
$\ket{0qrstu}$, $(q,r)\neq(0,0)$ & $\mathbf{8}$ & fixed & $M_8^{\,y}$ \\
\hline
$\ket{1qr101}$ & $\mathbf{3}$ & fixed & $M_3^{\,y}$ \\
$\ket{2qr011}$ & $\mathbf{3}^{*}$ & fixed & $M_3^{\,y}$ \\
$\ket{1qr0tu}$ & $\mathbf{6}$ & fixed & $M_6^{\,y}$ \\
$\ket{2qr1tu}$ & $\mathbf{6}^{*}$ & fixed & $M_6^{\,y}$ \\
$\ket{1qr1tu}$, $(t,u)\neq(0,1)$ & $\mathbf{3},\mathbf{3},\mathbf{3}$ & orbit & $a=t(2-u)$, $W$ \\
$\ket{2qr0tu}$, $(t,u)\neq(1,1)$ & $\mathbf{3}^{*},\mathbf{3}^{*},\mathbf{3}^{*}$ & orbit & $a=2u+t$, $W$ \\
\hline\hline
\end{tabular}
\end{table}

$\mathfrak{U}_{FFT}$ is determined by its action: it block-diagonalizes the left and right regular representations, one block per copy of each irrep. Its matrix elements agree with $\sqrt{d_\rho/|G|}\,\rho(g)_{ij}$ only up to a fixed unitary on the multiplicity index and a diagonal phase on the output states, and some blocks carry their column index on $\tilde r=r+q$. Neither affects $\mathfrak{U}_{FFT}^{\dagger}f(\rho)\,\mathfrak{U}_{FFT}$ for a function of the irrep label alone, which is all that the kinetic term requires.

Table~\ref{tab:uft_assembled_het} gives the cost of every level in both architectures; the gate-level circuits are collected in App.~\ref{app:per_block_rho_d}. Sector flags are computed from the mixed entanglers and Toffoli gates of App.~\ref{app:toffolistaircase}, and a cube-root phase on a flag qubit is applied through a clean qutrit (Fig.~\ref{fig:fft_flag_gadgets}). The basis changes $Y$, $U_3$, $B$ and $H_3$ are applied to all states and undone afterwards; outside the sector the two copies cancel, so neither needs a control.

Table~\ref{tab:uft_assembled_het} gives the totals in both architectures, and Tab.~\ref{tab:qutritcosts} the conversion of the mixed $T_{(3)}$ and qutrit rotations to $R$ at the rate of Sec.~\ref{sec:quantumgates}. Flags are counted per sector without sharing, so both totals are upper bounds.

\begin{table}[!ht]
\caption{Cost of $\mathfrak{U}_{FFT}^{\dsb}$ by level of Eq.~(\ref{eq:fftchain}). Mixed columns: $T_{(3)}$, mixed entanglers, qubit $T$, executed qubit $R_Z$ ($1.15\log_2(1/\epsilon)+9.2\,T$ each), and two-level qutrit rotations $R_Z^{(a,b)}$. Qubit columns: exact $T$ and $R_Z$ syntheses.}
\label{tab:uft_assembled_het}
\begin{tabular}{@{}l|cccccc|cc@{}}
\hline\hline
Level & $T_{(3)}$ & $E_{(2,3)}$ & $E_{(3,2)}$ & $T$ & $R_Z$ & $R_Z^{(a,b)}$ & $T$ & $R_Z$ \\\hline
$\Delta(27)$ & 0 & 0 & 0 & 0 & 0 & 0 & 9 & 38 \\
$V^2$ & 81 & 6 & 9 & 21 & 0 & 0 & 210 & 2 \\
$V$ & 63 & 11 & 14 & 138 & 0 & 12 & 450 & 8 \\
$X$ & 150 & 28 & 58 & 557 & 0 & 24 & 4177 & 14 \\
$D$ & 177 & 187 & 86 & 1850 & 28 & 176 & 4832 & 364 \\\hline
\textbf{Total} & $\mathbf{471}$ & $\mathbf{232}$ & $\mathbf{167}$ & $\mathbf{2566}$ & $\mathbf{28}$ & $\mathbf{212}$ & $\mathbf{9678}$ & $\mathbf{426}$ \\
\hline\hline
\end{tabular}
\end{table}

\subsection{Implementation of \texorpdfstring{$\mathfrak{U}_{Ph}^{\dsb}$}{UPh}}
\label{sec:uph_impl}

In the irrep basis, $\mathfrak{U}_{Ph}^{\dsb}(\tau)$ acts as $e^{-i\tau\,\lambda_\rho}$ on $\ket{\rho,i,j}$, with $\lambda_\rho$ from Tab.~\ref{tab:eigenvalues}; the 24 irreps fall into ten nonzero phase classes, and the trivial irrep needs no rotation. Which output state of $\mathfrak{U}_{FFT}^{\dsb}$ carries which class follows from the transform itself,
\begin{equation}
\label{eq:lambdadiag}
\mathfrak{U}_{FFT}^{\dsb}\,H_E\,\mathfrak{U}_{FFT}^{\dsb\,\dagger}=\mathrm{diag}\big(\lambda(p,s,t,u,y,[q{=}r{=}0])\big),
\end{equation}
so the eigenvalue depends only on $p$, $s$, $t$, $u$, $y$ and on whether $q=r=0$.

We decode and rotate on these bits (Figs.~\ref{fig:uph_rotations} and \ref{fig:uph_flags}). The value flags $[p{=}v]$, $[y{=}v]$, $[q{=}0]$, $[r{=}0]$ and a few shared products are computed once with $17$ Toffoli gates. Each phase class is the parity of up to six of these flags, so one parity-conjugated rotation $\mathcal{P}(\lambda_\rho\tau)$ of Fig.~\ref{fig:utr_parity_gadget} applies the phase, and the flags are then uncomputed. In $\mathcal{P}$, \textsc{cnot}s collect the parity on one qubit, $R_Z$ acts on it, and the \textsc{cnot}s are undone; $\mathcal{P}^{\prime}$ rotates the complementary parity. The circuit is classical reversible logic plus ten rotations, and it reproduces $\mathrm{diag}(e^{-i\lambda\tau})$ on all $648$ basis states.

\begin{figure}[!ht]
\centering
\includegraphics[width=\linewidth]{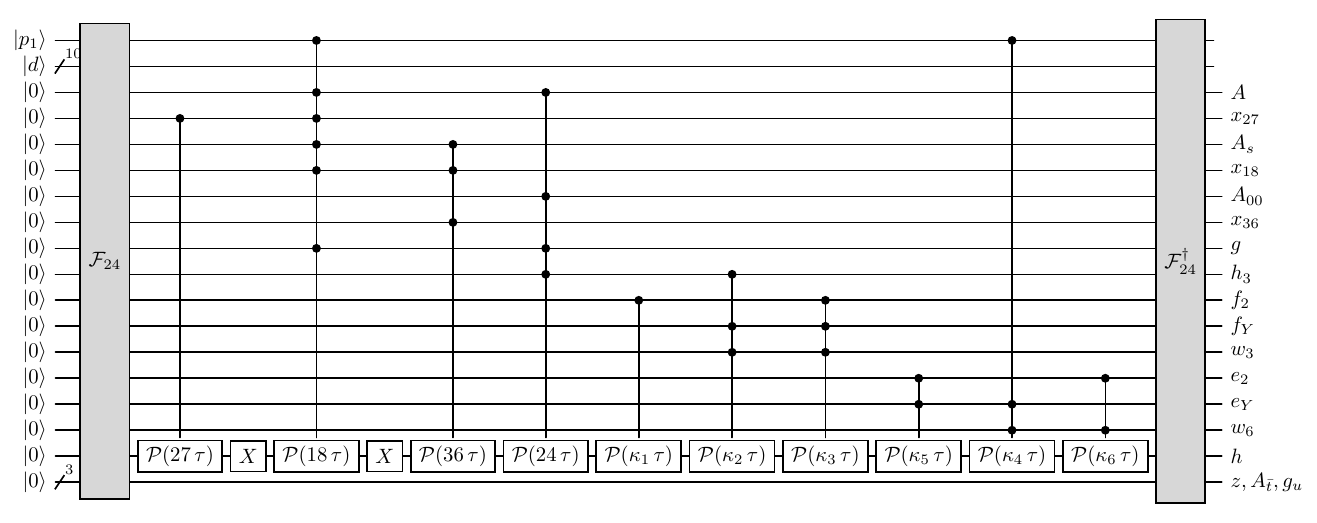}
\caption{Qubit circuit for $\mathfrak{U}_{Ph}^{\dsb}$: each box $\mathcal{P}(\theta)$ is the parity-phase gadget of Fig.~\ref{fig:utr_parity_gadget} with the angles of Tab.~\ref{tab:eigenvalues}; $\ket{d}=\ket{p_0q_0q_1r_0r_1stuy_0y_1}$ bundles the ten data qubits no rotation reads, and the bottom bundle the three unused ancillas.}
\label{fig:uph_rotations}
\end{figure}

\begin{figure}[!ht]
\centering
\includegraphics[scale=0.533]{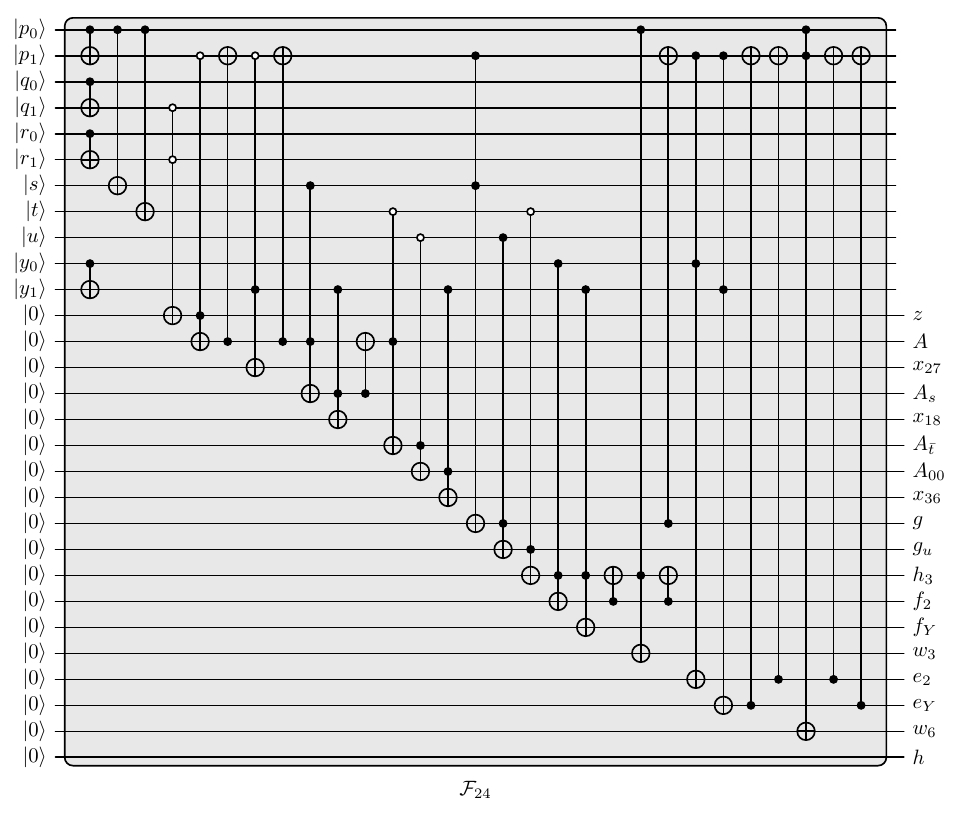}
\caption{The flag block of Fig.~\ref{fig:uph_rotations}.}
\label{fig:uph_flags}
\end{figure}

Table~\ref{tab:uph_assembled_het} gives the cost. In the qubit encoding the value flags are single bits; in the mixed architecture only the six value tests change, becoming $12$ mixed entanglers $E_{(3,2)}$, while the rotations remain qubit $R_Z$, so $\mathfrak{U}_{Ph}^{\dsb}$ needs no $T_{(3)}$ and no metaplectic $R$.

\begin{table}[!ht]
\caption{Cost of $\mathfrak{U}_{Ph}^{\dsb}$ in the two architectures: qutrit $T_{(3)}$, mixed entanglers, qubit $T$, $R_Z$ syntheses, and clean ancillas.}
\label{tab:uph_assembled_het}
\begin{tabular}{@{}l|ccccc|c@{}}
\hline\hline
Architecture & $T_{(3)}$ & $E_{(2,3)}$ & $E_{(3,2)}$ & $T$ & $R_Z$ & clean ancillas \\\hline
qubit & 0 & 0 & 0 & 238 & 10 & 18 \\
mixed & 0 & 0 & 12 & 238 & 10 & 24 \\
\hline\hline
\end{tabular}
\end{table}

\section{Resource Costs}
\label{sec:resources}
Tables~\ref{tab:qubitcosts} and \ref{tab:qutritcosts} summarize the primitive costs in the qubit and mixed resource models of Sec.~\ref{sec:quantumgates}.

\begin{table}[!ht]
\caption{Resource costs of the five primitive gates in the qubit encoding.}
\label{tab:qubitcosts}
\begin{tabular}{cccc}
\hline\hline 
Gate & $T$ gates & $T$ width & clean ancillas\\\hline
$\mathfrak{U}_{-1}$ & 2065 & 2 & 4\\
$\mathfrak{U}_{\times}$ & 2548 & 3 & 5 \\
$\mathfrak{U}_{\rm Tr}$ & $30924+17.25\log_2(1/\epsilon)$ & 1 & 10 \\
$\mathfrak{U}_{FFT}$ & $13597.2 + 489.9\log_2(1/\epsilon)$ & 1 & 9 \\
$\mathfrak{U}_{Ph}$ & $330 + 11.5\log_2(1/\epsilon)$ & 1 & 18 \\\hline\hline
\end{tabular}
\end{table}

\begin{table*}
\caption{Resource costs for the mixed architecture. The $T$ column is the qubit $T$ side channel; $R$ gates come from synthesizing the qutrit $T_{(3)}$ and $R_Z^{(a,b)}$ at the rate of Tab.~\ref{tab:qubitbasicresources}, and $E_{(2,3)}$, $E_{(3,2)}$ are the mixed entanglers of Sec.~\ref{sec:quantumgates}. \textnormal{\textsc{ccsum}} and \textnormal{\textsc{cccsum}} gates are already reduced to these primitives (App.~\ref{app:toffolistaircase}) and are not tabulated separately.}
\label{tab:qutritcosts}
\begin{tabular}{cccccccc}
\hline\hline
Gate & $T$ gates & $T$ width & $R$ gates & $R$ width & $E_{(2,3)}$ & $E_{(3,2)}$ & clean ancillas \\\hline
$\mathfrak{U}_{-1}$ & 7 & 1 & $2035.2 + 3268.404\log_3(1/\epsilon)$ & 2 & 90 & 0 & 4 \\
$\mathfrak{U}_{\times}$ & 3 & 1 & $10140.8 + 16285.491\log_3(1/\epsilon)$ & 3 & 197 & 3 & 5 \\
$\mathfrak{U}_{\rm Tr}$ & $2428.8+303.60\log_2(1/\epsilon)$ & 1 & $0$ & --- & $0$ & $108$ & 1q \\
$\mathfrak{U}_{FFT}$ & $2823.6 + 32.2\log_2(1/\epsilon)$ & 1 & $2185.6 + 3509.937\log_3(1/\epsilon)$ & 1 & 232 & 167 & 9+1q \\
$\mathfrak{U}_{Ph}$ & $330 + 11.5\log_2(1/\epsilon)$ & 1 & $0$ & --- & 0 & 12 & 24 \\\hline\hline
\end{tabular}
\end{table*}

We estimate the cost of a fiducial shear-viscosity calculation~\cite{Rhodes:2024zbr,Davoudi:2026pcu,Kan:2021xfc,Gustafson:2022xdt,Gustafson:2023kvd,Gustafson:2024kym} on an $L^3=10^3$ lattice evolved for $N_t=50$ steps, with total synthesis error $\epsilon_T=10^{-8}$. Thermal-state preparation is neglected; its cost may be significant~\cite{Xie:2022jgj,Davoudi:2022uzo,Avkhadiev:2019niu,Gustafson:2022hjf,peruzzo2014variational,Gustafson:2019mpk,Gustafson:2019vsd,Harmalkar:2020mpd,Gustafson:2020yfe,Jordan:2017lea,PhysRevLett.108.080402,motta2020determining,Clemente:2020lpr,motta2020determining,deJong:2021wsd,Gustafson:2023ayr,Kane:2023jdo}, although thermal observables can also be extracted from real-time evolution without preparing the thermal state~\cite{Lamm:2026cfq}. We use first-order Trotterization with step size $\tau$ for the Kogut--Susskind $H_{KS}$~\cite{PhysRevD.11.395} and Symanzik-improved $H_I$~\cite{Carena:2022kpg} Hamiltonians. Table~\ref{tab:primcost} lists the required primitive gates per link per time step.

\begin{table}[ht]
   \caption{Number of primitive gates per link per $\tau$ neglecting boundary effects as a function of $d$ for $H_{KS}$ and $H_{I}$.}
    \label{tab:primcost}
    \begin{tabular}{c|c|c}
    \hline\hline
    Gate & $N[H_{KS}]$&$N[H_{I}]$\\
    \hline
    $\mathfrak{U}_{FFT}$ & 2  &4\\
    $\mathfrak U_{Ph}$ & 1 & 2\\
    $\mathfrak{U}_{\rm Tr}$&$\frac{1}{2}(d-1)$ &$\frac{3}{2}(d-1)$\\
    $\mathfrak U_{-1}$& $3(d-1)$ & $2+11(d-1)$\\
    $\mathfrak U_{\times}$& $6(d-1)$ &$4+26(d-1)$\\
    \hline\hline
    \end{tabular}
\end{table}

The per-link-per-$\tau$ gate counts follow from the per-gate costs (Tabs.~\ref{tab:qubitcosts}, \ref{tab:qutritcosts}) and the per-link multiplicities (Tab.~\ref{tab:primcost}); Tab.~\ref{tab:costsummary} collects them with the synthesis-error budgets for both encodings under the Kogut--Susskind ($H_{KS}$) and Symanzik-improved ($H_I$) Hamiltonians. For the qubit encoding the whole cost
sits on the $T$ channel. For the mixed encoding it splits into the
qubit-$T$ side channel, the qutrit $R$ channel (the $T_{(3)}$ content converted
at the synthesis cost), and the two mixed entanglers, counted separately under an
exact-implementation assumption.

\begin{table*}[t]
\caption{Per-link-per-$\tau$ resource costs and total synthesis-error budgets
for the qubit and mixed encodings under both the Kogut--Susskind
($H_{KS}$) and Symanzik-improved ($H_I$) Hamiltonians. The cost column is the per-link-per-$\tau$ count $C$; the
total count is $C\times dL^dN_t$. The synthesis-error column is the total accumulated
error in terms of the per-synthesis precision $\epsilon$.}
\label{tab:costsummary}
\centering
\footnotesize
\setlength{\tabcolsep}{3pt}
\begin{tabular}{lllllr}
\hline\hline
Encoding & $H$ & Resource & Cost per link per $\tau$ & Synthesis error & Fiducial total \\
\hline
Qubit & $H_{KS}$ & $T$ & $36945\,d-9420.6+(8.625\,d+982.675)\log_2(1/\epsilon)$ & $(7.5\,d+854.5)\,dL^dN_t\,\epsilon$ & $2.3\times10^{10}$ \\
Qubit & $H_{I}$  & $T$ & $135349\,d-65978.2+(25.875\,d+1956.725)\log_2(1/\epsilon)$ & $(22.5\,d+1701.5)\,dL^dN_t\,\epsilon$ & $6.8\times10^{10}$ \\
\hline
Het.  & $H_{KS}$ & qubit $T$   & $1253.4\,d+4723.8+(151.8\,d-75.9)\log_2(1/\epsilon)$ & $(132\,d-66)\,dL^dN_t\,\epsilon$ & $4.6\times10^{9}$ \\
      &          & qutrit $R$  & $66950.4\,d-62579.2+(107518.158\,d-100498.284)\log_3(1/\epsilon)$ & $(20922\,d-19556)\,dL^dN_t\,\epsilon$ & $1.3\times10^{12}$ \\
      &          & $E_{(2,3)}$ & $1452\,d-988$ & --- & $5.1\times10^{8}$ \\
      &          & $E_{(3,2)}$ & $72\,d+274$ & --- & $7.4\times10^{7}$ \\
\hline
Het.  & $H_{I}$  & qubit $T$   & $3798.2\,d+8182.2+(455.4\,d-303.6)\log_2(1/\epsilon)$ & $(396\,d-264)\,dL^dN_t\,\epsilon$ & $1.3\times10^{10}$ \\
      &          & qutrit $R$  & $286048\,d-232672+(459375.21\,d-373656.69)\log_3(1/\epsilon)$ & $(89390\,d-72710)\,dL^dN_t\,\epsilon$ & $5.9\times10^{12}$ \\
      &          & $E_{(2,3)}$ & $6112\,d-4216$ & --- & $2.1\times10^{9}$ \\
      &          & $E_{(3,2)}$ & $240\,d+464$ & --- & $1.8\times10^{8}$ \\
\hline\hline
\end{tabular}
\end{table*}

Table~\ref{tab:fiducial} collects the total \ac{FT} resource counts for this fiducial calculation. The per-synthesis precision $\epsilon$ is set by requiring the summed synthesis error to equal $\epsilon_T$. For the qubit encoding, $\epsilon=\epsilon_T/N_{R_Z}$, where $N_{R_Z}$ is the total number of $R_Z$ rotations. For the mixed encoding, the budget is spread over the qubit $R_Z$ and qutrit $T_{(3)}$ syntheses, $\epsilon=\epsilon_T/(N_{R_Z}+N_{T_{(3)}})$.

For the fiducial calculation with the improved Hamiltonian, our $\dsb$
construction requires $6.8\times10^{10}$ $T$ gates, compared with the
FFT-based estimate of $1.2\times10^{10}$ for $\dsd$ reported by
Murairi et al.~\cite{Murairi:2024xpc}.

On the mixed architecture the qubit $T$ side channel is $4.6\times10^{9}$ to $1.3\times10^{10}$, but the qutrit synthesis dominates the accounting.
The $1.3\times10^{12}$ ($H_{KS}$) to $5.9\times10^{12}$ ($H_I$) metaplectic $R$
gates come almost entirely from the $T_{(3)}$ content of
$\mathfrak{U}_{\times}$ (Fig.~\ref{fig:pie}), converted at the rate of Tab.~\ref{tab:qubitbasicresources}.

The mixed entanglers
$E_{(2,3)}$ and $E_{(3,2)}$ contribute $5.1\times10^{8}$ to $2.1\times10^{9}$ and $7.4\times10^{7}$ to $1.8\times10^{8}$ (Tab.~\ref{tab:fiducial}). Whether the mixed encoding is preferable therefore depends
on the hardware cost ratio of a qutrit $R$ gate to a qubit $T$ gate: the
mixed $R$-count exceeds the qubit $T$-count by a factor of $55$ to $90$,
so the mixed route wins only when a native qutrit $R$ is roughly two
orders of magnitude cheaper than a qubit $T$.

\begin{table}[H]
\caption{Total \ac{FT} resources for the fiducial calculation of Sec.~\ref{sec:resources}. The lower row gives the $\dsd$ FFT benchmark; the dash indicates an unreported value.}
\label{tab:fiducial}
\begin{tabular}{llcc}
\hline\hline
Architecture & Resource & $H_{KS}$ & $H_I$ \\
\hline
Qubit         & total $T$   & $2.3\times10^{10}$ & $6.8\times10^{10}$ \\
\hline
Mixed & qubit $T$   & $4.6\times10^{9}$  & $1.3\times10^{10}$ \\
              & qutrit $R$  & $1.3\times10^{12}$ & $5.9\times10^{12}$ \\
              & $E_{(2,3)}$ & $5.1\times10^{8}$  & $2.1\times10^{9}$  \\
              & $E_{(3,2)}$ & $7.4\times10^{7}$  & $1.8\times10^{8}$  \\
\hline
\multicolumn{2}{l}{$\dsd$ qubit $T$ (FFT)~\cite{Murairi:2024xpc}} & --- & $1.2\times10^{10}$ \\
\hline\hline
\end{tabular}
\end{table}

\begin{figure}[H]
\centering
\includegraphics[width=0.9\linewidth]{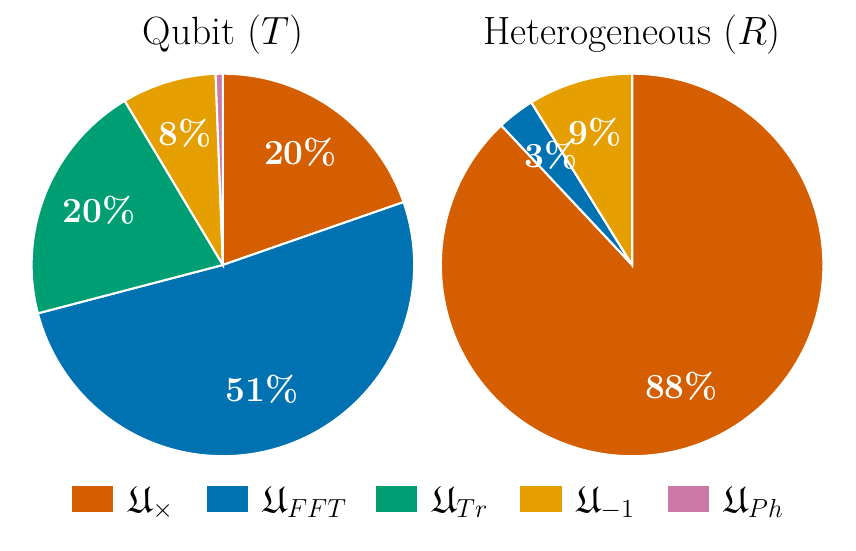}
\caption{Per-primitive share of the fiducial \ac{FT} cost ($H_{KS}$):
qubit $T$ (left) and mixed qutrit $R$ (right).}
\label{fig:pie}
\end{figure}

\section{Conclusion and Outlook}
\label{sec:outlook}

In this work we constructed the complete set of five primitive gates for $\dsb$, the
second-largest crystallographic subgroup of $SU(3)$: the inversion, multiplication,
trace, \ac{FFT}, and kinetic-phase gates, in both a qubit and a
mixed encoding. Simulating the fiducial shear-viscosity
calculation on the qubit architecture requires $2.3\times10^{10}$ ($H_{KS}$) to
$6.8\times10^{10}$ ($H_I$) $T$ gates, roughly $38$ orders of magnitude below the
truncated-field estimate of Kan and Nam~\cite{Kan:2021xfc} and comparable to the
block-encoding estimates of Rhodes~\emph{et al.}~\cite{Rhodes:2024zbr} and the improved product-formula estimates of Davoudi and Stryker~\cite{Davoudi:2026pcu}.

The improved-Hamiltonian qubit cost is nearly six times the FFT-based
estimate for $\dsd$~\cite{Murairi:2024xpc}, extending the construction to a group
with six times as many elements. The Fourier transform carries about half of the
qubit cost, the trace and multiplication gates a fifth each; the controlled-sum
arithmetic of its $X$ and $D$ levels and the non-Clifford $6\times6$ basis change $B$ of
Eq.~(\ref{eq:toptwiddles}) are the natural targets for further optimization.

The mixed encoding introduced here removes nearly all of
the qubit cost of the group multiplication and inversion, whose Toffoli
decompositions are replaced by native qutrit permutations, while the trace gate keeps continuous rotations on its qubit trace-class register and the
Fourier transform keeps them in the $6\times6$ basis change $B$ of Eq.~(\ref{eq:toptwiddles}). Whether the mixed route is ultimately cheaper depends
on the \ac{FT} cost of a native qutrit gate: under the metaplectic-$R$ synthesis adopted here it wins only when a native $R$ is about $10^{2}$ cheaper than a $T$ (Sec.~\ref{sec:resources}). Native $T_{(3)}$ magic-state injection, or compilation into Clifford+$\mathcal{D}$ with $\mathcal{D}$ a family of cyclotomic diagonal gates~\cite{Kalra:2025synthesis,Evra:2024arithmeticity}, may shift that balance.

At the circuit level, group multiplication accounts for roughly $88\%$ of the mixed $R$-gate cost and is therefore the main optimization target. Sharing intermediate computations between primitive gates and measurement-assisted uncomputation~\cite{Gidney:2018add} could reduce this cost. Their benefits depend on the circuit layout and required feed-forward corrections and remain to be quantified.

The \ac{SGS} encoding also permits grouping several registers into larger qudits~\cite{Wang:2020qhd}. Exploring such encodings together with compatible error-correcting codes could reveal further resource tradeoffs~\cite{Wills:2026kom,1715533,Grassl:1999qrs}.

With $\dsb$ complete, only $\dsc$, the largest crystallographic subgroup of $SU(3)$,
remains before the discrete-subgroup tower of $SU(3)$ is exhausted, and the binary
icosahedral group similarly completes the $SU(2)$ series. Beyond the pure-gauge
sector, faithful QCD needs fermion fields~\cite{Florio:2023kel,Zohar:2016iic,Zohar:2018nvl}, discretized as staggered or Wilson fermions; a full estimate must compare the cost of these discretizations and include the state-preparation cost neglected here. The systematic and renormalization uncertainties intrinsic to the
discrete-group approximation remain the central open questions.

\begin{acknowledgements}
EG is a KBR employee working under the Intelligent
Systems Research and Development-3 (ISRDS-3) Contract
80ARC020D0010. During the early course of this work E.G was an employee of USRA working under the NASA Academic Mission
Services, Contract No. NNA16BD14C and as a subcontractor for (ISRDS-3) Contract
80ARC020D0010. EG, HL, and EM were supported by the U.S. Department of Energy, Office of Science, National Quantum Information Science Research Centers, Superconducting Quantum Materials and Systems Center (SQMS), under Contract No. 89243024CSC000002. Fermilab is operated by Fermi Forward Discovery Group, LLC under Contract No. 89243024CSC000002 with the U.S. Department of Energy, Office of Science, Office of High Energy Physics.
EG’s part is under the NASA-DOE interagency agreement SAA2-403956.
 SO was supported by the Department of Energy through the Fermilab QuantiSED program in the area of ``Toward Lattice QCD on Quantum Computers".
Claude and ChatGPT were used to assist with circuit implementation, drafting, and editorial reviews.
\end{acknowledgements}

\bibliography{ref}

\appendix
\section{Staircase decompositions of multi-controlled Gates}
\label{app:toffolistaircase}

This appendix collects decompositions of the controlled-\textnormal{\textsc{sum}} gates defined in Sec.~\ref{sec:quantumgates}. Figure~\ref{fig:csumstaircase} shows the compute--use--uncompute construction with one clean ancilla. Uncomputation requires the inverse gate; \textnormal{\textsc{ccsum}} is self-inverse only for target dimension two.
\begin{figure}[!ht]
    \centering
    \includegraphics[width=0.430\linewidth]{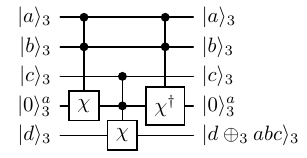}
    \caption{Staircase decomposition of \textnormal{\textsc{c}}$^3$\textnormal{\textsc{sum}}.}
    \label{fig:csumstaircase}
\end{figure}

\begin{table}[!ht]
\caption{Toffoli and $T$ gate counts for various mixed \textnormal{\textsc{c}}$^n$\textnormal{\textsc{sum}} gates. Gate names follow the convention $\textnormal{\textsc{c}}^{n}\textnormal{\textsc{sum}}_{d_1,\ldots,d_n\to d_t}$, where the control qudit dimensions $d_1,\ldots,d_n$ precede the arrow and the target dimension $d_t$ follows it (e.g.\ \CnSum{ccsum}{2,3}{3} has a qubit control, a qutrit control, and a qutrit target).}
\label{tab:csumcosts}
\begin{tabular}{ccc}
\hline\hline Gate & Toffoli gates & $T$-gates\\\hline
\CnSum{csum}{2}{3}     & 2  & 14\\
\CnSum{csum}{3}{3}     & 4  & 28\\
\CnSum{ccsum}{2,2}{3}  & 4  & 28\\
\CnSum{ccsum}{2,3}{3}  & 8  & 56\\
\CnSum{ccsum}{3,3}{3}  & 12 & 84\\
\CnSum{cccsum}{2,2,2}{3} & 6  & 42\\
\CnSum{cccsum}{3,3,2}{3} & 16 & 112\\\hline\hline
\end{tabular}
\end{table}

On qubit hardware each qutrit is a qubit pair. A qutrit control selects among $\mathbbm{1}_3$, $\chi$, and $\chi^2$, requiring more Toffoli gates than a qubit control. Figure~\ref{fig:qubitdecompositioncsum} gives the circuits and Tab.~\ref{tab:csumcosts} their costs.

\begin{figure}[h]
    \centering
\includegraphics[width=0.602\linewidth]{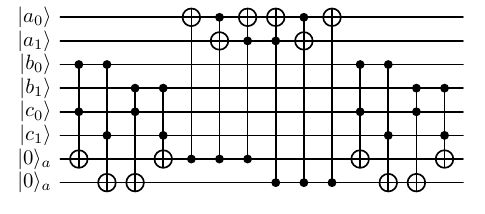}
\includegraphics[width=0.430\linewidth]{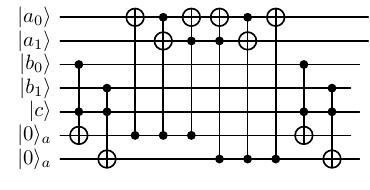}
\includegraphics[width=0.688\linewidth]{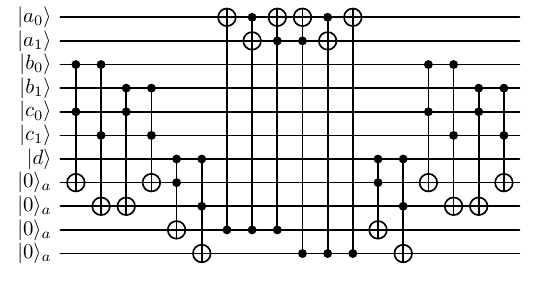}
    \caption{(top) qubit decomposition of a three-qutrit $\textnormal{\textsc{ccsum}}$ using 2 ancilla qubits. (middle) qubit decomposition of a $\textnormal{\textsc{ccsum}}$ gate with one qubit and one qutrit control.
    (bottom) a \textnormal{\textsc{c}}$^3$\textnormal{\textsc{sum}} gate with one qubit and two qutrit controls.}
    \label{fig:qubitdecompositioncsum}
\end{figure}

On native qutrit hardware we count the three resources of Sec.~\ref{sec:quantumgates}: qutrit $T_{(3)}$ gates~\cite{Yeh:2022eow} and mixed entanglers $E_{(2,3)}$, $E_{(3,2)}$. Single-qudit Cliffords, \textnormal{\textsc{cnot}}, and pure-qutrit \textnormal{\textsc{csum}} are free.

Non-Clifford cost enters through the increments. With $X_{+1}=\chi$ (Eq.~(\ref{eq:chival})) and $X_{-1}=\chi^{-1}$, a \textnormal{\textsc{ccsum}} fires one increment per control pair with nonzero product,
\begin{multline}
\label{eq:ccsumladder}
\CnSum{ccsum}{a,b}{c} = \big(C_{a,b}^{1,1} X_{+1}\big)\big(C_{a,b}^{2,2} X_{+1}\big)\\
\times\big(C_{a,b}^{1,2} X_{-1}\big)\big(C_{a,b}^{2,1} X_{-1}\big),
\end{multline}
so two qutrit controls give four increments, a mixed pair two, and two qubit controls one.

A singly controlled qutrit permutation costs $3\,T_{(3)}$~\cite{Yeh:2022eow}, so a doubly controlled one built from five (the ancilla-free Sleator--Weinfurter block $\mathcal{E}_\alpha(V)$ of Fig.~\ref{fig:cctransposition}) costs $15\,T_{(3)}$. The block is the identity unless $a=\alpha$, then fires $V$ at $b=2$ and $V^2$ at $b=1$. With $\alpha=2$, $V=X_c^{(0,1)}$ it is the doubly controlled transposition
\begin{multline}
\label{eq:ccsw}
C_{a,b}^{2,2} X_c^{(0,1)} = C_b^1 X_c^{(0,1)}\, C_a^2 X_{-1,b}\, C_b^1 X_c^{(0,1)}\\
\times C_a^2 X_{+1,b}\, C_a^2 X_c^{(0,1)},
\end{multline}
the $b=1$ case contributing nothing because $V$ is an involution.
\begin{figure}[!ht]
    \centering
    \includegraphics[width=0.731\linewidth]{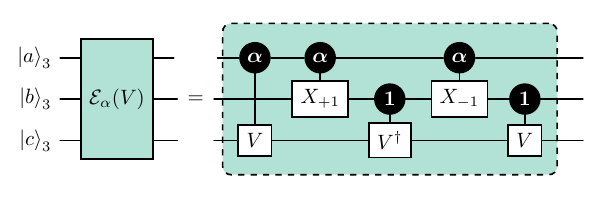}
    \caption{Ancilla-free five-gate block $\mathcal{E}_{\alpha}(V)$ (Lemma~1 of Ref.~\cite{Yeh:2022eow}), with three legs controlled by $a$ and two by $b$, applying $V$, $V^{\dagger}$, and $V$ to the target.}
    \label{fig:cctransposition}
\end{figure}
A doubly controlled increment uses two transpositions, costing $30\,T_{(3)}$. The constructions below reduce this cost for mixed controls.

\begin{table}[!ht]
\caption{Native controlled-\textnormal{\textsc{sum}} costs in the resource model $(T_{(3)},E_{(2,3)},E_{(3,2)})$. Entries give the lowest-cost constructions found; where constructions trade non-Clifford gates against mixed entanglers, we select fewer mixed entanglers.}
\label{tab:csumcostsqutrit}
\begin{tabular}{lcccc}
\hline\hline Gate & $T_{(3)}$ & $E_{(2,3)}$ & $E_{(3,2)}$ & Circuit \\\hline
\CnSum{csum}{2}{3}       & 0   & 2  & 0 & Fig.~\ref{fig:het_csum_2}  \\
\CnSum{csum}{3}{3}       & 0   & 0  & 0 & Fig.~\ref{fig:het_csum_3}  \\
\CnSum{csum}{3}{2}     & 0   & 0  & 2 & Fig.~\ref{fig:het_csum_3_2}  \\
\CnSum{ccsum}{2,2}{3}    & 15  & 4  & 0 & Fig.~\ref{fig:het_ccsum_2_2} \\
\CnSum{ccsum}{2,3}{3}    & 15  & 3  & 0 & Fig.~\ref{fig:het_ccsum_2_3} \\
\CnSum{ccsum}{3,3}{3}    & 30  & 0  & 0 & Fig.~\ref{fig:het_ccsum_3_3} \\
\CnSum{cccsum}{2,2,2}{3} & 45  & 10 & 0 & Fig.~\ref{fig:het_cccsum_staircase} \\
\CnSum{cccsum}{3,3,2}{3} & 75  & 3  & 0 & Fig.~\ref{fig:het_cccsum_staircase} \\
\hline\hline
\end{tabular}
\end{table}

\subsection{\textnormal{\textsc{c}}\texorpdfstring{$^n$}{n}\textnormal{\textsc{sum}} circuits}
\label{app:cheapest_csum_circuits}

The qubit circuits of Fig.~\ref{fig:qubitdecompositioncsum} were verified on the physical subspace. Each native circuit below was checked against its target permutation $\ket{x_1,\ldots,x_n,y}\to\ket{x_1,\ldots,x_n,y+x_1\cdots x_n\bmod d_y}$ on the full Hilbert space, with ancilla restoration confirmed on all physical inputs.

\emph{Singly controlled \textnormal{\textsc{sum}}.}---Figure~\ref{fig:het_csum_base} shows the three base cases. \CnSum{csum}{2}{3} drives two transposition legs from its qubit control; \CnSum{csum}{3}{3} is the native qutrit entangler; and \CnSum{csum}{3}{2} fires its $i=1,2$ branches on the qubit target.

\begin{figure}[!ht]
  \centering
  \subfloat[]{%
    \includegraphics[scale=0.86]{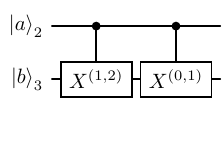}%
    \label{fig:het_csum_2}%
  }\hfill
  \subfloat[]{%
    \includegraphics[scale=0.86]{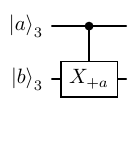}%
    \label{fig:het_csum_3}%
  }\hfill
  \subfloat[]{%
    \includegraphics[scale=0.86]{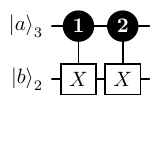}%
    \label{fig:het_csum_3_2}%
  }
  \caption{Singly controlled \textnormal{\textsc{sum}} gates: (a) \CnSum{csum}{2}{3}, (b) \CnSum{csum}{3}{3}, and (c) \CnSum{csum}{3}{2}.}
  \label{fig:het_csum_base}
\end{figure}

\emph{Doubly controlled \textnormal{\textsc{ccsum}}.}---\CnSum{ccsum}{2,3}{3} (Fig.~\ref{fig:het_ccsum_2_3}) uses $\mathcal{E}_{1}(X_{-1})$, with $\alpha=1$ because the qubit $a$ has no $\ket{2}$, covering both control values $(a,b)=(1,1),(1,2)$. \CnSum{ccsum}{3,3}{3} (Fig.~\ref{fig:het_ccsum_3_3}) uses $\mathcal{E}_{2}(X_{+1})$ and $\mathcal{E}_{1}(X_{-1})$ on disjoint values of $a$ to cover all four control values. \CnSum{ccsum}{2,2}{3} (Fig.~\ref{fig:het_ccsum_2_2}) instead borrows a qutrit ancilla $d$: the controls raise it $\ket{0}\!\to\!\ket{1}\!\to\!\ket{2}$, so $d=2$ exactly when $a=b=1$; the target increments under $d$, and the climb is undone in five legs, four of them qubit-controlled.

\begin{figure}[!ht]
  \centering
  \includegraphics[width=0.860\linewidth]{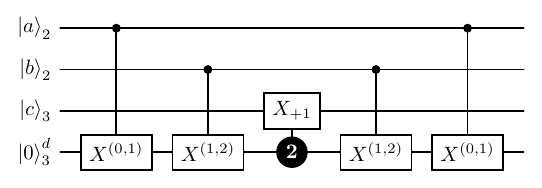}
  \caption{\CnSum{ccsum}{2,2}{3} with a borrowed qutrit ancilla.}
  \label{fig:het_ccsum_2_2}
\end{figure}

\begin{figure}[!ht]
  \centering
  \includegraphics[width=0.860\linewidth]{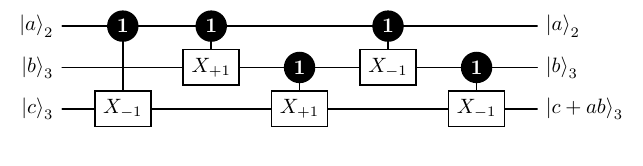}
  \caption{\CnSum{ccsum}{2,3}{3} using $\mathcal{E}_{1}(X_{-1})$.}
  \label{fig:het_ccsum_2_3}
\end{figure}

\begin{figure}[!ht]
  \centering
  \includegraphics[scale=0.86]{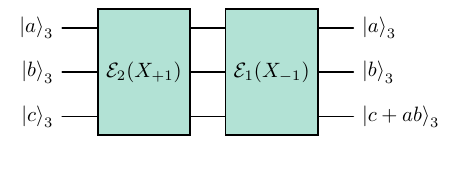}
  \caption{\CnSum{ccsum}{3,3}{3} using $\mathcal{E}_\alpha(V)$ of Fig.~\ref{fig:cctransposition}.}
  \label{fig:het_ccsum_3_3}
\end{figure}

\emph{Triply controlled \textnormal{\textsc{cccsum}}.}---Both triply controlled gates use the compute--use--uncompute staircase of Fig.~\ref{fig:het_cccsum_staircase}. \CnSum{cccsum}{2,2,2}{3} fills each block with the borrowed-ancilla climb: $d=ab$ at $(15,4,0)$, then $abc$ added to the target at $(15,2,0)$ (one fewer mixed entangler, its qutrit control being the ancilla), then uncompute, for a total of $(45,10,0)$. \CnSum{cccsum}{3,3,2}{3} pairs the two qutrit controls first, $2\times(30,0,0)+(15,3,0)=(75,3,0)$. Pairing the qubit first instead gives $(60,6,0)$, so we tabulate pair-qutrits-first for fewer mixed entanglers.

\begin{figure}[!ht]
  \centering
  \includegraphics[scale=0.86]{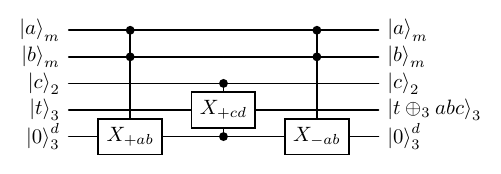}
  \caption{The \textnormal{\textsc{cccsum}}$_{m,m,2\rightarrow3}$ for control dimension $m$.}
  \label{fig:het_cccsum_staircase}
\end{figure}

\section{Mixed-register decomposition of \texorpdfstring{$\mathfrak{U}_{\times}^{\Sigma(72\times3)}$}{Utimes}}
\label{app:het_sigma72x3}

We give the mixed circuit for the inner multiplication block $\mathfrak{U}_{\times}^{\dsa}$ of Fig.~\ref{fig:dsbmultiplyqutrit}. The kernel in Fig.~\ref{fig:utimes_two_register_kernel} left-multiplies the $g$ register by $h\in\dsa$, applying the \ac{SGS} factors from right to left.

\begin{figure}[!ht]
\centering
\includegraphics[width=0.860\linewidth]{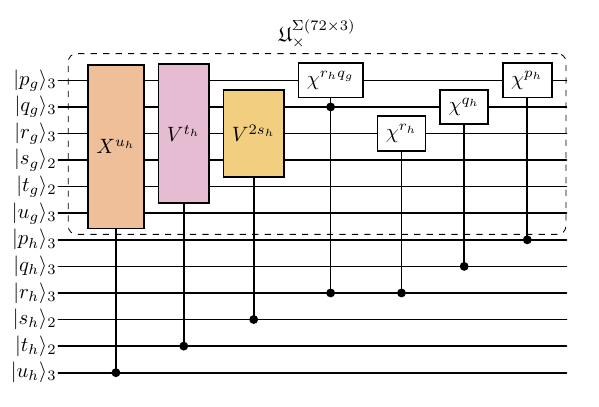}
\caption{Mixed circuit for $\mathfrak{U}_{\times}^{\Sigma(72\times3)}$. $X^{u_h}$, $V^{t_h}$, and $V^{2s_h}$ are in
Figs.~\ref{fig:utimes_stage6_breakout_compressed}, \ref{fig:utimes_stage5_breakout},
and~\ref{fig:utimes_stage4_breakout} respectively.}
\label{fig:utimes_two_register_kernel}
\end{figure}

The kernel uses the controlled-\textnormal{\textsc{sum}} primitives of App.~\ref{app:toffolistaircase}, gated on a shared $\ket{y_h}$ flag. Computing and uncomputing this flag requires two qubit-target \textnormal{\textsc{ccsum}}$_{2,2}$ gates. Table~\ref{tab:het_sigma72x3_breakdown} gives the costs by \ac{SGS} stage. Adding the outer $y_h$-controlled $D$ stages gives $(2974,176,3)+(195,21,0)=(3169,197,3)$ in $(T_{(3)},E_{(2,3)},E_{(3,2)})$, plus $3$ qubit $T$ gates, as listed in Tab.~\ref{tab:qutritcosts}.

\begin{figure}[!ht]
\centering
\includegraphics[width=0.860\linewidth]{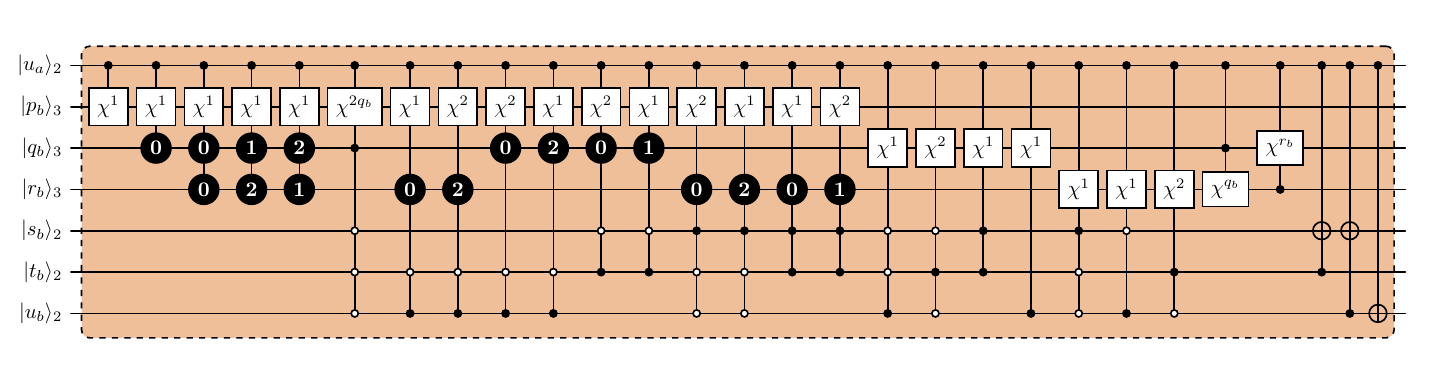}
\caption{$X^{u_h}$ subcircuit of Fig.~\ref{fig:utimes_two_register_kernel}.}
\label{fig:utimes_stage6_breakout_compressed}
\end{figure}

\begin{figure}[H]
\centering
\includegraphics[width=0.860\linewidth]{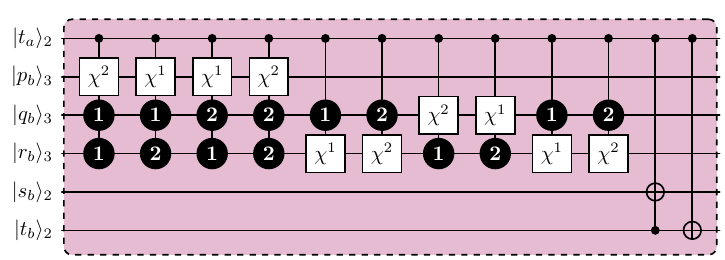}
\caption{$V^{t_h}$ subcircuit of Fig.~\ref{fig:utimes_two_register_kernel}.}
\label{fig:utimes_stage5_breakout}
\end{figure}

\begin{figure}[H]
\centering
\includegraphics[width=0.387\linewidth]{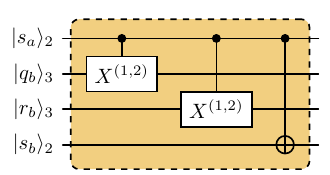}
\caption{$V^{2s_h}$ subcircuit of Fig.~\ref{fig:utimes_two_register_kernel}.}
\label{fig:utimes_stage4_breakout}
\end{figure}

\onecolumngrid
\vspace{2pt}

\begin{figure}[H]
    \centering
    \includegraphics[width=0.860\linewidth]{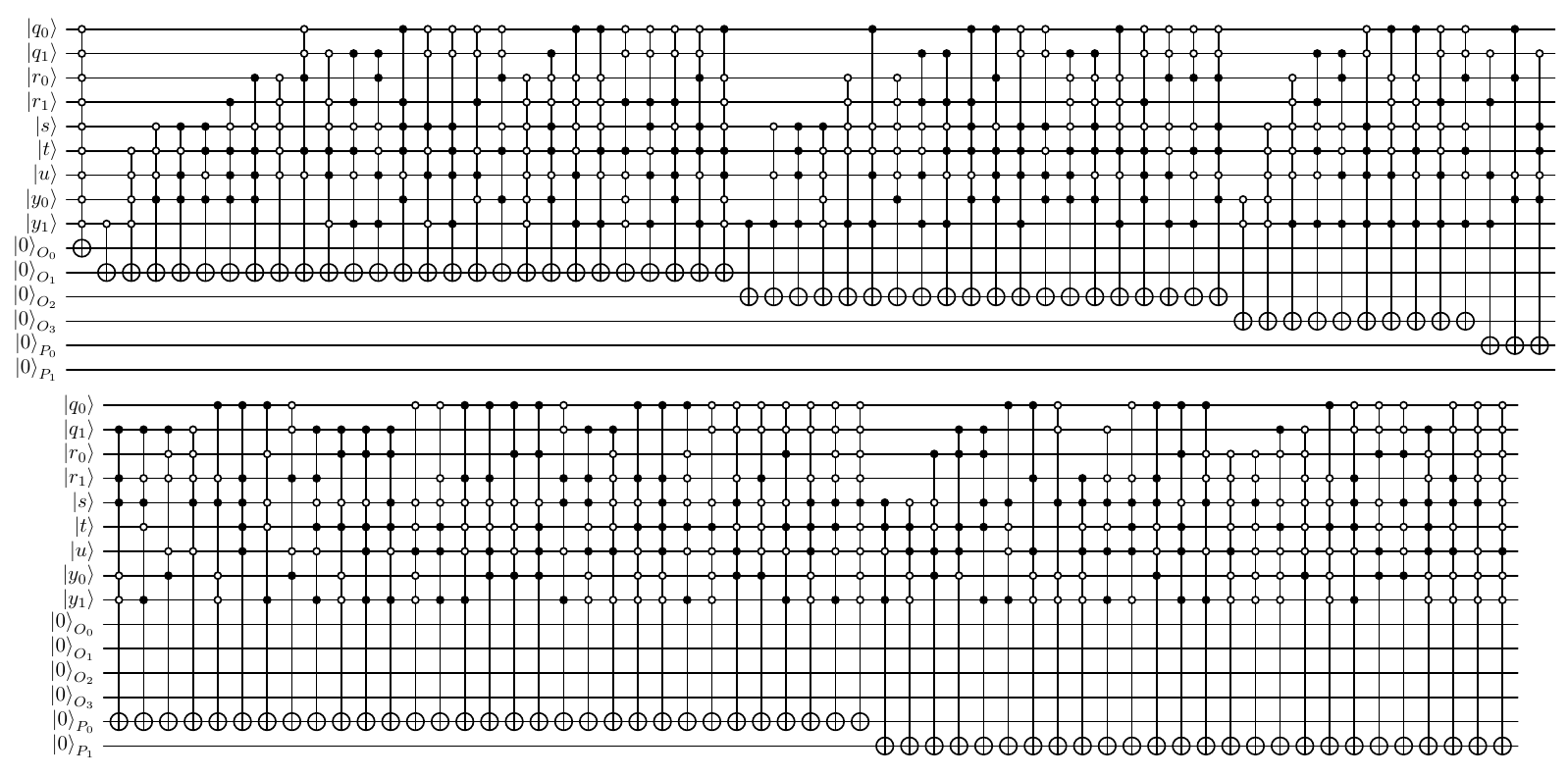}
    \caption{Classifier $\mathrm{Cls}$ writes the orbit/position scratch $(O_0O_1O_2O_3,P_0P_1)$ from the nine register bits of $g'=(q,r,s,t,u,y)$. $\mathrm{Cls}^\dagger$ is the same gates in reverse. This is the first block of $\mathfrak{U}_{\rm squish}$ (Fig.~\ref{fig:s216x3tracesquishdiagram}).}
    \label{fig:squish_classifier}
\end{figure}
\twocolumngrid

\begin{table}[!ht]
\caption{App.~\ref{app:toffolistaircase} cost of the mixed $\mathfrak{U}_{\times}^{\Sigma(72\times3)}$ block across the AND-compress wrapper and the six \ac{SGS} stages.}
\label{tab:het_sigma72x3_breakdown}
\begin{tabular}{@{}lccccl@{}}
\hline\hline
Subcircuit & $T_{(3)}$ & $E_{(2,3)}$ & $E_{(3,2)}$ & qubit $T$ & Figure \\\hline
AND-compress wrapper                & 18   & 4   & 0 & 0 & App.~\ref{app:toffolistaircase} \\
$\omega^{p_h}$                      & 15   & 3   & 0 & 0 & Fig.~\ref{fig:utimes_two_register_kernel} \\
$C^{q_h}$                           & 15   & 3   & 0 & 0 & Fig.~\ref{fig:utimes_two_register_kernel} \\
$E^{r_h}$                           & 90   & 6   & 0 & 0 & Fig.~\ref{fig:utimes_two_register_kernel} \\
$V^{2s_h}$                          & 25   & 4   & 0 & 0 & Fig.~\ref{fig:utimes_stage4_breakout} \\
$V^{t_h}$                           & 781  & 40  & 0 & 0 & Fig.~\ref{fig:utimes_stage5_breakout} \\
$X^{u_h}$                           & 2030 & 116 & 3 & 3 & Fig.~\ref{fig:utimes_stage6_breakout_compressed} \\\hline
\textbf{Total} & $\mathbf{2974}$ & $\mathbf{176}$ & $\mathbf{3}$ & $\mathbf{3}$ & Fig.~\ref{fig:utimes_two_register_kernel} \\\hline\hline
\end{tabular}
\end{table}

\section{Subcircuits for the qubit \texorpdfstring{$\mathfrak{U}_{\rm squish}$}{Usquish}}
\label{app:tracecircuits}

The qubit subcircuits of $\mathfrak{U}_{\rm squish}$ are shown in Figs.~\ref{fig:squish_classifier}, \ref{fig:squish_encoder}, and \ref{fig:squish_adder}; their assembly is described in Sec.~\ref{sec:trace}. An open circle denotes a control on $0$, a filled dot a control on $1$, and gates are not decomposed into $T$.

\begin{figure}[ht]
    \centering
    \includegraphics[width=0.9\linewidth]{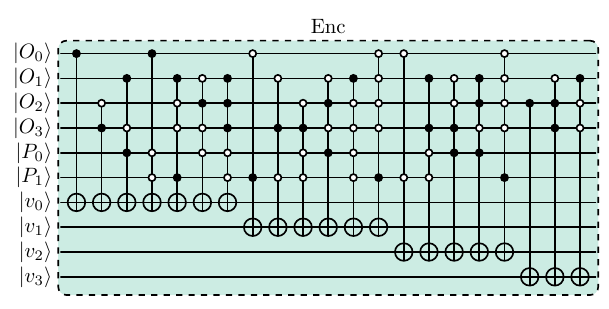}
    \caption{Encoder $\mathrm{Enc}$ that writes the trace code $v_0v_1v_2v_3$ from the orbit/position scratch $(O,P)$ found in  Fig.~\ref{fig:s216x3tracesquishdiagram}.}
    \label{fig:squish_encoder}
\end{figure}

\begin{figure}[ht]
    \centering
    \includegraphics[width=0.4\linewidth]{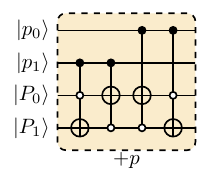}
    \caption{Phase adder $+p$. $-p$ is the reverse (Fig.~\ref{fig:s216x3tracesquishdiagram}).}
    \label{fig:squish_adder}
\end{figure}

\section{Gate-level circuits of \texorpdfstring{$\mathfrak{U}_{FFT}^{\dsb}$ and $\mathfrak{U}_{Ph}^{\dsb}$}{UFFT and UPh}}
\setlength{\textfloatsep}{12pt plus 2pt minus 2pt}\setlength{\floatsep}{8pt plus 2pt minus 2pt}\setlength{\dbltextfloatsep}{12pt plus 2pt minus 2pt}\setlength{\intextsep}{8pt plus 2pt minus 2pt}
\begin{table}[!t]
\caption{Cost of subcircuits of $\mathfrak{U}_{FFT}^{\dsb}$ and of $\mathfrak{U}_{Ph}^{\dsb}$, by level. Sectors are written $(p,s)$ or $(p,s,t)$.}
\label{tab:fft_pieces}
\centering
\scriptsize
\setlength{\tabcolsep}{1.8pt}
\begin{tabular}{@{}lc|crrr|rr@{}}
\hline\hline
 & & \multicolumn{4}{c|}{mixed} & \multicolumn{2}{c}{qubit} \\
Piece & Fig. & $T_{(3)},E_{(2,3)},E_{(3,2)}$ & $T$ & $R_Z$ & $R_Z^{(a,b)}$ & $T$ & $R_Z$ \\\hline
$\Delta(27)$ & \ref{fig:fft_chain} & $0,0,0$ & 0 & 0 & 0 & 9 & 38 \\\hline
$V^2$ & \ref{fig:fft_l3}, \ref{fig:fft_l3_qubit} & $81,6,9$ & 21 & 0 & 0 & 210 & 2 \\\hline
$V$ & & & & & & & \\
$p\neq0$ & \ref{fig:fft_l4}, \ref{fig:fft_l4_qubit}, \ref{fig:fft_l4_qubit_y} & $48,6,8$ & 12 & 0 & 12 & 268 & 6 \\
$p=0$ & \ref{fig:fft_l4b}, \ref{fig:fft_l4_qubit} & $15,5,6$ & 126 & 0 & 0 & 182 & 2 \\\hline
$X$ & & & & & & & \\
a: $\mathbf{3}$, $(1,1)$ & \ref{fig:fft_l5_h3} & $6,2,6$ & 45 & 0 & 0 & 297 & 0 \\
b: $\mathbf{3}$, $(2,0)$ & \ref{fig:fft_l5_h3} & $36,2,6$ & 45 & 0 & 0 & 745 & 0 \\
c: swap, $(1,0,0)$ & \ref{fig:fft_l5_tr} & $3,2,2$ & 70 & 0 & 0 & 77 & 0 \\
d: swap, $(2,1,0)$ & \ref{fig:fft_l5_tr} & $21,2,2$ & 70 & 0 & 0 & 413 & 0 \\
e: $W$, $(1,0,1)$ & \ref{fig:fft_l5_h3e} & $12,2,10$ & 48 & 0 & 0 & 636 & 0 \\
f: $W$, $(2,1,1)$ & \ref{fig:fft_l5_h3e} & $42,2,10$ & 48 & 0 & 0 & 972 & 0 \\
shared, $p\neq0$ & \ref{fig:fft_l5_merged} & $0,0,0$ & 0 & 0 & 24 & 64 & 12 \\
g: $\mathbf{1}$, $p=0$ & \ref{fig:fft_l5_g} & $0,0,6$ & 77 & 0 & 0 & 77 & 0 \\
h: $\mathbf{4}$, $p=0$ & \ref{fig:fft_l5_p0} & $0,6,8$ & 84 & 0 & 0 & 210 & 0 \\
i: $W$ on $\mathbf{4}$ & \ref{fig:fft_l5_p0i} & $30,10,8$ & 63 & 0 & 0 & 455 & 0 \\
shared, $p=0$ & \ref{fig:fft_l5_merged0} & $0,0,0$ & 0 & 0 & 0 & 224 & 2 \\
Z-fix & & $0,0,0$ & 7 & 0 & 0 & 7 & 0 \\\hline
$D$ & & & & & & & \\
$\mathcal{T}_1$: $W$ & \ref{fig:fft_top_p1} & $48,8,4$ & 224 & 0 & 0 & 1910 & 22 \\
$\mathcal{T}_2$: $\mathbf{1},\mathbf{1},\mathbf{1}$ & \ref{fig:fft_top_p2} & $0,5,10$ & 168 & 0 & 0 & 203 & 0 \\
$\mathcal{T}_2$: $\mathbf{3},\mathbf{3},\mathbf{3}$ & \ref{fig:fft_top_p2} & $0,13,6$ & 224 & 0 & 0 & 259 & 2 \\
$\mathcal{T}_2$: $\mathbf{3}^*,\mathbf{3}^*,\mathbf{3}^*$ & \ref{fig:fft_top_p2} & $0,13,6$ & 224 & 0 & 0 & 259 & 2 \\
$\mathcal{T}_3$: $\mathbf{2}$ & \ref{fig:fft_top_p3} & $0,0,10$ & 114 & 0 & 0 & 114 & 0 \\
$\mathcal{T}_4$: $\mathbf{8}$ & \ref{fig:fft_top_p4} & $0,32,10$ & 588 & 0 & 0 & 588 & 8 \\
$\mathcal{T}_5$: $\mathbf{3}$ & \ref{fig:fft_top_p5} & $35,8,4$ & 84 & 0 & 0 & 202 & 24 \\
$\mathcal{T}_5$: $\mathbf{3}^*$ & \ref{fig:fft_top_p5} & $31,8,4$ & 84 & 0 & 0 & 538 & 28 \\
$\mathcal{T}_6$: $\mathbf{6}$ & \ref{fig:fft_top_p6} & $32,50,16$ & 70 & 14 & 88 & 210 & 132 \\
$\mathcal{T}_6$: $\mathbf{6}^*$ & \ref{fig:fft_top_p6} & $31,50,16$ & 70 & 14 & 88 & 546 & 136 \\
$H_3$ on $y$ & \ref{fig:fft_chain} & $0,0,0$ & 0 & 0 & 0 & 3 & 10 \\\hline
$\mathfrak{U}_{Ph}$ & \ref{fig:uph_rotations}, \ref{fig:uph_flags} & $0,0,12$ & 238 & 10 & 0 & 238 & 10 \\
\hline\hline
\end{tabular}
\end{table}

\label{app:per_block_rho_d}
This appendix draws every level of Eq.~(\ref{eq:fftchain}) and the kinetic-phase gate. Every circuit was checked numerically on its full register, ancillas included, and the levels compose to $\mathfrak{U}_{FFT}^{\dsb}$. Mixed circuits are drawn for every piece; qubit circuits are drawn where they differ in structure ($V^2$, $V$, the $X$ level, $\mathfrak{U}_{Ph}$), and elsewhere only their costs are listed. Each circuit sits in a group filled with the color of its box in Fig.~\ref{fig:fft_chain}: green $\mathfrak{U}_{V^2}$, blue $\mathfrak{U}_{V}$, orange $\mathfrak{U}_{X}$, red $W$, purple $\mathcal{T}$. The yellow boxes are the phase gadgets of Fig.~\ref{fig:fft_flag_gadgets}, and the blue and grey blocks inside Figs.~\ref{fig:fft_top_p4} and \ref{fig:uph_rotations} are expanded in the group of the same color.

\begin{figure}[!htbp]
\centering
\includegraphics[width=\linewidth]{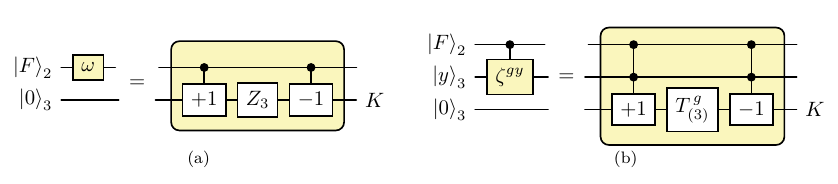}
\caption{Cube- and ninth-root phases on a flag qubit $F$ through a clean qutrit $K$: (a) $\omega^{F}$, (b) $\zeta^{gFy}$.}
\label{fig:fft_flag_gadgets}
\end{figure}

Table~\ref{tab:fft_pieces} collects the cost of every piece as $(T_{(3)},E_{(2,3)},E_{(3,2)})$ plus qubit $T$ and rotations, in the notation of Tab.~\ref{tab:uft_assembled_het}, whose rows are the per-level sums. The $\Delta(27)$ level is free on qutrits; in the qubit encoding each $H_3$ costs $3\,T+10\,R_Z$ and each $\omega^{qr}$ diagonal $4\,R_Z$.

\begin{figure}[!htbp]
\centering
\includegraphics[scale=0.55]{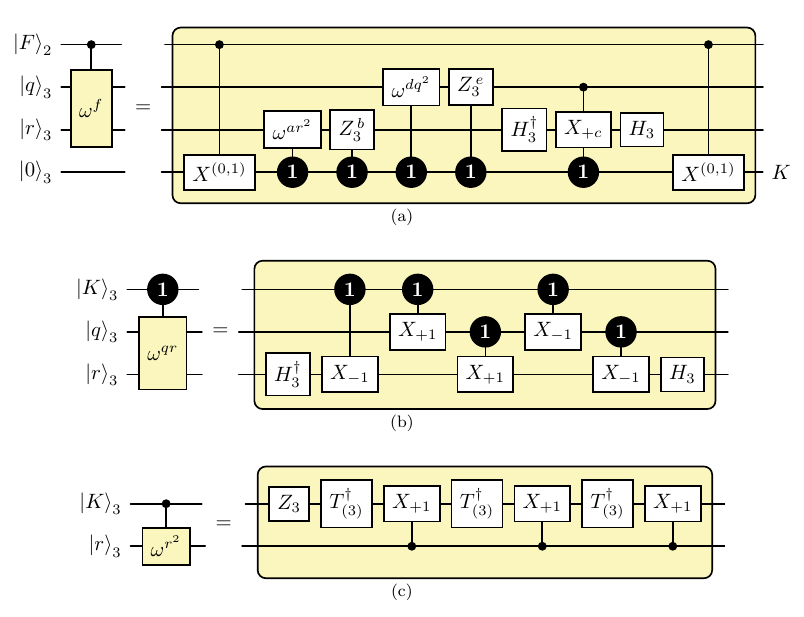}
\caption{Cube-root phases controlled by a flag $F$: (a) $\omega^{Ff}$ for a quadratic $f$ of $q$ and $r$, with $F$ copied into $K$ and back; (b) its cross term; (c) its square.}
\label{fig:fft_cphase_gadgets}
\end{figure}

Every piece follows the same pattern, and the figures are laid out to show it. A dashed box on the left, numbered $\mathcal{F}_1,\mathcal{F}_2,\dots$ in the order the figures appear, computes the sector flags of the piece from the data wires it reads, with mixed entanglers $E_{(3,2)}$ and Toffoli gates. The piece then acts under those flags, and the box $\mathcal{F}_n^{\dagger}$ on the right is the exact mirror that returns the flag wires to $\ket{0}$.

Where a piece is simplest in another basis, a window encloses it: a basis change applied before and undone after, so that only the diagonal part in the middle needs controls. A level ends with its Z-fix, the sign on the induced orbits after $H_2$, applied as a $\mathrm{CZ}$ on a flag that is kept computed across $H_2$.

\begin{figure}[!htbp]
\centering
\includegraphics[width=\linewidth]{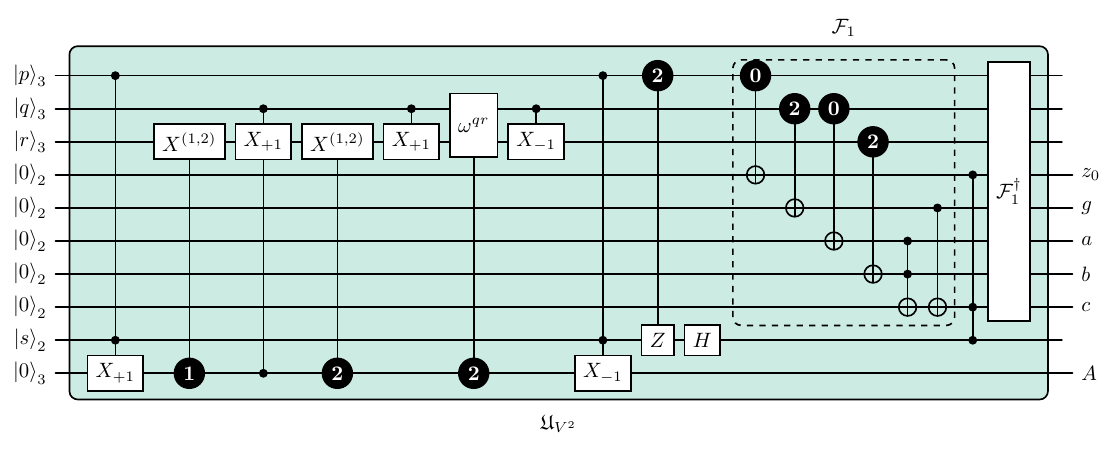}
\caption{Mixed-register circuit for $\mathfrak{U}_{V^2}$.}
\label{fig:fft_l3}
\end{figure}

The control conventions are those of Sec.~\ref{sec:quantumgates}; $X_{\pm k}$ on a qutrit adds $\pm k$ to it under the drawn controls (App.~\ref{app:toffolistaircase}), and a box $\omega^{f}$ with a control applies $\omega^{vf}$, $v$ the control value. Cube-root phases on a flag qubit go through a clean qutrit $K$ (Fig.~\ref{fig:fft_flag_gadgets}), and a phase controlled by a qubit flag is paid for as in Fig.~\ref{fig:fft_cphase_gadgets}, which fixes the price of each flag-controlled phase in Tab.~\ref{tab:fft_pieces}; in the qubit encoding $K$ is a pair that accumulates the phases. The relabeling $\tilde r=r+q$ is one $\textsc{CSum}$, free on qutrits and $28\,T$ in the qubit encoding.

\begin{figure}[!htbp]
\centering
\includegraphics[width=\linewidth]{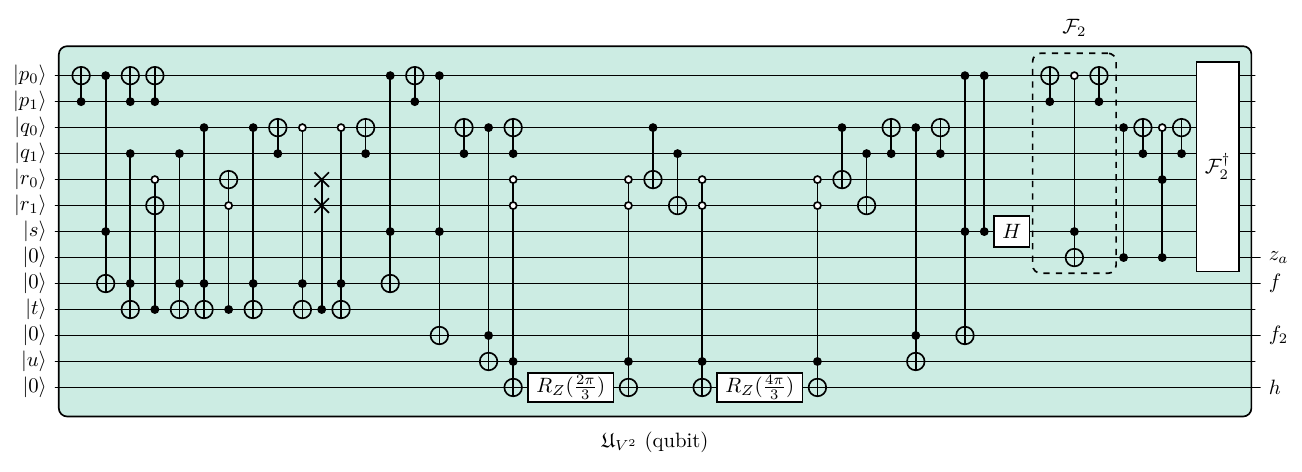}
\caption{Qubit circuit for $\mathfrak{U}_{V^2}$.}
\label{fig:fft_l3_qubit}
\end{figure}

The chain opens with the $V^2$ level (Fig.~\ref{fig:fft_l3}), which negates $\tilde r$ on the two three-dimensional irreps. A qutrit flag $A=s\,p$ selects the copy, a transposition and a controlled addition carry out the negation, and the $p{=}2$ copy picks up its twiddle phase on the way. The qubit circuit (Fig.~\ref{fig:fft_l3_qubit}) realizes the same permutation through three Toffoli branches and the phases by decode-and-rotate, and is read the same way.

\begin{figure}[!htbp]
\centering
\includegraphics[width=\linewidth]{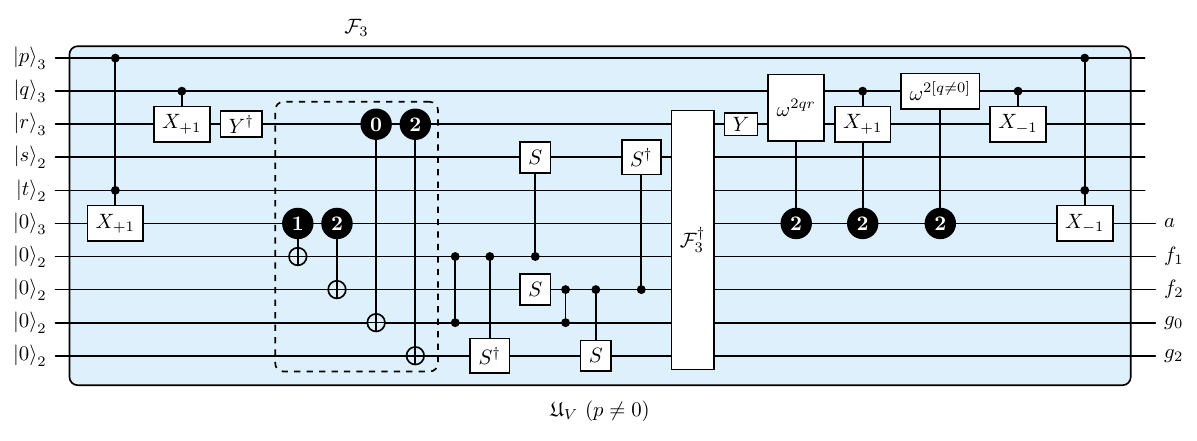}
\caption{Mixed-register circuit for $\mathfrak{U}_{V}$, the $p\neq0$ part.}
\label{fig:fft_l4}
\end{figure}

The $V$ level comes next (Figs.~\ref{fig:fft_l4} and \ref{fig:fft_l4b}), a controlled qutrit Hadamard. For $p\neq0$ a qutrit flag $a=t\,p$ selects $H_3$ or $H_3^{\dagger}$ on $\tilde r$. Both are diagonal in the same basis $Y$, so the window $Y^{\dagger}\cdots Y$ needs no control and encloses only controlled fourth-root phases; the $p{=}2$ copy adds the cube-root phases and the relabeling that turn $H_3$ into $H_3^{\dagger}$ together with its twiddle (Tab.~\ref{tab:dd_blockmap}).

The $p{=}0$ part follows in Fig.~\ref{fig:fft_l4b}: the basis change $W$ and the twiddle act under the flag $w$, which stays computed across $H_2$ so that the Z-fix is free. In the qubit encoding (Fig.~\ref{fig:fft_l4_qubit}) the structure is the same with pairs in place of qutrits: $Y$ becomes two Givens rotations (Fig.~\ref{fig:fft_l4_qubit_y}), the flags are Toffolis, the negations of $q$ and $r$ are controlled swaps of their bits, and the cube-root phases are applied by decoding the needed value tests into ancillas and rotating.

\begin{figure}[!htbp]
\centering
\includegraphics[scale=0.52]{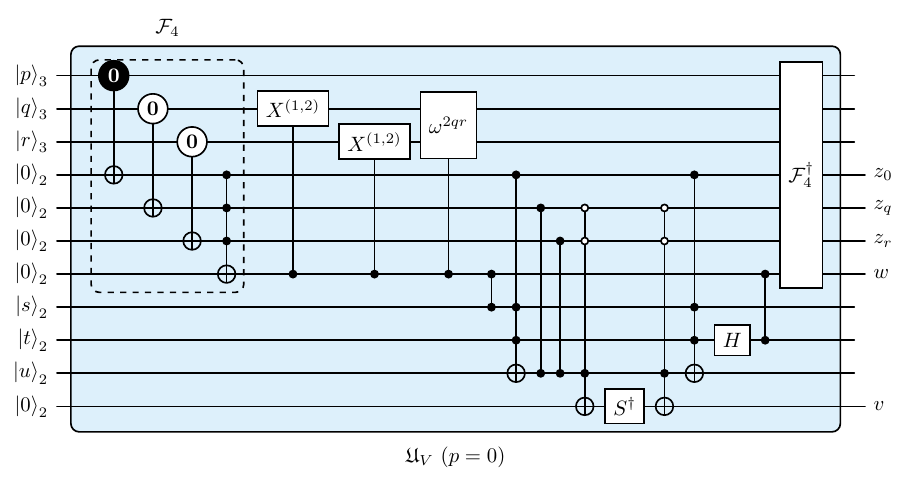}
\caption{Mixed-register circuit for the $p=0$ part of $\mathfrak{U}_{V}$.}
\label{fig:fft_l4b}
\end{figure}

The $X$ level is the first with several sectors: nine pieces, selected by $p$, $s$, $t$ and, for the twiddle, $u$, in six figures. Pieces a, b, e, f are qutrit Fourier gates with quadratic cube-root phases before and after. Inside the shared window $Y^{\dagger}\cdots Y$ each reduces to one diagonal $\Lambda_Y^{\pm1}$ (Figs.~\ref{fig:fft_l5_h3} and \ref{fig:fft_l5_h3e}), drawn as a single box whose interior is Fig.~\ref{fig:fft_lam_gadget}. Pieces c and d swap two values of $\tilde r$ (Fig.~\ref{fig:fft_l5_tr}).

\begin{figure}[!tbp]
\centering
\includegraphics[width=\linewidth]{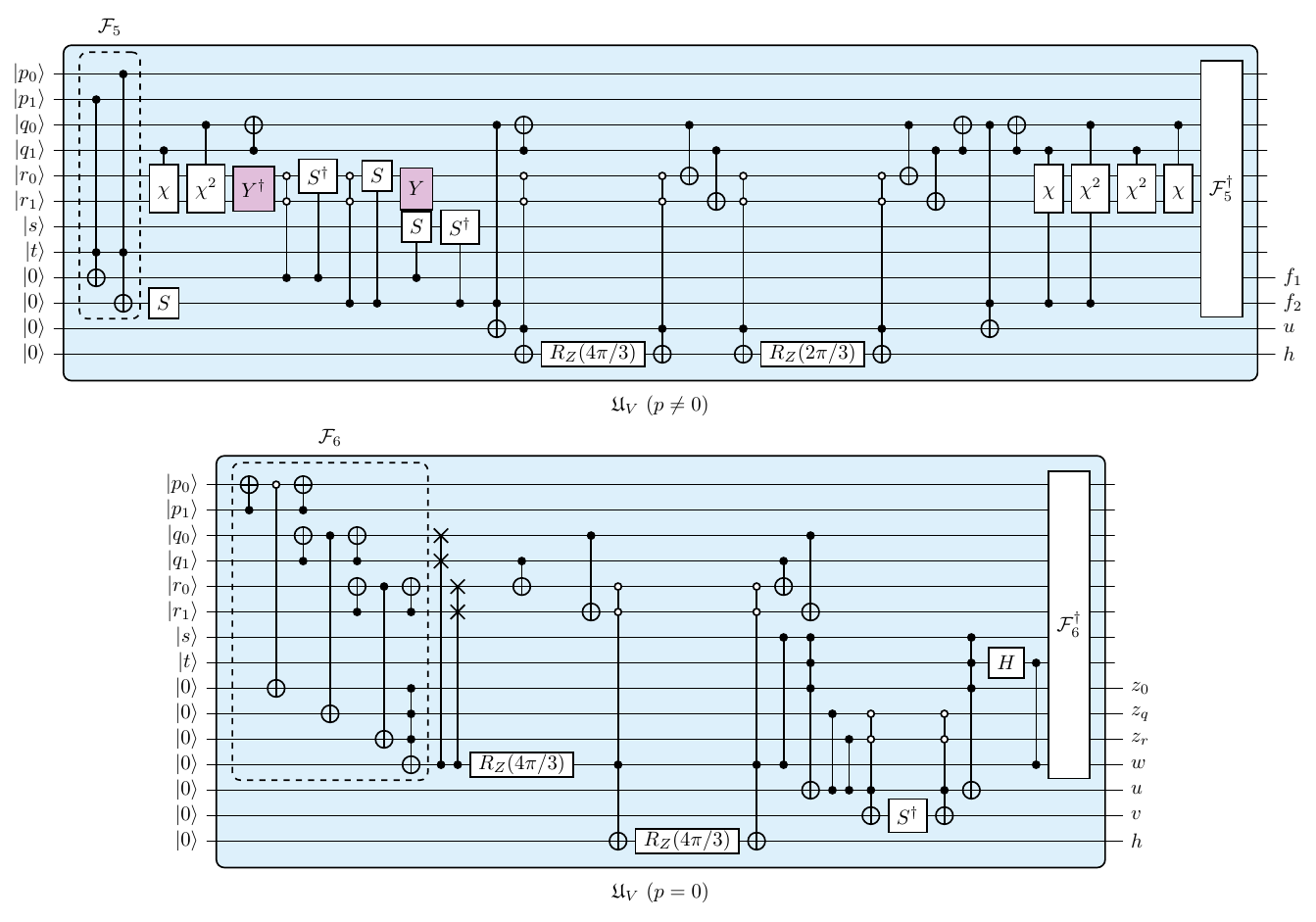}
\caption{Qubit circuit for $\mathfrak{U}_{V}$, the $p\neq0$ part (top) and the $p=0$ part (bottom); $Y$ is Fig.~\ref{fig:fft_l4_qubit_y}.}
\label{fig:fft_l4_qubit}
\end{figure}

\begin{figure}[!htbp]
\centering
\includegraphics[scale=0.533]{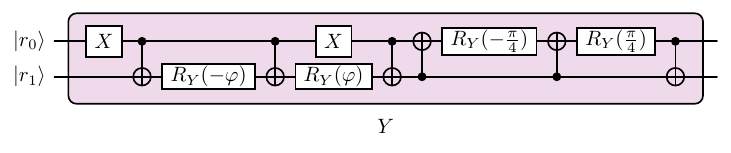}
\caption{Qubit circuit for $Y$, used in Figs.~\ref{fig:fft_l4_qubit}, \ref{fig:fft_l5_merged} and \ref{fig:fft_l5_merged0}.}
\label{fig:fft_l4_qubit_y}
\end{figure}

\begin{figure}[!htbp]
\centering
\includegraphics[scale=0.65]{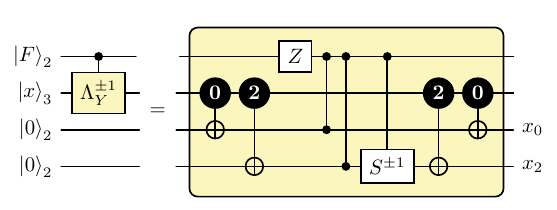}
\caption{The window phase $\Lambda_Y^{\pm1}$ on $x$ controlled by $F$.}
\label{fig:fft_lam_gadget}
\end{figure}

On $p{=}0$ the three remaining pieces are simpler: g is a sign on the one-dimensional state, and h and i are permutations conjugated by $H\otimes H$, with cube-root phases on a qutrit flag (Figs.~\ref{fig:fft_l5_g}, \ref{fig:fft_l5_p0} and \ref{fig:fft_l5_p0i}).

\begin{figure}[!htbp]
\centering
\includegraphics[scale=0.533]{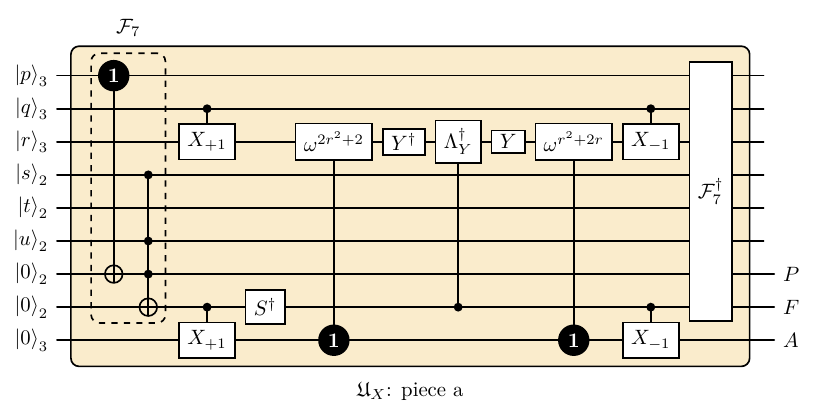}
\caption{Mixed-register circuit for piece a of $\mathfrak{U}_{X}$; piece b is the same on $(2,0)$ with $\Lambda_Y$ in place of $\Lambda_Y^{\dagger}$.}
\label{fig:fft_l5_h3}
\end{figure}

\begin{figure}[!htbp]
\centering
\includegraphics[width=\linewidth]{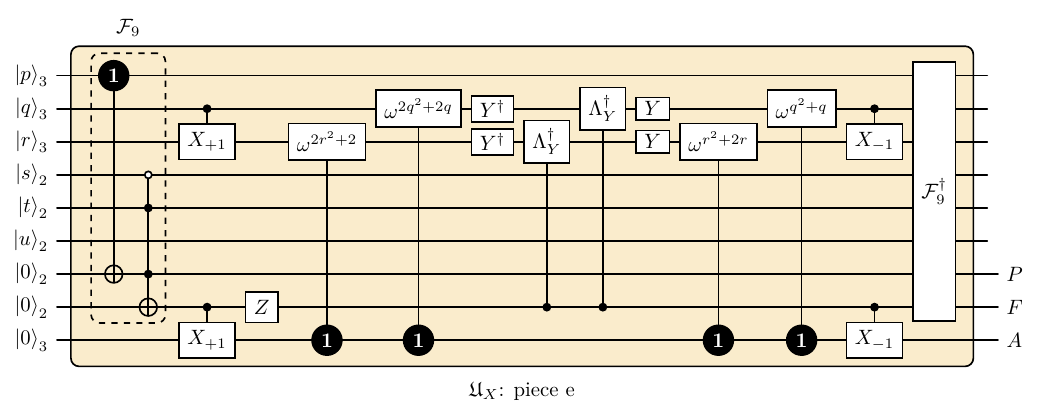}
\caption{Mixed-register circuit for piece e of $\mathfrak{U}_{X}$; piece f is the same on $(2,1,1)$ with $\Lambda_Y$ in place of $\Lambda_Y^{\dagger}$.}
\label{fig:fft_l5_h3e}
\end{figure}

\begin{figure}[!htbp]
\centering
\includegraphics[scale=0.533]{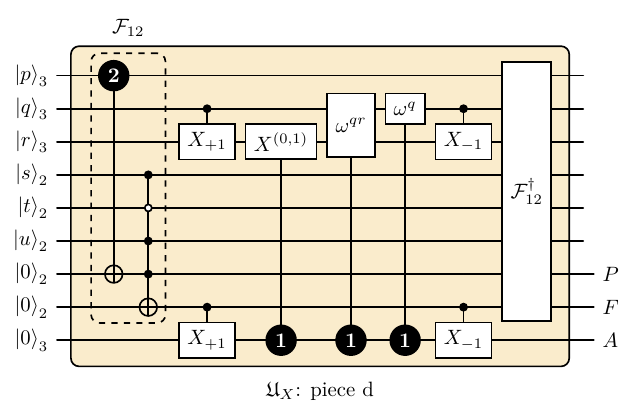}
\caption{Mixed-register circuit for piece d of $\mathfrak{U}_{X}$; piece c is d without the phase, with $X^{(0,2)}$ on $(1,0,0)$.}
\label{fig:fft_l5_tr}
\end{figure}

\begin{figure}[!htbp]
\centering
\includegraphics[scale=0.533]{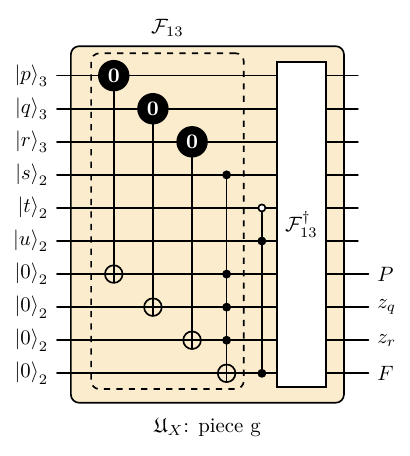}
\caption{Mixed-register circuit for piece g of $\mathfrak{U}_{X}$.}
\label{fig:fft_l5_g}
\end{figure}

\begin{figure*}[!t]
\centering
\includegraphics[width=0.93\textwidth]{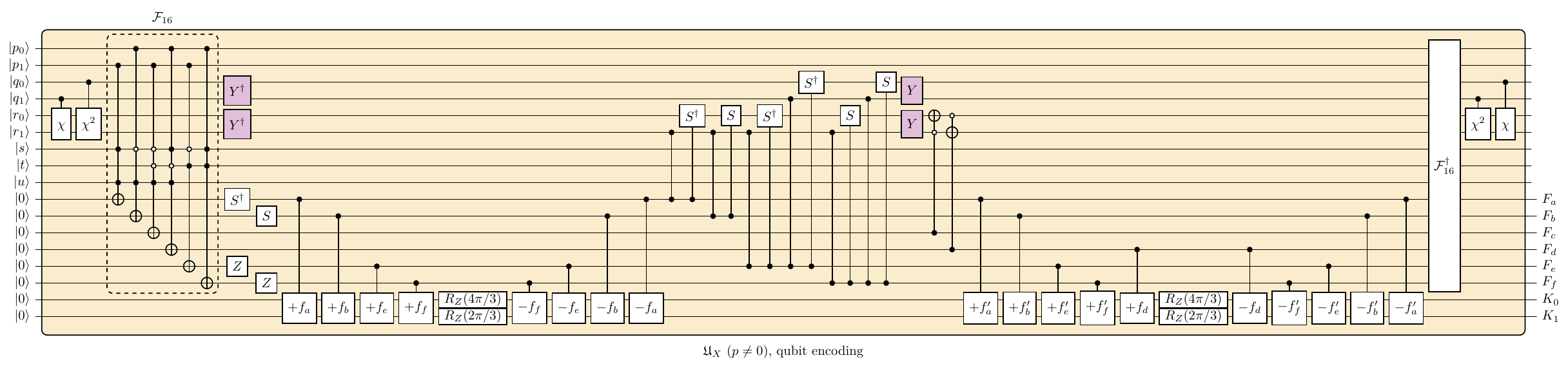}
\caption{Qubit circuit for the $p\neq0$ pieces of $\mathfrak{U}_{X}$: $p$, $q$, $r$ and $K$ are the pairs $x_0x_1$, and a box $\pm f_F$ adds $\pm F\,f$ to the phase register $K$ as in Fig.~\ref{fig:fft_kadd_gadget}, with $f$ the quadratic of the piece's phase box before the window and $f'$ the one after (Figs.~\ref{fig:fft_l5_h3} and \ref{fig:fft_l5_h3e}).}
\label{fig:fft_l5_merged}
\end{figure*}

\begin{figure*}[!t]
\centering
\includegraphics[width=0.93\textwidth]{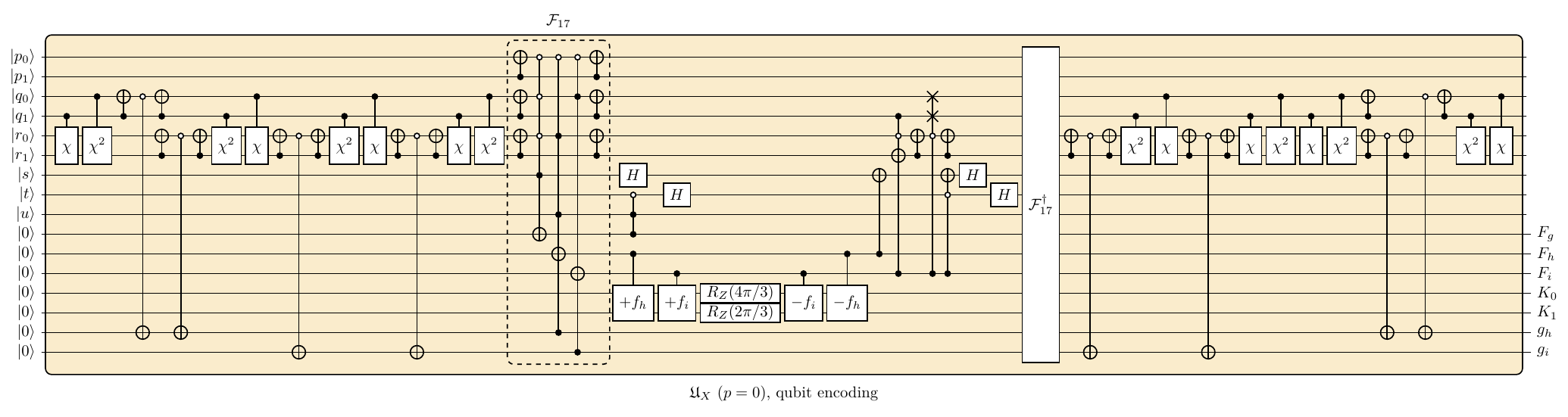}
\caption{Qubit circuit for the $p=0$ pieces of $\mathfrak{U}_{X}$; conventions as in Fig.~\ref{fig:fft_l5_merged}.}
\label{fig:fft_l5_merged0}
\end{figure*}

\begin{figure}[!htbp]
\centering
\includegraphics[width=\linewidth]{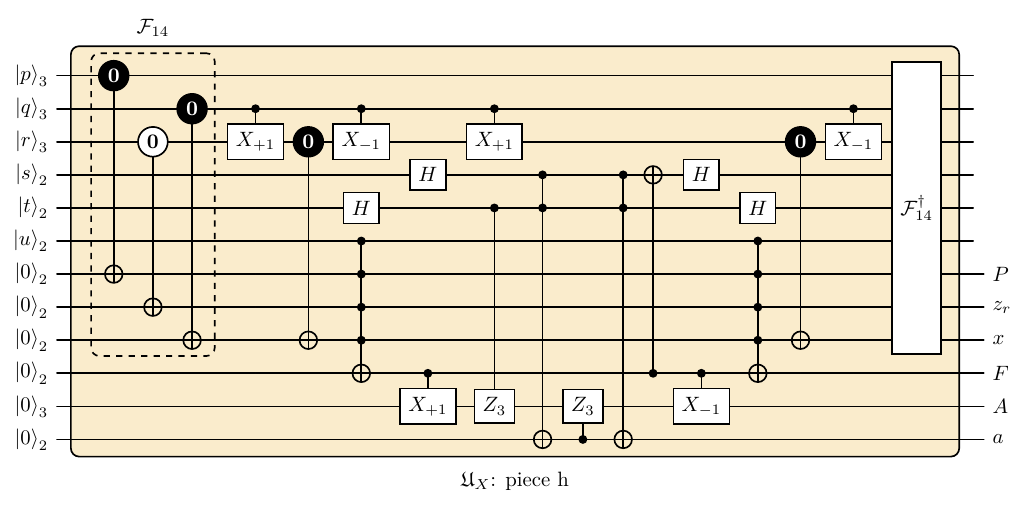}
\caption{Mixed-register circuit for piece h of $\mathfrak{U}_{X}$.}
\label{fig:fft_l5_p0}
\end{figure}

\begin{figure}[ht]
\centering
\includegraphics[width=\linewidth]{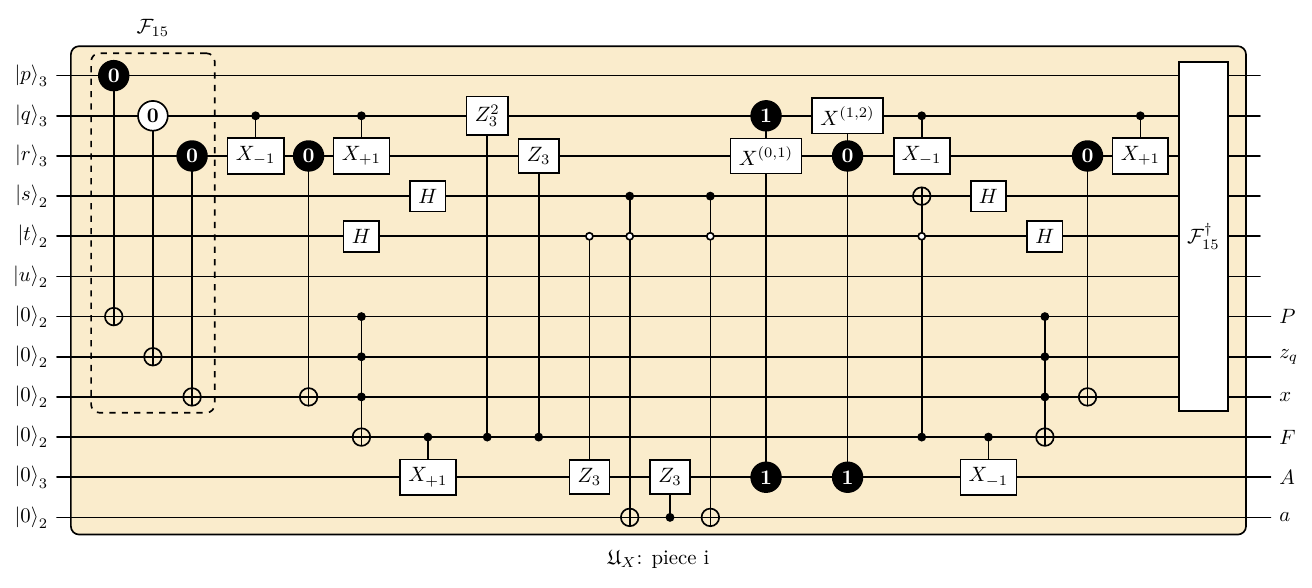}
\caption{Mixed-register circuit for piece i of $\mathfrak{U}_{X}$: the basis change $W$ on the other four $p=0$ states.}
\label{fig:fft_l5_p0i}
\end{figure}

\begin{figure}[!htbp]
\centering
\includegraphics[scale=0.65]{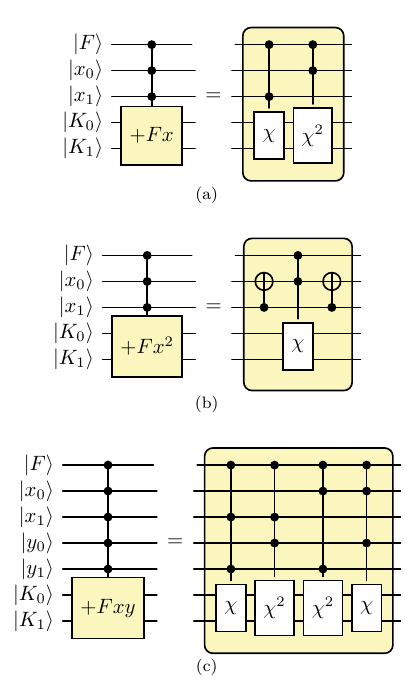}
\caption{Qubit additions into the phase register $K$ for a flag $F$ and pairs $x=2x_0+x_1$: (a) $K\mathrel{+}=Fx$, bit $x_1$ weighs one and $x_0$ two; (b) $K\mathrel{+}=Fx^2$ through $[x\neq0]=x_0\oplus x_1$; (c) $K\mathrel{+}=Fxy$. A coefficient multiplies the powers of $\chi$, with $2\equiv-1$; each box is a controlled sum of Tab.~\ref{tab:csumcosts}.}
\label{fig:fft_kadd_gadget}
\end{figure}

In the qubit encoding the nine pieces are not drawn one by one. Their flags stay computed together, and Figs.~\ref{fig:fft_l5_merged} and \ref{fig:fft_l5_merged0} show the six $p\neq0$ pieces and the three $p{=}0$ pieces as two circuits that share what they can: one relabeling $\tilde r$, one window ($Y^{\dagger}\cdots Y$ on $q$ and $\tilde r$, or $H\otimes H$), and a phase register $K$ that accumulates all cube-root phases of a window before a single $Z_3$. Value tests on a pair are single bits or one \textsc{cnot}.

The top step $\mathcal{T}$ is drawn piece by piece, in the order applied. $\mathcal{T}_1$, the basis change $W$ on the four orbit partners, is Fig.~\ref{fig:fft_top_p1} in Sec.~\ref{sec:fouriertransform}. Figures~\ref{fig:fft_top_p2}--\ref{fig:fft_top_p6} show the three orbit relabelings ($\mathcal{T}_2$), the fixed $\mathbf{2}$ and $\mathbf{8}$ ($\mathcal{T}_3$, $\mathcal{T}_4$) and the fixed $\mathbf{3}$'s and $\mathbf{6}$'s ($\mathcal{T}_5$, $\mathcal{T}_6$); $H_3$ on $\ket{y}$ closes the step. In each orbit piece the relabeling $y\to a-y$ is a transposition on $\ket{y}$ under the sector flag, with the flag-controlled additions after it and the closing phase before it; in the fixed pieces the flag controls only the phases, and the windows $U_3$ and $B$ are uncontrolled.

The window $B$ of $\mathcal{T}_6$ is the only block of the mixed chain with generic rotation angles. It enters through its cosine--sine decomposition with $u$ as the outer index,
\begin{equation}
\label{eq:csdA}
B=(L_1\oplus L_2)\,R_Y(\theta_{\tilde r})\,(R_1\oplus R_2),
\end{equation}
two uncontrolled $U(3)$ on $\tilde r$ ($R_1$, $L_1$), two $u$-controlled $U(3)$ ($R_2R_1^{\dagger}$, $L_2L_1^{\dagger}$), and the rotation $R_Y(\theta_{\tilde r})$ of $u$ controlled on the value of $\tilde r$. The controlled $U(3)$ is opened in Fig.~\ref{fig:fft_p6_gadgets}, and Tab.~\ref{tab:csd_angles} lists the angles. In the qubit encoding the same pieces use the costs of App.~\ref{app:toffolistaircase}, with the cube-root phases on a shared phase register and the windows $B^{\dagger}$, $B$ as quantum Shannon decompositions of their $8\times8$ embedding.

\begin{figure}[!htbp]
\centering
\includegraphics[width=\linewidth]{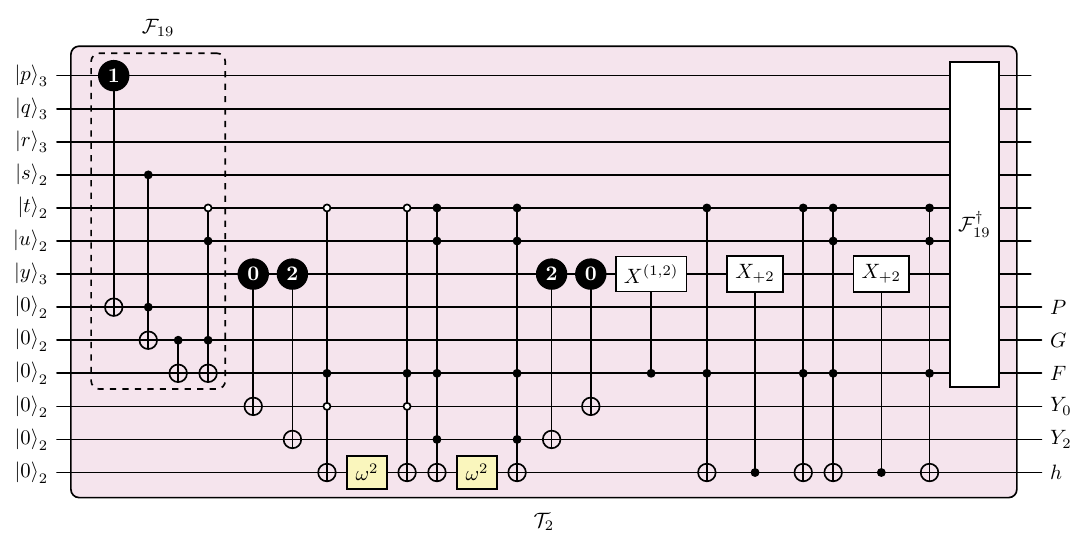}
\caption{Mixed-register circuit for $\mathcal{T}_2$, the orbit of $\mathbf{3}$'s; the orbits of $\mathbf{1}$'s and $\mathbf{3}^{*}$'s differ only in the sector flag and in $a(t,u)$ (Tab.~\ref{tab:dd_blockmap}).}
\label{fig:fft_top_p2}
\end{figure}

\begin{figure}[!htbp]
\centering
\includegraphics[scale=0.533]{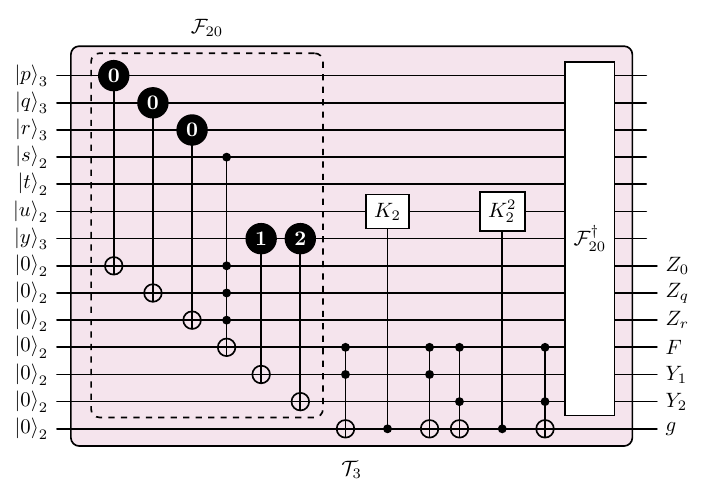}
\caption{Mixed-register circuit for $\mathcal{T}_3$, the $\mathbf{2}$.}
\label{fig:fft_top_p3}
\end{figure}

\begin{figure}[!htbp]
\centering
\includegraphics[width=\linewidth]{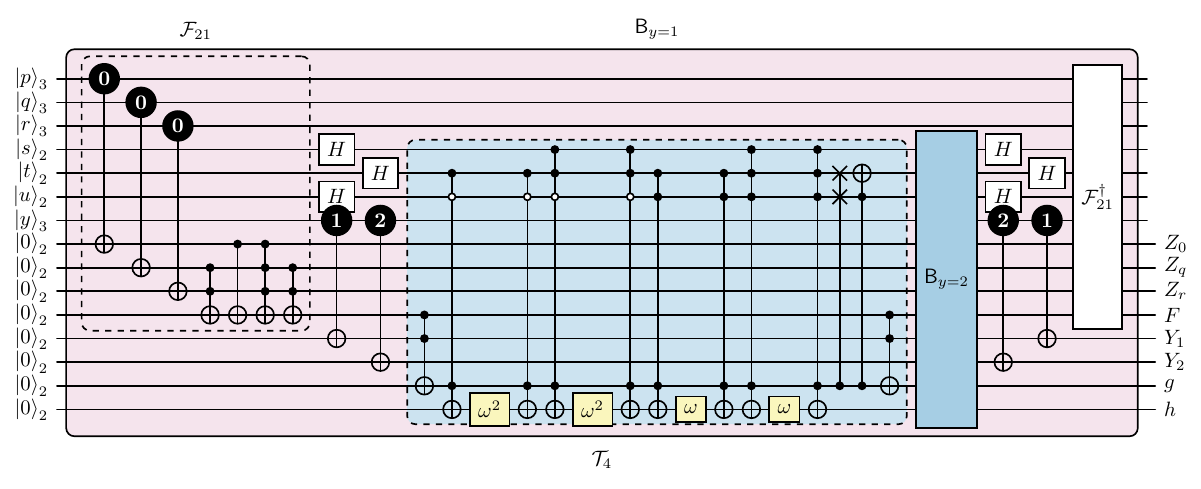}
\caption{Mixed-register circuit for $\mathcal{T}_4$, the $\mathbf{8}$; $\mathsf{B}_{y=2}$ repeats $\mathsf{B}_{y=1}$ with the phases of $M_8^{2}$ and the cycle applied twice.}
\label{fig:fft_top_p4}
\end{figure}

\begin{figure}[!htbp]
\centering
\includegraphics[scale=0.533]{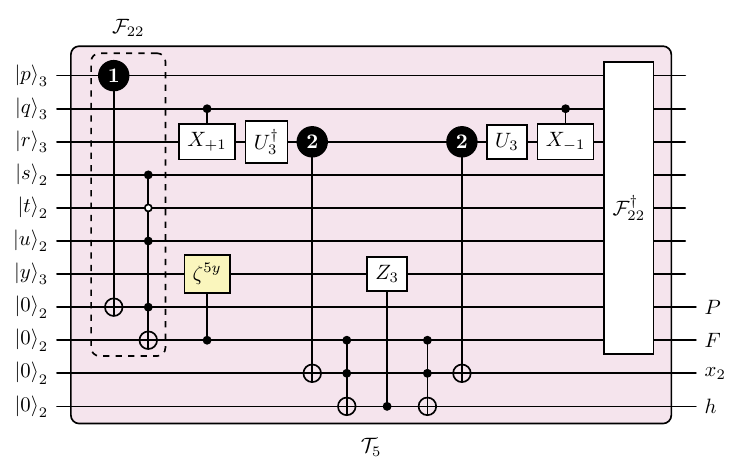}
\caption{Mixed-register circuit for $\mathcal{T}_5$, the fixed $\mathbf{3}$; the $\mathbf{3}^{*}$ is the same on $p=2$, after the phase $\omega^{2q\tilde r}$, with $\zeta^{y}$ and $\omega^{2y[\tilde r=2]}$.}
\label{fig:fft_top_p5}
\end{figure}

\begin{figure}[!htbp]
\centering
\includegraphics[width=\linewidth]{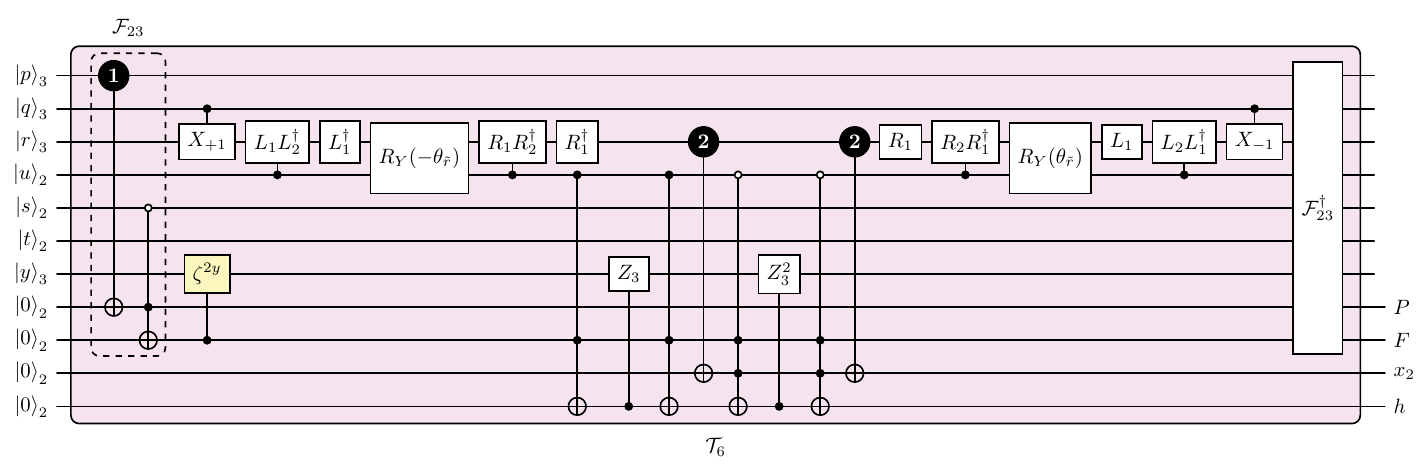}
\caption{Mixed-register circuit for $\mathcal{T}_6$, the fixed $\mathbf{6}$ ($B$: Eq.~(\ref{eq:csdA}), Tab.~\ref{tab:csd_angles}); the $\mathbf{6}^{*}$ is the same on $p=2$ after $\omega^{2q\tilde r}$.}
\label{fig:fft_top_p6}
\end{figure}

\begin{figure}[!htbp]
\centering
\includegraphics[width=\linewidth]{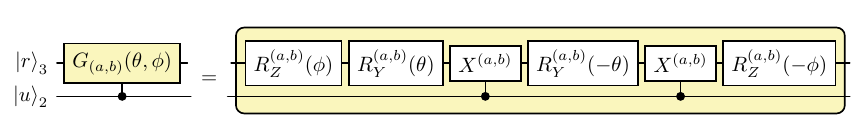}
\caption{The $u$-controlled two-level rotation $G_{(a,b)}(\theta,\phi)$ of Fig.~\ref{fig:fft_top_p6}; the rotation of $u$ controlled on $\tilde r=k$ is the usual controlled rotation with $E_{(3,2)}$ in place of the \textsc{cnot}.}
\label{fig:fft_p6_gadgets}
\end{figure}

\input{csd_angles_table}

The kinetic-phase gate is drawn in the qubit encoding in Figs.~\ref{fig:uph_rotations} and \ref{fig:uph_flags} of Sec.~\ref{sec:uph_impl}: the flag block computes the value tests once, each rotation is controlled on the flags of its class, and the mirror block clears them. On the mixed architecture the tests become six value flags on six extra qubits, with the control polarities on $[p{=}0]$ adjusted; the Toffoli count and the ten rotations are unchanged.

\FloatBarrier
\setlength{\textfloatsep}{20pt plus 2pt minus 4pt}\setlength{\floatsep}{12pt plus 2pt minus 2pt}\setlength{\dbltextfloatsep}{20pt plus 2pt minus 4pt}\setlength{\intextsep}{12pt plus 2pt minus 2pt}
\onecolumngrid
\section{Native-schedule tables for \texorpdfstring{$\mathfrak{U}_{\rm Tr}^{\dsb}$}{UTr}}
\label{app:utr-tables}
\input{utr_angle_table}
\twocolumngrid

\clearpage

\begin{turnpage}

\input{rep_tables2}
\end{turnpage}

\end{document}

%% file: csd_angles_table.tex
\begin{table}[!htbp]
\caption{Angles of the cosine--sine form of $B$ for the fixed $\mathbf{6}$ and $\mathbf{6}^{*}$ (Eq.~(\ref{eq:csdA})), rounded to $10^{-4}$. Each $U(3)$ block is $G_{(1,2)}(\theta_1,\phi_1)\,G_{(0,1)}(\theta_2,\phi_2)\,G_{(1,2)}(\theta_3,\phi_3)$ followed by the phases $e^{i\alpha_0},e^{i\alpha_1},e^{i\alpha_2}$, with $G_{(a,b)}(\theta,\phi)$ the two-level rotation of Fig.~\ref{fig:fft_p6_gadgets}; $\vartheta_k$ is the rotation of $u$ controlled on $\tilde r=k$, $\vartheta_{0,1,2}=0.1631,\ 0.6316,\ 1.3653$ for both.}
\label{tab:csd_angles}
\scriptsize
\setlength{\tabcolsep}{2.2pt}
\begin{tabular}{l|rrrr|rrrr}
\hline\hline
 & \multicolumn{4}{c|}{$\mathbf{6}$} & \multicolumn{4}{c}{$\mathbf{6}^{*}$} \\
 & $R_1$ & $R_2R_1^{\dagger}$ & $L_1$ & $L_2L_1^{\dagger}$ & $R_1$ & $R_2R_1^{\dagger}$ & $L_1$ & $L_2L_1^{\dagger}$ \\\hline
$\theta_1$ & 0.4243 & 1.1615 & 0.9830 & 0.7054 & 0.3510 & 0.9711 & 0.5878 & 0.8654 \\
$\phi_1$ & -3.1416 & -0.5203 & 2.4032 & 3.0719 & -3.1416 & -1.5899 & 2.4032 & 3.0719 \\
$\theta_2$ & 0.9804 & 0.5744 & 1.0827 & 1.4424 & 1.0985 & 1.2361 & 1.0827 & 1.4424 \\
$\phi_2$ & -3.1416 & -0.9735 & 2.4101 & 3.0257 & -3.1416 & -0.5398 & 1.4699 & 0.1856 \\
$\theta_3$ & 1.2569 & 1.0055 & 0.4521 & 1.5085 & 1.0913 & 0.3447 & 0.4521 & 0.0623 \\
$\phi_3$ & -0.8801 & -0.7126 & 0.5926 & -0.4512 & -1.5879 & 1.5058 & 1.0722 & 0.3119 \\\hline
$\alpha_0$ & -3.1416 & -1.4610 & -1.4216 & 2.3004 & -3.1416 & 2.1177 & 1.2221 & -2.3004 \\
$\alpha_1$ & 0.7436 & 0.1398 & -1.6292 & 3.0098 & -1.1533 & -0.6385 & -1.8171 & -2.1622 \\
$\alpha_2$ & 2.8474 & -2.2362 & 1.9959 & 2.5437 & -1.2271 & -2.1113 & 3.0502 & 2.8918 \\
\hline\hline
\end{tabular}
\end{table}

%% file: utr_angle_table.tex

\noindent Every rotation in the schedule is exact: with $c_1=\cos(\pi/9)$ and
$c_2=\cos(2\pi/9)$,
\[
R_Z\!\left(\tfrac{\theta}{24}\,(n_0 + n_1 c_1 + n_2 c_2)\right),
\qquad n_i \in \mathbb{Z},\ |n_i| \le 6 .
\]
Table~\ref{tab:utr-gadgets} lists the class-gadget labels and rotation counts by stage. Table~\ref{tab:utr-angles} specifies the angles through the integer triples $(n_0,n_1,n_2)$. Stages A, B, and C correspond to the flags $[\ell=1]$, $[\ell\neq0]$, and $[\ell=2]$, respectively, with the class index $\ell=p+m_1q+m_2r+m_3y\bmod3$ folded onto $p$.
\begin{table}[!ht]
\caption{Class-gadget index $G$, label $m=m_1m_2m_3$ with $m_i\in\mathbb{Z}_3$, and rotation counts $a_m,b_m,c_m$ for stages A, B, C. The gadget couples $p$ to the qutrits with $m_i\neq0$.}\label{tab:utr-gadgets}
\centering
\setlength{\tabcolsep}{4pt}
\begin{tabular}{rlrrr@{\hspace{7pt}}rlrrr@{\hspace{7pt}}rlrrr}
\toprule
$G$ & $m$ & $a_m$ & $b_m$ & $c_m$ & $G$ & $m$ & $a_m$ & $b_m$ & $c_m$ & $G$ & $m$ & $a_m$ & $b_m$ & $c_m$\\
\midrule
1 & $000$ & 2 & 2 & 4 & 10 & $100$ & 3 & 1 & 4 & 19 & $200$ & 2 & 3 & 3\\
2 & $001$ & 3 & 3 & 2 & 11 & $101$ & 3 & 3 & 2 & 20 & $201$ & 4 & 0 & 4\\
3 & $002$ & 4 & 2 & 2 & 12 & $102$ & 4 & 0 & 4 & 21 & $202$ & 3 & 3 & 2\\
4 & $010$ & 3 & 3 & 6 & 13 & $110$ & 4 & 2 & 4 & 22 & $210$ & 4 & 1 & 5\\
5 & $011$ & 6 & 1 & 5 & 14 & $111$ & 4 & 1 & 5 & 23 & $211$ & 5 & 0 & 5\\
6 & $012$ & 6 & 1 & 5 & 15 & $112$ & 4 & 2 & 4 & 24 & $212$ & 5 & 0 & 5\\
7 & $020$ & 4 & 4 & 4 & 16 & $120$ & 3 & 4 & 3 & 25 & $220$ & 3 & 3 & 4\\
8 & $021$ & 5 & 1 & 6 & 17 & $121$ & 5 & 0 & 5 & 26 & $221$ & 5 & 0 & 5\\
9 & $022$ & 6 & 0 & 6 & 18 & $122$ & 5 & 0 & 5 & 27 & $222$ & 5 & 0 & 5\\
\bottomrule
\end{tabular}
\end{table}

\onecolumngrid

\begingroup
\newcommand{\Tstrut}{\rule{0pt}{2.4ex}}
\newcommand{\Bstrut}{\rule[-0.8ex]{0pt}{0pt}}
\setlength{\tabcolsep}{9pt}
\renewcommand{\arraystretch}{1.0}
\setlength{\LTcapwidth}{\textwidth}
\needspace{16\baselineskip}
\begin{longtable}{r cccccc}
\caption{Rotation-angle triples $(n_0,n_1,n_2)$ by gadget $G$ and pair support $S_b$ (the \textsc{cnot}-dressing wires; $\varnothing$ denotes the pure-qutrit pair). Each cell contains the conjugate pair's two stage-tagged triples; a dash denotes an unused support. Every rotation acts on ancilla $a_0$.}\label{tab:utr-angles}\\
\hline\hline
 & \multicolumn{6}{c}{support $S_b$}\\
$G$ & $\varnothing$ & $s$ & $u$ & $st$ & $su$ & $stu$\\
\hline
\endfirsthead
\multicolumn{7}{l}{\emph{Table~\ref{tab:utr-angles}, continued}}\\
\hline\hline
 & \multicolumn{6}{c}{support $S_b$}\\
$G$ & $\varnothing$ & $s$ & $u$ & $st$ & $su$ & $stu$\\
\hline
\endhead
\multicolumn{7}{r}{\emph{continued}}\\
\endfoot
\hline\hline
\endlastfoot
\multirow{2}{*}{$1$} & \multirow{2}{*}{---} & \Tstrut A\,$(-1,0,6)$ & \multirow{2}{*}{---} & B\,$(3,0,-2)$ & B\,$(1,0,2)$ & A\,$(1,0,-6)$\\
 &  & \Bstrut C\,$(2,0,2)$ &  & C\,$(-1,0,-4)$ & C\,$(1,0,-4)$ & C\,$(-2,0,-2)$\\
\hline
\multirow{2}{*}{$2$} & \multirow{2}{*}{---} & \Tstrut A\,$(-3,-4,0)$ & \multirow{2}{*}{---} & B\,$(3,2,0)$ & A\,$(-1,-4,0)$ & A\,$(1,6,0)$\\
 &  & \Bstrut B\,$(2,-2,0)$ &  & C\,$(-1,4,0)$ & B\,$(2,2,0)$ & C\,$(-2,2,0)$\\
\hline
\multirow{2}{*}{$3$} & \multirow{2}{*}{---} & \Tstrut A\,$(-1,6,-6)$ & \multirow{2}{*}{---} & A\,$(1,4,-4)$ & A\,$(-1,4,-4)$ & A\,$(1,-6,6)$\\
 &  & \Bstrut C\,$(2,2,-2)$ &  & B\,$(2,-6,6)$ & B\,$(2,-2,2)$ & C\,$(-2,-2,2)$\\
\hline
\multirow{2}{*}{$4$} & A\,$(-1,0,-2)$ & \Tstrut A\,$(0,0,2)$ & A\,$(1,0,2)$ & B\,$(1,0,0)$ & B\,$(2,0,-2)$ & B\,$(1,0,0)$\\
 & C\,$(-2,0,2)$ & \Bstrut C\,$(1,0,0)$ & C\,$(2,0,-2)$ & C\,$(0,0,-2)$ & C\,$(-1,0,-2)$ & C\,$(0,0,-2)$\\
\hline
\multirow{2}{*}{$5$} & A\,$(-1,2,0)$ & \Tstrut A\,$(0,-2,0)$ & A\,$(-1,-4,0)$ & A\,$(1,0,0)$ & A\,$(2,2,0)$ & A\,$(1,0,0)$\\
 & C\,$(-2,-2,0)$ & \Bstrut C\,$(1,0,0)$ & B\,$(2,2,0)$ & C\,$(1,2,0)$ & C\,$(1,4,0)$ & C\,$(1,2,0)$\\
\hline
\multirow{2}{*}{$6$} & A\,$(-1,-2,2)$ & \Tstrut A\,$(0,2,-2)$ & A\,$(1,2,-2)$ & A\,$(1,0,0)$ & A\,$(1,2,-2)$ & A\,$(1,0,0)$\\
 & C\,$(-2,2,-2)$ & \Bstrut C\,$(1,0,0)$ & C\,$(2,-2,2)$ & C\,$(1,-2,2)$ & B\,$(1,-4,4)$ & C\,$(1,-2,2)$\\
\hline
\multirow{2}{*}{$7$} & A\,$(1,0,2)$ & \Tstrut A\,$(-1,0,2)$ & A\,$(-1,0,-2)$ & B\,$(1,0,0)$ & A\,$(1,0,2)$ & B\,$(1,0,0)$\\
 & C\,$(2,0,-2)$ & \Bstrut B\,$(1,0,0)$ & C\,$(-2,0,2)$ & C\,$(0,0,-2)$ & B\,$(1,0,-4)$ & C\,$(0,0,-2)$\\
\hline
\multirow{2}{*}{$8$} & A\,$(1,-2,0)$ & \Tstrut A\,$(0,-2,0)$ & A\,$(-1,2,0)$ & A\,$(1,0,0)$ & B\,$(2,2,0)$ & A\,$(1,0,0)$\\
 & C\,$(2,2,0)$ & \Bstrut C\,$(1,0,0)$ & C\,$(-2,-2,0)$ & C\,$(1,2,0)$ & C\,$(-1,2,0)$ & C\,$(1,2,0)$\\
\hline
\multirow{2}{*}{$9$} & A\,$(1,2,-2)$ & \Tstrut A\,$(0,2,-2)$ & A\,$(-1,-2,2)$ & A\,$(1,0,0)$ & A\,$(2,-2,2)$ & A\,$(1,0,0)$\\
 & C\,$(2,-2,2)$ & \Bstrut C\,$(1,0,0)$ & C\,$(-2,2,-2)$ & C\,$(1,-2,2)$ & C\,$(1,-4,4)$ & C\,$(1,-2,2)$\\
\hline
\multirow{2}{*}{$10$} & \multirow{2}{*}{---} & \Tstrut A\,$(0,0,2)$ & \multirow{2}{*}{---} & B\,$(2,0,2)$ & A\,$(2,0,-2)$ & A\,$(0,0,-2)$\\
 &  & \Bstrut C\,$(1,0,0)$ &  & C\,$(1,0,-6)$ & C\,$(1,0,-4)$ & C\,$(-1,0,0)$\\
\hline
\multirow{2}{*}{$11$} & \multirow{2}{*}{---} & \Tstrut A\,$(-1,-2,0)$ & \multirow{2}{*}{---} & B\,$(2,-2,0)$ & A\,$(2,2,0)$ & A\,$(1,2,0)$\\
 &  & \Bstrut B\,$(1,0,0)$ &  & C\,$(1,6,0)$ & C\,$(1,4,0)$ & B\,$(-1,0,0)$\\
\hline
\multirow{2}{*}{$12$} & \multirow{2}{*}{---} & \Tstrut A\,$(0,2,-2)$ & \multirow{2}{*}{---} & A\,$(2,2,-2)$ & A\,$(2,-2,2)$ & A\,$(0,-2,2)$\\
 &  & \Bstrut C\,$(1,0,0)$ &  & C\,$(3,-4,4)$ & C\,$(1,-4,4)$ & C\,$(-1,0,0)$\\
\hline
\multirow{2}{*}{$13$} & A\,$(3,0,-6)$ & \Tstrut B\,$(2,0,0)$ & A\,$(-1,0,4)$ & A\,$(-1,0,2)$ & \multirow{2}{*}{---} & A\,$(-1,0,2)$\\
 & C\,$(0,0,-6)$ & \Bstrut C\,$(0,0,-4)$ & B\,$(2,0,-2)$ & C\,$(0,0,2)$ &  & C\,$(0,0,2)$\\
\hline
\multirow{2}{*}{$14$} & A\,$(3,6,0)$ & \Tstrut A\,$(2,0,0)$ & B\,$(1,-2,0)$ & A\,$(-1,-2,0)$ & \multirow{2}{*}{---} & A\,$(-1,-2,0)$\\
 & C\,$(0,6,0)$ & \Bstrut C\,$(2,4,0)$ & C\,$(1,4,0)$ & C\,$(0,-2,0)$ &  & C\,$(0,-2,0)$\\
\hline
\multirow{2}{*}{$15$} & A\,$(3,-6,6)$ & \Tstrut A\,$(2,0,0)$ & B\,$(1,2,-2)$ & A\,$(-1,0,0)$ & \multirow{2}{*}{---} & A\,$(-1,2,-2)$\\
 & C\,$(0,-6,6)$ & \Bstrut C\,$(2,-4,4)$ & C\,$(1,-4,4)$ & B\,$(0,2,-2)$ &  & C\,$(0,2,-2)$\\
\hline
\multirow{2}{*}{$16$} & A\,$(3,0,0)$ & \Tstrut B\,$(0,0,-4)$ & A\,$(-2,0,2)$ & A\,$(0,0,2)$ & \multirow{2}{*}{---} & B\,$(1,0,0)$\\
 & C\,$(3,0,-6)$ & \Bstrut C\,$(-2,0,4)$ & B\,$(1,0,2)$ & B\,$(1,0,-2)$ &  & C\,$(0,0,-2)$\\
\hline
\multirow{2}{*}{$17$} & A\,$(3,0,0)$ & \Tstrut A\,$(0,4,0)$ & A\,$(-1,-4,0)$ & A\,$(1,0,0)$ & \multirow{2}{*}{---} & A\,$(1,0,0)$\\
 & C\,$(3,6,0)$ & \Bstrut C\,$(-2,0,0)$ & C\,$(1,-2,0)$ & C\,$(1,2,0)$ &  & C\,$(1,2,0)$\\
\hline
\multirow{2}{*}{$18$} & A\,$(3,0,0)$ & \Tstrut A\,$(0,-4,4)$ & A\,$(-1,4,-4)$ & A\,$(1,0,0)$ & \multirow{2}{*}{---} & A\,$(1,0,0)$\\
 & C\,$(3,-6,6)$ & \Bstrut C\,$(-2,0,0)$ & C\,$(1,2,-2)$ & C\,$(1,-2,2)$ &  & C\,$(1,-2,2)$\\
\hline
\multirow{2}{*}{$19$} & \multirow{2}{*}{---} & \Tstrut B\,$(3,-4,4)$ & \multirow{2}{*}{---} & B\,$(1,0,0)$ & A\,$(-2,2,-2)$ & A\,$(1,0,0)$\\
 &  & \Bstrut C\,$(-2,-2,2)$ &  & C\,$(0,-2,2)$ & B\,$(1,2,-2)$ & C\,$(1,-2,2)$\\
\hline
\multirow{2}{*}{$20$} & \multirow{2}{*}{---} & \Tstrut A\,$(3,0,-4)$ & \multirow{2}{*}{---} & A\,$(1,0,0)$ & A\,$(-1,0,4)$ & A\,$(1,0,0)$\\
 &  & \Bstrut C\,$(1,0,-6)$ &  & C\,$(1,0,-2)$ & C\,$(1,0,2)$ & C\,$(1,0,-2)$\\
\hline
\multirow{2}{*}{$21$} & \multirow{2}{*}{---} & \Tstrut A\,$(2,-2,0)$ & \multirow{2}{*}{---} & A\,$(0,-2,0)$ & A\,$(-1,-4,0)$ & B\,$(1,0,0)$\\
 &  & \Bstrut B\,$(1,6,0)$ &  & B\,$(1,2,0)$ & C\,$(1,-2,0)$ & C\,$(0,2,0)$\\
\hline
\multirow{2}{*}{$22$} & A\,$(1,2,-2)$ & \Tstrut A\,$(0,-4,4)$ & A\,$(1,2,-2)$ & B\,$(3,-2,2)$ & \multirow{2}{*}{---} & A\,$(-1,2,-2)$\\
 & C\,$(2,-2,2)$ & \Bstrut C\,$(-2,0,0)$ & C\,$(2,-2,2)$ & C\,$(-1,-4,4)$ &  & C\,$(0,2,-2)$\\
\hline
\multirow{2}{*}{$23$} & A\,$(1,0,2)$ & \Tstrut A\,$(0,0,-4)$ & A\,$(1,0,2)$ & A\,$(3,0,-2)$ & \multirow{2}{*}{---} & A\,$(-1,0,2)$\\
 & C\,$(2,0,-2)$ & \Bstrut C\,$(-2,0,0)$ & C\,$(2,0,-2)$ & C\,$(2,0,-6)$ &  & C\,$(0,0,2)$\\
\hline
\multirow{2}{*}{$24$} & A\,$(1,-2,0)$ & \Tstrut A\,$(0,4,0)$ & A\,$(1,-2,0)$ & A\,$(3,2,0)$ & \multirow{2}{*}{---} & A\,$(-1,-2,0)$\\
 & C\,$(2,2,0)$ & \Bstrut C\,$(-2,0,0)$ & C\,$(2,2,0)$ & C\,$(2,6,0)$ &  & C\,$(0,-2,0)$\\
\hline
\multirow{2}{*}{$25$} & A\,$(2,-2,2)$ & \Tstrut B\,$(2,0,0)$ & B\,$(2,-2,2)$ & A\,$(-3,2,-2)$ & \multirow{2}{*}{---} & A\,$(0,-2,2)$\\
 & C\,$(1,-4,4)$ & \Bstrut C\,$(0,-4,4)$ & C\,$(-1,-2,2)$ & B\,$(1,4,-4)$ &  & C\,$(-1,0,0)$\\
\hline
\multirow{2}{*}{$26$} & A\,$(2,0,-2)$ & \Tstrut A\,$(2,0,0)$ & A\,$(2,0,-2)$ & A\,$(-2,0,6)$ & \multirow{2}{*}{---} & A\,$(0,0,-2)$\\
 & C\,$(1,0,-4)$ & \Bstrut C\,$(2,0,-4)$ & C\,$(1,0,-4)$ & C\,$(1,0,4)$ &  & C\,$(-1,0,0)$\\
\hline
\multirow{2}{*}{$27$} & A\,$(2,2,0)$ & \Tstrut A\,$(2,0,0)$ & A\,$(2,2,0)$ & A\,$(-2,-6,0)$ & \multirow{2}{*}{---} & A\,$(0,2,0)$\\
 & C\,$(1,4,0)$ & \Bstrut C\,$(2,4,0)$ & C\,$(1,4,0)$ & C\,$(1,-4,0)$ &  & C\,$(-1,0,0)$\\
\hline
\end{longtable}
\endgroup

\twocolumngrid

%% file: rep_tables2.tex
\newpage
\begin{table}[ht]
    \caption{Matrix representations of the generators used for digitization of $\dsb$ ($\phi = e^{i\pi/4}$)}
    \label{tab:irrepss648p1}
    \centering\begin{adjustbox}{width=1.25\textwidth}
    \begin{tabular}{c|cccccc}
    g & $\omega$ & $C$ & $E$ & $V$ & $X$ & $D$\\
    \hline \hline
    $\mathbf{1}^{(x)}$ & $1$ & $1$ & $1$ & $1$ & $1$ & $\omega^x$\\
    
    $\mathbf{2}^{(x)}$ & $\mathbbm{1}_2$ & $\begin{pmatrix} 1 & 0 \\ 0 & 1 \end{pmatrix}$ & $\begin{pmatrix} 1 & 0 \\ 0 & 1 \end{pmatrix}$ & $\begin{pmatrix} 0 & i \\ i & 0 \end{pmatrix}$ & $\begin{pmatrix} -i & 0 \\ 0 & i \end{pmatrix}$ & $\frac{\omega^x}{\sqrt{2}}\begin{pmatrix} -\phi & \phi \\ -\phi^* & -\phi^* \end{pmatrix}$\\
    
    $\mathbf{3}^{(a)}$ & $\mathbbm{1}_3$ & $\begin{pmatrix} 1 & 0 & 0 \\ 0 & 1 & 0 \\ 0 & 0 & 1 \end{pmatrix}$ & $\begin{pmatrix} 1 & 0 & 0 \\ 0 & 1 & 0 \\ 0 & 0 & 1 \end{pmatrix}$ & $\begin{pmatrix} 1 & 0 & 0 \\ 0 & -1 & 0 \\ 0 & 0 & -1 \end{pmatrix}$ & $\begin{pmatrix} -1 & 0 & 0 \\ 0 & -1 & 0 \\ 0 & 0 & 1 \end{pmatrix}$ & $\begin{pmatrix} 0 & 1 & 0 \\ 0 & 0 & -1 \\ -1 & 0 & 0 \end{pmatrix}$\\
    
    $\mathbf{3}^{(x)}$ & $\omega\mathbbm{1}_3$ & $\begin{pmatrix} 1 & 0 & 0 \\ 0 & \omega & 0 \\ 0 & 0 & \omega^2 \end{pmatrix}$ & $\begin{pmatrix} 0 & 1 & 0 \\ 0 & 0 & 1 \\ 1 & 0 & 0 \\
    \end{pmatrix}$ & $\frac{1}{\sqrt{3}i}\begin{pmatrix} 1 & 1 & 1 \\ 1 & \omega & \omega^2 \\ 1 & \omega^2 & \omega \\ \end{pmatrix}$ & $\frac{1}{\sqrt{3}i} \begin{pmatrix} 1 & 1 & \omega^2 \\ 1 & \omega & \omega \\ \omega & 1 & \omega \\ \end{pmatrix}$ & $\omega^x\begin{pmatrix} \epsilon & 0 & 0 \\ 0 & \epsilon & 0 \\ 0 & 0 & \epsilon\omega \\ \end{pmatrix}$\\
    
    $\mathbf{3}^{(x)*}$ & $\omega^2\mathbbm{1}_3$ & $\begin{pmatrix} 1 & 0 & 0 \\ 0 & \omega^2 & 0 \\ 0 & 0 & \omega \end{pmatrix}$ & $\begin{pmatrix} 0 & 1 & 0 \\ 0 & 0 & 1 \\ 1 & 0 & 0 \\
    \end{pmatrix}$ & $-\frac{1}{\sqrt{3}i}\begin{pmatrix} 1 & 1 & 1 \\ 1 & \omega^2 & \omega \\ 1 & \omega & \omega^2 \\ \end{pmatrix}$ & $-\frac{1}{\sqrt{3}i} \begin{pmatrix} 1 & 1 & \omega \\ 1 & \omega^2 & \omega^2 \\ \omega^2 & 1 & \omega^2 \\ \end{pmatrix}$ & $\omega^{2x}\begin{pmatrix} \epsilon^2\omega & 0 & 0 \\ 0 & \epsilon^2\omega & 0 \\ 0 & 0 & \epsilon^2 \\ \end{pmatrix}$\\

    $\mathbf{6}^{(x)}$ & $\omega\mathbbm{1}_6$& $\text{Diag}(1,\omega,\omega^2,\omega^2,\omega,1)$ & $\begin{pmatrix} 0 & 1 & 0 & 0 & 0 & 0 \\ 0 & 0 & 1 & 0 & 0 & 0 \\ 1 & 0 & 0 & 0 & 0 & 0 \\ 0 & 0 & 0 & 0 & 0 & 1 \\ 0 & 0 & 0 & 1 & 0 & 0 \\ 0 & 0 & 0 & 0 & 1 & 0 \\ \end{pmatrix}$ & $-\frac{1}{3}\begin{pmatrix} 1 & 1 & 1 & \sqrt{2} & \sqrt{2} & \sqrt{2} \\ 1 & \omega & \omega^2 & \sqrt{2}\omega^2 & \sqrt{2}\omega & \sqrt{2} \\ 1 & \omega^2 & \omega & \sqrt{2}\omega & \sqrt{2}\omega^2 & \sqrt{2} \\ \sqrt{2} & \sqrt{2}\omega^2 & \sqrt{2}\omega & 1+\omega^2 & 1+\omega & -1 \\ \sqrt{2} & \sqrt{2}\omega & \sqrt{2}\omega^2 & 1+\omega & 1+\omega^2 & -1 \\ \sqrt{2} & \sqrt{2} & \sqrt{2} & -1 & -1 & -1 \\ \end{pmatrix}$ & $-\frac{1}{3}\begin{pmatrix} 1 & 1 & \omega^2 & \sqrt{2} & \sqrt{2} \omega & \sqrt{2} \omega \\ 1 & \omega & \omega & \sqrt{2} \omega^2 & \sqrt{2} \omega^2 & \sqrt{2} \omega \\ \omega & 1 & \omega & \sqrt{2} \omega^2 & \sqrt{2} \omega & \sqrt{2} \omega^2 \\ \sqrt{2} & \sqrt{2} \omega^2 & \sqrt{2} & \omega^2+1 & -1 & -\omega \\ \sqrt{2} \omega^2 & \sqrt{2} & \sqrt{2} & \omega^2+1 & \omega^2+1 & -1 \\ \sqrt{2} \omega^2 & \sqrt{2} \omega^2 & \sqrt{2} \omega & -\omega^2 & -1 & -1 \\ \end{pmatrix}$ & $\epsilon\omega^x\times\text{Diag}(\omega,\omega,\omega^2,\omega,1,1)$\\

    $\mathbf{6}^{(x)*}$ & $\omega^2\mathbbm{1}_6$& $\text{Diag}(1,\omega^2,\omega,\omega,\omega^2,1)$  & $\begin{pmatrix} 0 & 1 & 0 & 0 & 0 & 0 \\ 0 & 0 & 1 & 0 & 0 & 0 \\ 1 & 0 & 0 & 0 & 0 & 0 \\ 0 & 0 & 0 & 0 & 0 & 1 \\ 0 & 0 & 0 & 1 & 0 & 0 \\ 0 & 0 & 0 & 0 & 1 & 0 \\ \end{pmatrix}$ & $-\frac{1}{3}\begin{pmatrix} 1 & 1 & 1 & \sqrt{2} & \sqrt{2} & \sqrt{2} \\ 1 & \omega^2 & \omega & \sqrt{2} \omega & \sqrt{2} \omega^2 & \sqrt{2} \\ 1 & \omega & \omega^2 & \sqrt{2} \omega^2 & \sqrt{2} \omega & \sqrt{2} \\ \sqrt{2} & \sqrt{2} \omega & \sqrt{2} \omega^2 & \omega+1 & \omega^2+1 & -1 \\ \sqrt{2} & \sqrt{2} \omega^2 & \sqrt{2} \omega & \omega^2+1 & \omega+1 & -1 \\ \sqrt{2} & \sqrt{2} & \sqrt{2} & -1 & -1 & -1 \\ \end{pmatrix}$ & $-\frac{1}{3}\begin{pmatrix} 1 & 1 & \omega & \sqrt{2} & \sqrt{2} \omega^2 & \sqrt{2} \omega^2 \\ 1 & \omega^2 & \omega^2 & \sqrt{2} \omega & \sqrt{2} \omega & \sqrt{2} \omega^2 \\ \omega^2 & 1 & \omega^2 & \sqrt{2} \omega & \sqrt{2} \omega^2 & \sqrt{2} \omega \\ \sqrt{2} & \sqrt{2} \omega & \sqrt{2} & \omega+1 & -1 & -\omega^2 \\ \sqrt{2} \omega & \sqrt{2} & \sqrt{2} & \omega+1 & \omega+1 & -1 \\ \sqrt{2} \omega & \sqrt{2} \omega & \sqrt{2} \omega^2 & -\omega & -1 & -1 \\ \end{pmatrix}$ & $\epsilon^2\omega^{2x}\times\text{Diag}(1,1,\omega^2,1,\omega,\omega)$\\
    $\mathbf{8}^{(x)}$ & $\mathbbm{1}_8$ & $\begin{pmatrix} -\frac{1}{2} & -\frac{\sqrt{3}}{2} & 0 & 0 & 0 & 0 & 0 & 0 \\ \frac{\sqrt{3}}{2} & -\frac{1}{2} & 0 & 0 & 0 & 0 & 0 & 0 \\ 0 & 0 & 1 & 0 & 0 & 0 & 0 & 0 \\ 0 & 0 & 0 & -\frac{1}{2} & \frac{\sqrt{3}}{2} & 0 & 0 & 0 \\ 0 & 0 & 0 & -\frac{\sqrt{3}}{2} & -\frac{1}{2} & 0 & 0 & 0 \\ 0 & 0 & 0 & 0 & 0 & -\frac{1}{2} & -\frac{\sqrt{3}}{2} & 0 \\ 0 & 0 & 0 & 0 & 0 & \frac{\sqrt{3}}{2} & -\frac{1}{2} & 0 \\ 0 & 0 & 0 & 0 & 0 & 0 & 0 & 1 \\ \end{pmatrix}$ & $\begin{pmatrix} 0 & 0 & 0 & 0 & 0 & 1 & 0 & 0 \\ 0 & 0 & 0 & 0 & 0 & 0 & 1 & 0 \\ 0 & 0 & -\frac{1}{2} & 0 & 0 & 0 & 0 & \frac{\sqrt{3}}{2} \\ 1 & 0 & 0 & 0 & 0 & 0 & 0 & 0 \\ 0 & -1 & 0 & 0 & 0 & 0 & 0 & 0 \\ 0 & 0 & 0 & 1 & 0 & 0 & 0 & 0 \\ 0 & 0 & 0 & 0 & -1 & 0 & 0 & 0 \\ 0 & 0 & -\frac{\sqrt{3}}{2} & 0 & 0 & 0 & 0 & -\frac{1}{2} \\ \end{pmatrix}$ & $\begin{pmatrix} \frac{1}{6} & -\frac{1}{2 \sqrt{3}} & \frac{1}{2} & \frac{1}{6} & \frac{1}{2 \sqrt{3}} & -\frac{1}{3} & \frac{1}{\sqrt{3}} & \frac{1}{2 \sqrt{3}} \\ \frac{1}{2 \sqrt{3}} & -\frac{1}{2} & -\frac{1}{2 \sqrt{3}} & -\frac{1}{2 \sqrt{3}} & -\frac{1}{2} & 0 & 0 & \frac{1}{2} \\ \frac{1}{2} & \frac{1}{2 \sqrt{3}} & 0 & \frac{1}{2} & -\frac{1}{2 \sqrt{3}} & \frac{1}{2} & \frac{1}{2 \sqrt{3}} & 0 \\ \frac{1}{6} & \frac{1}{2 \sqrt{3}} & \frac{1}{2} & \frac{1}{6} & -\frac{1}{2 \sqrt{3}} & -\frac{1}{3} & -\frac{1}{\sqrt{3}} & \frac{1}{2 \sqrt{3}} \\ -\frac{1}{2 \sqrt{3}} & -\frac{1}{2} & \frac{1}{2 \sqrt{3}} & \frac{1}{2 \sqrt{3}} & -\frac{1}{2} & 0 & 0 & -\frac{1}{2} \\ -\frac{1}{3} & 0 & \frac{1}{2} & -\frac{1}{3} & 0 & \frac{2}{3} & 0 & \frac{1}{2 \sqrt{3}} \\ -\frac{1}{\sqrt{3}} & 0 & -\frac{1}{2 \sqrt{3}} & \frac{1}{\sqrt{3}} & 0 & 0 & 0 & \frac{1}{2} \\ \frac{1}{2 \sqrt{3}} & -\frac{1}{2} & 0 & \frac{1}{2 \sqrt{3}} & \frac{1}{2} & \frac{1}{2 \sqrt{3}} & -\frac{1}{2} & 0 \\ \end{pmatrix}$ & $ \begin{pmatrix} \frac{1}{6} & -\frac{1}{2 \sqrt{3}} & \frac{1}{2} & -\frac{1}{3} & 0 & -\frac{1}{3} & -\frac{1}{\sqrt{3}} & \frac{1}{2 \sqrt{3}} \\ \frac{1}{2 \sqrt{3}} & -\frac{1}{2} & -\frac{1}{2 \sqrt{3}} & \frac{1}{\sqrt{3}} & 0 & 0 & 0 & \frac{1}{2} \\ \frac{1}{2} & \frac{1}{2 \sqrt{3}} & 0 & 0 & \frac{1}{\sqrt{3}} & -\frac{1}{2} & \frac{1}{2 \sqrt{3}} & 0 \\ \frac{1}{6} & \frac{1}{2 \sqrt{3}} & -\frac{1}{2} & -\frac{1}{3} & -\frac{1}{\sqrt{3}} & -\frac{1}{3} & 0 & \frac{1}{2 \sqrt{3}} \\ \frac{1}{2 \sqrt{3}} & \frac{1}{2} & \frac{1}{2 \sqrt{3}} & 0 & 0 & \frac{1}{\sqrt{3}} & 0 & \frac{1}{2} \\ \frac{2}{3} & 0 & 0 & \frac{1}{6} & -\frac{1}{2 \sqrt{3}} & \frac{1}{6} & -\frac{1}{2 \sqrt{3}} & -\frac{1}{\sqrt{3}} \\ 0 & 0 & \frac{1}{\sqrt{3}} & \frac{1}{2 \sqrt{3}} & -\frac{1}{2} & -\frac{1}{2 \sqrt{3}} & \frac{1}{2} & 0 \\ \frac{1}{2 \sqrt{3}} & -\frac{1}{2} & 0 & -\frac{1}{\sqrt{3}} & 0 & \frac{1}{2 \sqrt{3}} & \frac{1}{2} & 0 \\ \end{pmatrix}$ & $\omega^x\begin{pmatrix} 1 & 0 & 0 & 0 & 0 & 0 & 0 & 0 \\ 0 & 1 & 0 & 0 & 0 & 0 & 0 & 0 \\ 0 & 0 & 1 & 0 & 0 & 0 & 0 & 0 \\ 0 & 0 & 0 & -\frac{1}{2} & -\frac{\sqrt{3}}{2} & 0 & 0 & 0 \\ 0 & 0 & 0 & \frac{\sqrt{3}}{2} & -\frac{1}{2} & 0 & 0 & 0 \\ 0 & 0 & 0 & 0 & 0 & -\frac{1}{2} & -\frac{\sqrt{3}}{2} & 0 \\ 0 & 0 & 0 & 0 & 0 & \frac{\sqrt{3}}{2} & -\frac{1}{2} & 0 \\ 0 & 0 & 0 & 0 & 0 & 0 & 0 & 1 \\ \end{pmatrix}$\\
    $\mathbf{9}$ & $\omega\mathbbm{1}_9$ & $\text{Diag}(1,1,1,\omega,\omega,\omega,\omega^2,\omega^2,\omega^2)$ & $\begin{pmatrix} 0 & 0 & 0 & 1 & 0 & 0 & 0 & 0 & 0 \\ 0 & 0 & 0 & 0 & 1 & 0 & 0 & 0 & 0 \\ 0 & 0 & 0 & 0 & 0 & 1 & 0 & 0 & 0 \\ 0 & 0 & 0 & 0 & 0 & 0 & 1 & 0 & 0 \\ 0 & 0 & 0 & 0 & 0 & 0 & 0 & 1 & 0 \\ 0 & 0 & 0 & 0 & 0 & 0 & 0 & 0 & 1 \\ 1 & 0 & 0 & 0 & 0 & 0 & 0 & 0 & 0 \\ 0 & 1 & 0 & 0 & 0 & 0 & 0 & 0 & 0 \\ 0 & 0 & 1 & 0 & 0 & 0 & 0 & 0 & 0 \\ \end{pmatrix}$ & $\frac{1}{\sqrt{3}i}\begin{pmatrix} 1 & 0 & 0 & 1 & 0 & 0 & 1 & 0 & 0 \\ 0 & -1 & 0 & 0 & -1 & 0 & 0 & -1 & 0 \\ 0 & 0 & -1 & 0 & 0 & -1 & 0 & 0 & -1 \\ 1 & 0 & 0 & \omega & 0 & 0 & \omega^2 & 0 & 0 \\ 0 & -1 & 0 & 0 & -\omega & 0 & 0 & -\omega^2 & 0 \\ 0 & 0 & -1 & 0 & 0 & -\omega & 0 & 0 & -\omega^2 \\ 1 & 0 & 0 & \omega^2 & 0 & 0 & \omega & 0 & 0 \\ 0 & -1 & 0 & 0 & -\omega^2 & 0 & 0 & -\omega & 0 \\ 0 & 0 & -1 & 0 & 0 & -\omega^2 & 0 & 0 & -\omega \\ \end{pmatrix}$ & $\frac{1}{\sqrt{3}i}\begin{pmatrix} -1 & 0 & 0 & -1 & 0 & 0 & -\omega^2 & 0 & 0 \\ 0 & -1 & 0 & 0 & -1 & 0 & 0 & -\omega^2 & 0 \\ 0 & 0 & 1 & 0 & 0 & 1 & 0 & 0 & \omega^2 \\ -1 & 0 & 0 & -\omega & 0 & 0 & -\omega & 0 & 0 \\ 0 & -1 & 0 & 0 & -\omega & 0 & 0 & -\omega & 0 \\ 0 & 0 & 1 & 0 & 0 & \omega & 0 & 0 & \omega \\ -\omega & 0 & 0 & -1 & 0 & 0 & -\omega & 0 & 0 \\ 0 & -\omega & 0 & 0 & -1 & 0 & 0 & -\omega & 0 \\ 0 & 0 & \omega & 0 & 0 & 1 & 0 & 0 & \omega \\ \end{pmatrix}$ & $\epsilon\begin{pmatrix} 0 & 1 & 0 & 0 & 0 & 0 & 0 & 0 & 0 \\ 0 & 0 & -1 & 0 & 0 & 0 & 0 & 0 & 0 \\ -1& 0 & 0 & 0 & 0 & 0 & 0 & 0 & 0 \\ 0 & 0 & 0 & 0 & 1 & 0 & 0 & 0 & 0 \\ 0 & 0 & 0 & 0 & 0 & -1 & 0 & 0 & 0 \\ 0 & 0 & 0 & -1 & 0 & 0 & 0 & 0 & 0 \\ 0 & 0 & 0 & 0 & 0 & 0 & 0 & \omega & 0 \\ 0 & 0 & 0 & 0 & 0 & 0 & 0 & 0 & -\omega \\ 0 & 0 & 0 & 0 & 0 & 0 & -\omega & 0 & 0 \\ \end{pmatrix}$\\

    $\mathbf{9^*}$ & $\omega^2\mathbbm{1}$ & $\text{Diag}(1,1,1,\omega^2,\omega^2,\omega^2,\omega,\omega,\omega)$ & $\begin{pmatrix} 0 & 0 & 0 & 1 & 0 & 0 & 0 & 0 & 0 \\ 0 & 0 & 0 & 0 & 1 & 0 & 0 & 0 & 0 \\ 0 & 0 & 0 & 0 & 0 & 1 & 0 & 0 & 0 \\ 0 & 0 & 0 & 0 & 0 & 0 & 1 & 0 & 0 \\ 0 & 0 & 0 & 0 & 0 & 0 & 0 & 1 & 0 \\ 0 & 0 & 0 & 0 & 0 & 0 & 0 & 0 & 1 \\ 1 & 0 & 0 & 0 & 0 & 0 & 0 & 0 & 0 \\ 0 & 1 & 0 & 0 & 0 & 0 & 0 & 0 & 0 \\ 0 & 0 & 1 & 0 & 0 & 0 & 0 & 0 & 0 \\ \end{pmatrix}$ & $-\frac{1}{\sqrt{3}i}\begin{pmatrix} 1 & 0 & 0 & 1 & 0 & 0 & 1 & 0 & 0 \\ 0 & -1 & 0 & 0 & -1 & 0 & 0 & -1 & 0 \\ 0 & 0 & -1 & 0 & 0 & -1 & 0 & 0 & -1 \\ 1 & 0 & 0 & \omega^2 & 0 & 0 & \omega & 0 & 0 \\ 0 & -1 & 0 & 0 & -\omega^2 & 0 & 0 & -\omega & 0 \\ 0 & 0 & -1 & 0 & 0 & -\omega^2 & 0 & 0 & -\omega \\ 1 & 0 & 0 & \omega & 0 & 0 & \omega^2 & 0 & 0 \\ 0 & -1 & 0 & 0 & -\omega & 0 & 0 & -\omega^2 & 0 \\ 0 & 0 & -1 & 0 & 0 & -\omega & 0 & 0 & -\omega^2 \\ \end{pmatrix}$ & 
    $\frac{1}{\sqrt{3}i}\begin{pmatrix} 1 & 0 & 0 & 1 & 0 & 0 & \omega & 0 & 0 \\ 0 & 1 & 0 & 0 & 1 & 0 & 0 & \omega & 0 \\ 0 & 0 & -1 & 0 & 0 & -1 & 0 & 0 & -\omega \\ 1 & 0 & 0 & \omega^2 & 0 & 0 & \omega^2 & 0 & 0 \\ 0 & 1 & 0 & 0 & \omega^2 & 0 & 0 & \omega^2 & 0 \\ 0 & 0 & -1 & 0 & 0 & -\omega^2 & 0 & 0 & -\omega^2 \\ \omega^2 & 0 & 0 & 1 & 0 & 0 & \omega^2 & 0 & 0 \\ 0 & \omega^2 & 0 & 0 & 1 & 0 & 0 & \omega^2 & 0 \\ 0 & 0 & -\omega^2 & 0 & 0 & -1 & 0 & 0 & -\omega^2 \\ \end{pmatrix}$ & $\epsilon^2\begin{pmatrix} 0 & \omega & 0 & 0 & 0 & 0 & 0 & 0 & 0 \\ 0 & 0 & -\omega & 0 & 0 & 0 & 0 & 0 & 0 \\ -\omega & 0 & 0 & 0 & 0 & 0 & 0 & 0 & 0 \\ 0 & 0 & 0 & 0 & \omega & 0 & 0 & 0 & 0 \\ 0 & 0 & 0 & 0 & 0 & -\omega & 0 & 0 & 0 \\ 0 & 0 & 0 & -\omega & 0 & 0 & 0 & 0 & 0 \\ 0 & 0 & 0 & 0 & 0 & 0 & 0 & 1 & 0 \\ 0 & 0 & 0 & 0 & 0 & 0 & 0 & 0 & -1 \\ 0 & 0 & 0 & 0 & 0 & 0 & -1 & 0 & 0 \\ \end{pmatrix}$\\
    \end{tabular}
    \end{adjustbox}
\end{table}

%% file: ref.bib
@misc{DiMeglio:2023nsa,
    author = "Di Meglio, Alberto and others",
    title = "{Quantum Computing for High-Energy Physics: State of the Art and Challenges. Summary of the QC4HEP Working Group}",
    eprint = "2307.03236",
    archivePrefix = "arXiv",
    primaryClass = "quant-ph",
    reportNumber = "FERMILAB-PUB-23-468-ETD",
    month = "7",
    year = "2023"
}

@article{Aarts_2016,
	Author = {Aarts, Gert},
	Doi = {10.1088/1742-6596/706/2/022004},
	Issn = {1742-6596},
	Journal = {Journal of Physics: Conference Series},
	Month = {Apr},
	Pages = {022004},
	Publisher = {IOP Publishing},
	Title = {Introductory lectures on lattice QCD at nonzero baryon number},
	Url = {http://dx.doi.org/10.1088/1742-6596/706/2/022004},
	Volume = {706},
	Year = {2016}}

@article{Zohar:2012xf,
      author         = "Zohar, Erez and Cirac, J. Ignacio and Reznik, Benni",
      title          = "{Cold-Atom Quantum Simulator for SU(2) Yang-Mills Lattice
                        Gauge Theory}",
      journal        = "Phys. Rev. Lett.",
      volume         = "110",
      year           = "2013",
      number         = "12",
      pages          = "125304",
      doi            = "10.1103/PhysRevLett.110.125304",
      eprint         = "1211.2241",
      archivePrefix  = "arXiv",
      primaryClass   = "quant-ph",
      SLACcitation   = "%%CITATION = ARXIV:1211.2241;%%"
}

@article{Illa:2024kmf,
    author = "Illa, Marc and Robin, Caroline E. P. and Savage, Martin J.",
    title = "{Qu8its for quantum simulations of lattice quantum chromodynamics}",
    eprint = "2403.14537",
    archivePrefix = "arXiv",
    primaryClass = "quant-ph",
    reportNumber = "IQuS@UW-21-074",
    doi = "10.1103/PhysRevD.110.014507",
    journal = "Phys. Rev. D",
    volume = "110",
    number = "1",
    pages = "014507",
    year = "2024"
}

@article{Singh:2019jog,
    author = "Singh, Hersh",
    title = "{Qubit regularized O(N) nonlinear sigma models}",
    eprint = "1911.12353",
    archivePrefix = "arXiv",
    primaryClass = "hep-lat",
    reportNumber = "IQuS@UW-21-021,INT-PUB-22-007",
    doi = "10.1103/PhysRevD.105.114509",
    journal = "Phys. Rev. D",
    volume = "105",
    number = "11",
    pages = "114509",
    year = "2022"
}

@article{Alexandru:2023qzd,
    author = "Alexandru, Andrei and Bedaque, Paulo F. and Carosso, Andrea and Cervia, Michael J. and Murairi, Edison M. and Sheng, Andy",
    title = "{Fuzzy gauge theory for quantum computers}",
    eprint = "2308.05253",
    archivePrefix = "arXiv",
    primaryClass = "hep-lat",
    doi = "10.1103/PhysRevD.109.094502",
    journal = "Phys. Rev. D",
    volume = "109",
    number = "9",
    pages = "094502",
    year = "2024"
}

@article{ORLAND1990647,
	author = {Peter Orland and Daniel Rohrlich},
	doi = {https://doi.org/10.1016/0550-3213(90)90646-U},
	issn = {0550-3213},
	journal = {Nuclear Physics B},
	number = {3},
	pages = {647-672},
	title = {Lattice gauge magnets: Local isospin from spin},
	url = {https://www.sciencedirect.com/science/article/pii/055032139090646U},
	volume = {338},
	year = {1990}
}

@article{Rajput:2021trn,
    author = "Rajput, Abhishek and Roggero, Alessandro and Wiebe, Nathan",
    title = "{Quantum error correction with gauge symmetries}",
    eprint = "2112.05186",
    archivePrefix = "arXiv",
    primaryClass = "quant-ph",
    reportNumber = "IQuS@UW-21-014",
    doi = "10.1038/s41534-023-00706-8",
    journal = "npj Quantum Inf.",
    volume = "9",
    number = "1",
    pages = "41",
    year = "2023"
}

@article{Spagnoli:2024mib,
    author = "Spagnoli, L. and Roggero, A. and Wiebe, N.",
    title = "{Fault-tolerant simulation of Lattice Gauge Theories with gauge covariant codes}",
    eprint = "2405.19293",
    archivePrefix = "arXiv",
    primaryClass = "quant-ph",
    doi = "10.22331/q-2026-01-16-1968",
    journal = "Quantum",
    volume = "10",
    pages = "1968",
    year = "2026"
}

@misc{Yao:2025cxs,
    author = "Yao, Xiaojun",
    title = "{Quantum Error Correction Codes for Truncated SU(2) Lattice Gauge Theories}",
    eprint = "2511.13721",
    archivePrefix = "arXiv",
    primaryClass = "quant-ph",
    reportNumber = "IQuS@UW-21-114",
    month = "11",
    year = "2025"
}

@misc{Pato:2026wow,
    author = "Pato, Balint and Klco, Natalie",
    title = "{Trade-offs in Gauss's law error correction for lattice gauge theory quantum simulations}",
    eprint = "2602.22121",
    archivePrefix = "arXiv",
    primaryClass = "quant-ph",
    month = "2",
    year = "2026"
}

@misc{Perez:2025cxl,
    author = "Perez, Sebastian Osorio and Murairi, Edison M. and Gustafson, Erik J. and Lamm, Henry",
    title = "{Primitive Quantum Gates for an $SU(3)$ Discrete Subgroup: $\Sigma(72\times3)$}",
    eprint = "2511.17437",
    archivePrefix = "arXiv",
    primaryClass = "hep-lat",
    reportNumber = "FERMILAB-PUB-25-0674-SQMS-STUDENT-T",
    month = "11",
    year = "2025"
}

@misc{Lamm:2026vqz,
    author = "Lamm, Henry",
    title = "{No Quantum Utility from Hadron Masses? No, Quantum Utility from Hadron Masses!}",
    eprint = "2603.00946",
    archivePrefix = "arXiv",
    primaryClass = "hep-lat",
    reportNumber = "FERMILAB-PUB-26-0133-SQMS-T",
    month = "3",
    year = "2026"
}

@misc{Osborne:2023rzx,
    author = "Osborne, Jesse and Yang, Bing and McCulloch, Ian P. and Hauke, Philipp and Halimeh, Jad C.",
    title = "{Spin-$S$$\mathrm{U}(1)$ Quantum Link Models with Dynamical Matter on a Quantum Simulator}",
    eprint = "2305.06368",
    archivePrefix = "arXiv",
    primaryClass = "cond-mat.quant-gas",
    month = "5",
    year = "2023"
}

@phdthesis{Kadam:2023gli,
    author = "Kadam, Saurabh Vasant",
    title = "{Theoretical Developments in Lattice Gauge Theory for Applications in Double-beta Decay Processes and Quantum Simulation}",
    eprint = "2312.00780",
    archivePrefix = "arXiv",
    primaryClass = "hep-lat",
    doi = "10.13016/dspace/cvbq-c4jt",
    school = "Maryland U., College Park",
    year = "2023"
}

@misc{Fromm:2023bit,
    author = "Fromm, Michael and Philipsen, Owe and Unger, Wolfgang and Winterowd, Christopher",
    title = "{Quantum Gate Sets for Lattice QCD in the strong coupling limit: $N_f=1$}",
    eprint = "2308.03196",
    archivePrefix = "arXiv",
    primaryClass = "hep-lat",
    month = "8",
    year = "2023"
}

@article{Davoudi:2024wyv,
    author = "Davoudi, Zohreh and Hsieh, Chung-Chun and Kadam, Saurabh V.",
    title = "{Scattering wave packets of hadrons in gauge theories: Preparation on a quantum computer}",
    eprint = "2402.00840",
    archivePrefix = "arXiv",
    primaryClass = "quant-ph",
    reportNumber = "UMD-PP-024-02, IQuS@UW-21-071",
    doi = "10.22331/q-2024-11-11-1520",
    journal = "Quantum",
    volume = "8",
    pages = "1520",
    year = "2024"
}

@article{Farrell:2024fit,
    author = "Farrell, Roland C. and Illa, Marc and Ciavarella, Anthony N. and Savage, Martin J.",
    title = "{Quantum simulations of hadron dynamics in the Schwinger model using 112 qubits}",
    eprint = "2401.08044",
    archivePrefix = "arXiv",
    primaryClass = "quant-ph",
    reportNumber = "IQuS@UW-21-069, NT@UW-24-1",
    doi = "10.1103/PhysRevD.109.114510",
    journal = "Phys. Rev. D",
    volume = "109",
    number = "11",
    pages = "114510",
    year = "2024"
}

@article{Calajo:2024qrc,
    author = "Calaj{\`o}, Giuseppe and Magnifico, Giuseppe and Edmunds, Claire and Ringbauer, Martin and Montangero, Simone and Silvi, Pietro",
    title = "{Digital Quantum Simulation of a (1+1)D SU(2) Lattice Gauge Theory with Ion Qudits}",
    eprint = "2402.07987",
    archivePrefix = "arXiv",
    primaryClass = "quant-ph",
    doi = "10.1103/PRXQuantum.5.040309",
    journal = "PRX Quantum",
    volume = "5",
    number = "4",
    pages = "040309",
    year = "2024"
}

@misc{Kan:2021xfc,
    author = "Kan, Angus and Nam, Yunseong",
    title = "{Lattice Quantum Chromodynamics and Electrodynamics on a Universal Quantum Computer}",
    eprint = "2107.12769",
    archivePrefix = "arXiv",
    primaryClass = "quant-ph",
    month = "7",
    year = "2021"
}

@book{nielsen_chuang_2010,
	place={Cambridge},
	title={Quantum Computation and Quantum Information: 10th Anniversary Edition},
	DOI={10.1017/CBO9780511976667},
	publisher={Cambridge University Press},
	author={Nielsen, Michael A. and Chuang, Isaac L.},
	year={2010}
}

@article{deJong:2021wsd,
    author = "de Jong, Wibe A. and Lee, Kyle and Mulligan, James and P{\l}osko{\'n}, Mateusz and Ringer, Felix and Yao, Xiaojun",
    title = "{Quantum simulation of nonequilibrium dynamics and thermalization in the Schwinger model}",
    eprint = "2106.08394",
    archivePrefix = "arXiv",
    primaryClass = "quant-ph",
    reportNumber = "MIT-CTP/5308",
    doi = "10.1103/PhysRevD.106.054508",
    journal = "Phys. Rev. D",
    volume = "106",
    number = "5",
    pages = "054508",
    year = "2022"
}

@inproceedings{Zohar:2018nvl,
    author = "Emonts, Patrick and Zohar, Erez",
    title = "{Gauss law, Minimal Coupling and Fermionic PEPS for Lattice Gauge Theories}",
    booktitle = "{Tensor Network and entanglement}",
    eprint = "1807.01294",
    archivePrefix = "arXiv",
    primaryClass = "quant-ph",
    doi = "10.21468/SciPostPhysLectNotes.12",
    month = "7",
    year = "2018"
}

@article{Gustafson:2023ayr,
    author = "Gustafson, Erik J. and Lamm, Henry and Unmuth-Yockey, Judah",
    title = "{Quantum mean estimation for lattice field theory}",
    eprint = "2303.00094",
    archivePrefix = "arXiv",
    primaryClass = "hep-lat",
    reportNumber = "FERMILAB-PUB-23-095-QIS-T",
    doi = "10.1103/PhysRevD.107.114511",
    journal = "Phys. Rev. D",
    volume = "107",
    number = "11",
    pages = "114511",
    year = "2023"
}

@misc{Davoudi:2020yln,
    author = "Davoudi, Zohreh and Raychowdhury, Indrakshi and Shaw, Andrew",
    title = "{Search for Efficient Formulations for Hamiltonian Simulation of non-Abelian Lattice Gauge Theories}",
    eprint = "2009.11802",
    archivePrefix = "arXiv",
    primaryClass = "hep-lat",
    reportNumber = "UMD-PP-020-6",
    month = "9",
    year = "2020"
}

@article{Davoudi:2022uzo,
    author = "Davoudi, Zohreh and Mueller, Niklas and Powers, Connor",
    title = "{Towards Quantum Computing Phase Diagrams of Gauge Theories with Thermal Pure Quantum States}",
    eprint = "2208.13112",
    archivePrefix = "arXiv",
    primaryClass = "hep-lat",
    doi = "10.1103/PhysRevLett.131.081901",
    journal = "Phys. Rev. Lett.",
    volume = "131",
    number = "8",
    pages = "081901",
    year = "2023"
}

@article{Avkhadiev:2019niu,
    author = "Avkhadiev, A. and Shanahan, P. E. and Young, R. D.",
    title = "{Accelerating Lattice Quantum Field Theory Calculations via Interpolator Optimization Using Noisy Intermediate-Scale Quantum Computing}",
    eprint = "1908.04194",
    archivePrefix = "arXiv",
    primaryClass = "hep-lat",
    reportNumber = "MIT-CTP/5138; ADP-19-17/T1097",
    doi = "10.1103/PhysRevLett.124.080501",
    journal = "Phys. Rev. Lett.",
    volume = "124",
    number = "8",
    pages = "080501",
    year = "2020"
}

@article{Singh:2019uwd,
    author = "Singh, Hersh and Chandrasekharan, Shailesh",
    title = "{Qubit regularization of the $O(3)$ sigma model}",
    eprint = "1905.13204",
    archivePrefix = "arXiv",
    primaryClass = "hep-lat",
    doi = "10.1103/PhysRevD.100.054505",
    journal = "Phys. Rev. D",
    volume = "100",
    number = "5",
    pages = "054505",
    year = "2019"
}

@article{Florio:2023kel,
    author = "Florio, Adrien and Weichselbaum, Andreas and Valgushev, Semeon and Pisarski, Robert D.",
    title = "{Mass gaps of a Z3 gauge theory with three fermion flavors in 1+1 dimensions}",
    eprint = "2310.18312",
    archivePrefix = "arXiv",
    primaryClass = "hep-th",
    doi = "10.1103/PhysRevD.110.045013",
    journal = "Phys. Rev. D",
    volume = "110",
    number = "4",
    pages = "045013",
    year = "2024"
}

@article{PhysRevD.107.114513,
  title = {Hamiltonians and gauge-invariant Hilbert space for lattice Yang-Mills-like theories with finite gauge group},
  author = {Mariani, A. and Pradhan, S. and Ercolessi, E.},
  journal = {Phys. Rev. D},
  volume = {107},
  issue = {11},
  pages = {114513},
  numpages = {15},
  year = {2023},
  month = {Jun},
  publisher = {American Physical Society},
  doi = {10.1103/PhysRevD.107.114513},
  url = {https://link.aps.org/doi/10.1103/PhysRevD.107.114513}
}

@article{Ciavarella:2024lsp,
    author = "Ciavarella, Anthony N.",
    title = "{String breaking in the heavy quark limit with scalable circuits}",
    eprint = "2411.05915",
    archivePrefix = "arXiv",
    primaryClass = "quant-ph",
    doi = "10.1103/PhysRevD.111.054501",
    journal = "Phys. Rev. D",
    volume = "111",
    number = "5",
    pages = "054501",
    year = "2025"
}

@article{Ciavarella:2021nmj,
    author = "Ciavarella, Anthony and Klco, Natalie and Savage, Martin J.",
    title = "{Trailhead for quantum simulation of SU(3) Yang-Mills lattice gauge theory in the local multiplet basis}",
    eprint = "2101.10227",
    archivePrefix = "arXiv",
    primaryClass = "quant-ph",
    reportNumber = "IQuS@UW-21-001",
    doi = "10.1103/PhysRevD.103.094501",
    journal = "Phys. Rev. D",
    volume = "103",
    number = "9",
    pages = "094501",
    year = "2021"
}

@article{Rhodes:2024zbr,
    author = "Rhodes, Mason L. and Kreshchuk, Michael and Pathak, Shivesh",
    title = "{Exponential Improvements in the Simulation of Lattice Gauge Theories Using Near-Optimal Techniques}",
    eprint = "2405.10416",
    archivePrefix = "arXiv",
    primaryClass = "quant-ph",
    doi = "10.1103/PRXQuantum.5.040347",
    journal = "PRX Quantum",
    volume = "5",
    number = "4",
    pages = "040347",
    year = "2024"
}

@article{Grimus:2010ak,
    author = "Grimus, W. and Ludl, P. O.",
    title = "{Principal series of finite subgroups of SU(3)}",
    eprint = "1006.0098",
    archivePrefix = "arXiv",
    primaryClass = "hep-ph",
    reportNumber = "UWTHPH-2010-10",
    doi = "10.1088/1751-8113/43/44/445209",
    journal = "J. Phys. A",
    volume = "43",
    pages = "445209",
    year = "2010"
}

@misc{Di:2011cvl,
    author = "Di, Yao-Min and Wei, Hai-Rui",
    title = "{Elementary gates for ternary quantum logic circuit}",
    eprint = "1105.5485",
    archivePrefix = "arXiv",
    primaryClass = "quant-ph",
    month = "5",
    year = "2011"
}

@article{Lamm:2019uyc,
    author = "Lamm, Henry and Lawrence, Scott and Yamauchi, Yukari",
    collaboration = "NuQS",
    title = "{Parton physics on a quantum computer}",
    eprint = "1908.10439",
    archivePrefix = "arXiv",
    primaryClass = "hep-lat",
    doi = "10.1103/PhysRevResearch.2.013272",
    journal = "Phys. Rev. Res.",
    volume = "2",
    number = "1",
    pages = "013272",
    year = "2020"
}

@article{Alexandru:2019nsa,
    author = "Alexandru, Andrei and Bedaque, Paulo F. and Harmalkar, Siddhartha and Lamm, Henry and Lawrence, Scott and Warrington, Neill C.",
    archivePrefix = "arXiv",
    collaboration = "NuQS",
    doi = "10.1103/PhysRevD.100.114501",
    eprint = "1906.11213",
    journal = "Phys.Rev.D",
    number = "11",
    pages = "114501",
    primaryClass = "hep-lat",
    title = "Gluon Field Digitization for Quantum Computers",
    volume = "100",
    year = "2019"
}

@article{Bhanot:1981pj,
      author         = "Bhanot, Gyan",
      title          = "{SU(3) Lattice Gauge Theory in Four-dimensions With a
                        Modified Wilson Action}",
      journal        = "Phys. Lett.",
      volume         = "108B",
      year           = "1982",
      pages          = "337-340",
      doi            = "10.1016/0370-2693(82)91207-2",
      reportNumber   = "Print-81-0779 (IAS,PRINCETON), BNL-30161",
      SLACcitation   = "%%CITATION = PHLTA,108B,337;%%"
}

@article{Creutz:1982dn,
      author         = "Creutz, Michael and Okawa, Masanori",
      title          = "{Generalized Actions in $Z(p$) Lattice Gauge Theory}",
      journal        = "Nucl. Phys.",
      volume         = "B220",
      year           = "1983",
      pages          = "149-166",
      doi            = "10.1016/0550-3213(83)90220-1",
      reportNumber   = "BNL-32052",
      SLACcitation   = "%%CITATION = NUPHA,B220,149;%%"
}

@article{Armon:2021uqr,
    author = "Armon, Tsafrir and Ashkenazi, Shachar and Garc{\'\i}a-Moreno, Gerardo and Gonz{\'a}lez-Tudela, Alejandro and Zohar, Erez",
    title = "{Photon-Mediated Stroboscopic Quantum Simulation of a Z2 Lattice Gauge Theory}",
    eprint = "2107.13024",
    archivePrefix = "arXiv",
    primaryClass = "quant-ph",
    doi = "10.1103/PhysRevLett.127.250501",
    journal = "Phys. Rev. Lett.",
    volume = "127",
    number = "25",
    pages = "250501",
    year = "2021"
}

@article{Lamm:2019bik,
      author         = "Lamm, Henry and Lawrence, Scott and Yamauchi, Yukari",
      title          = "{General Methods for Digital Quantum Simulation of Gauge
                        Theories}",
      collaboration  = "NuQS",
      journal        = "Phys. Rev.",
      volume         = "D100",
      year           = "2019",
      number         = "3",
      pages          = "034518",
      doi            = "10.1103/PhysRevD.100.034518",
      eprint         = "1903.08807",
      archivePrefix  = "arXiv",
      primaryClass   = "hep-lat",
      SLACcitation   = "%%CITATION = ARXIV:1903.08807;%%"
}

@article{Luo:2019vmi,
    author = "Luo, Di and Shen, Jiayu and Highman, Michael and Clark, Bryan K. and DeMarco, Brian and El-Khadra, Aida X. and Gadway, Bryce",
    title = "{Framework for simulating gauge theories with dipolar spin systems}",
    eprint = "1912.11488",
    archivePrefix = "arXiv",
    primaryClass = "quant-ph",
    doi = "10.1103/PhysRevA.102.032617",
    journal = "Phys. Rev. A",
    volume = "102",
    number = "3",
    pages = "032617",
    year = "2020"
}

@article{Brower:2020huh,
    author = "Brower, Richard C. and Berenstein, David and Kawai, Hiroki",
    title = "{Lattice Gauge Theory for a Quantum Computer}",
    eprint = "2002.10028",
    archivePrefix = "arXiv",
    primaryClass = "hep-lat",
    journal = "PoS",
    volume = "LATTICE2019",
    pages = "112",
    year = "2019"
}

@misc{Liu:2020eoa,
    author = "Liu, Junyu and Xin, Yuan",
    title = "{Quantum simulation of quantum field theories as quantum chemistry}",
    eprint = "2004.13234",
    archivePrefix = "arXiv",
    primaryClass = "hep-th",
    reportNumber = "CALT-TH-2020-009",
    month = "4",
    year = "2020"
}

@misc{Kreshchuk:2020kcz,
    author = "Kreshchuk, Michael and Jia, Shaoyang and Kirby, William M. and Goldstein, Gary and Vary, James P. and Love, Peter J.",
    title = "{Light-Front Field Theory on Current Quantum Computers}",
    eprint = "2009.07885",
    archivePrefix = "arXiv",
    primaryClass = "quant-ph",
    month = "9",
    year = "2020"
}

@misc{Kreshchuk:2020dla,
    author = "Kreshchuk, Michael and Kirby, William M. and Goldstein, Gary and Beauchemin, Hugo and Love, Peter J.",
    title = "{Quantum Simulation of Quantum Field Theory in the Light-Front Formulation}",
    eprint = "2002.04016",
    archivePrefix = "arXiv",
    primaryClass = "quant-ph",
    month = "2",
    year = "2020"
}

@article{Haase:2020kaj,
    author = "Haase, Jan F. and Dellantonio, Luca and Celi, Alessio and Paulson, Danny and Kan, Angus and Jansen, Karl and Muschik, Christine A.",
    title = "{A resource efficient approach for quantum and classical simulations of gauge theories in particle physics}",
    eprint = "2006.14160",
    archivePrefix = "arXiv",
    primaryClass = "quant-ph",
    reportNumber = "DESY-20-146",
    doi = "10.22331/q-2021-02-04-393",
    journal = "Quantum",
    volume = "5",
    pages = "393",
    year = "2021"
}

@article{Zhang:2018ufj,
      author         = "Zhang, Jin and Unmuth-Yockey, J. and Zeiher, J. and
                        Bazavov, A. and Tsai, S. -W. and Meurice, Y.",
      title          = "{Quantum simulation of the universal features of the
                        Polyakov loop}",
      journal        = "Phys. Rev. Lett.",
      volume         = "121",
      year           = "2018",
      number         = "22",
      pages          = "223201",
      doi            = "10.1103/PhysRevLett.121.223201",
      eprint         = "1803.11166",
      archivePrefix  = "arXiv",
      primaryClass   = "hep-lat",
      SLACcitation   = "%%CITATION = ARXIV:1803.11166;%%"
}

@article{Bazavov:2015kka,
      author         = "Bazavov, Alexei and Meurice, Yannick and Tsai, Shan-Wen
                        and Unmuth-Yockey, Judah and Zhang, Jin",
      title          = "{Gauge-invariant implementation of the Abelian Higgs
                        model on optical lattices}",
      journal        = "Phys. Rev.",
      volume         = "D92",
      year           = "2015",
      number         = "7",
      pages          = "076003",
      doi            = "10.1103/PhysRevD.92.076003",
      eprint         = "1503.08354",
      archivePrefix  = "arXiv",
      primaryClass   = "hep-lat",
      reportNumber   = "INT-PUB-15-008",
      SLACcitation   = "%%CITATION = ARXIV:1503.08354;%%"
}

@article{Unmuth-Yockey:2018xak,
    author = "Unmuth-Yockey, Judah F.",
    archivePrefix = "arXiv",
    doi = "10.1103/PhysRevD.99.074502",
    eprint = "1811.05884",
    journal = "Phys.\ Rev.\ D",
    number = "7",
    pages = "074502",
    primaryClass = "hep-lat",
    title = "{Gauge-invariant rotor Hamiltonian from dual variables of 3D $U(1)$ gauge theory}",
    volume = "99",
    year = "2019"
}

@article{Unmuth-Yockey:2018ugm,
      author         = "Unmuth-Yockey, Judah and Zhang, Jin and Bazavov, Alexei
                        and Meurice, Yannick and Tsai, Shan-Wen",
      title          = "{Universal features of the Abelian Polyakov loop in 1+1
                        dimensions}",
      journal        = "Phys. Rev.",
      volume         = "D98",
      year           = "2018",
      number         = "9",
      pages          = "094511",
      doi            = "10.1103/PhysRevD.98.094511",
      eprint         = "1807.09186",
      archivePrefix  = "arXiv",
      primaryClass   = "hep-lat",
      SLACcitation   = "%%CITATION = ARXIV:1807.09186;%%"
}

@article{Gustafson:2019mpk,
    author = "Gustafson, Erik and Meurice, Yannick and Unmuth-Yockey, Judah",
    title = "{Quantum simulation of scattering in the quantum Ising model}",
    eprint = "1901.05944",
    archivePrefix = "arXiv",
    primaryClass = "hep-lat",
    doi = "10.1103/PhysRevD.99.094503",
    journal = "Phys. Rev. D",
    volume = "99",
    number = "9",
    pages = "094503",
    year = "2019"
}

@article{Gustafson:2021qbt,
    author = "Gustafson, Erik",
    title = "{Prospects for Simulating a Qudit Based Model of (1+1)d Scalar QED}",
    eprint = "2104.10136",
    archivePrefix = "arXiv",
    primaryClass = "quant-ph",
    doi = "10.1103/PhysRevD.103.114505",
    journal = "Phys. Rev. D",
    volume = "103",
    number = "11",
    pages = "114505",
    year = "2021"
}

@misc{Kreshchuk:2020aiq,
    author = "Kreshchuk, Michael and Jia, Shaoyang and Kirby, William M. and Goldstein, Gary and Vary, James P. and Love, Peter J.",
    title = "{Simulating Hadronic Physics on NISQ devices using Basis Light-Front Quantization}",
    eprint = "2011.13443",
    archivePrefix = "arXiv",
    primaryClass = "quant-ph",
    month = "11",
    year = "2020"
}

@article{Klco:2019evd,
    author = "Klco, Natalie and Stryker, Jesse R. and Savage, Martin J.",
    title = "{SU(2) non-Abelian gauge field theory in one dimension on digital quantum computers}",
    eprint = "1908.06935",
    archivePrefix = "arXiv",
    primaryClass = "quant-ph",
    reportNumber = "INT-PUB-19-033",
    doi = "10.1103/PhysRevD.101.074512",
    journal = "Phys. Rev. D",
    volume = "101",
    number = "7",
    pages = "074512",
    year = "2020"
}

@article{Gustafson:2020yfe,
    author = "Gustafson, Erik J. and Lamm, Henry",
    title = "{Toward quantum simulations of $\mathbb{Z}_2$ gauge theory without state preparation}",
    eprint = "2011.11677",
    archivePrefix = "arXiv",
    primaryClass = "hep-lat",
    reportNumber = "FERMILAB-PUB-20-611-T",
    doi = "10.1103/PhysRevD.103.054507",
    journal = "Phys. Rev. D",
    volume = "103",
    number = "5",
    pages = "054507",
    year = "2021"
}

@article{Gustafson:2019vsd,
    author = "Gustafson, Erik and Dreher, Patrick and Hang, Zheyue and Meurice, Yannick",
    title = "{Benchmarking quantum computers for real-time evolution of a $(1+1)$ field theory with error mitigation}",
    eprint = "1910.09478",
    archivePrefix = "arXiv",
    primaryClass = "hep-lat",
    journal = "Quantum Sci. Technol.",
    volume = "6",
    pages = "045020",
    year = "2021"
}

@article{Yeh:2022eow,
    author = "Yeh, Lia and van de Wetering, John",
    title = "{Constructing all qutrit controlled Clifford+T gates in Clifford+T.}",
    eprint = "2204.00552",
    archivePrefix = "arXiv",
    primaryClass = "quant-ph",
    doi = "10.1007/978-3-031-09005-9_3",
    journal = "Lect. Notes Comput. Sci.",
    volume = "13354",
    pages = "28",
    year = "2022"
}

@article{Vatan:2004nmz,
    author = "Vatan, Farrokh and Williams, Colin",
    title = "{Optimal quantum circuits for general two-qubit gates}",
    eprint = "quant-ph/0308006",
    archivePrefix = "arXiv",
    doi = "10.1103/PhysRevA.69.032315",
    journal = "Phys. Rev. A",
    volume = "69",
    number = "3",
    pages = "032315",
    year = "2004"
}

@article{Glaudell:2022vvu,
    author = "Glaudell, Andrew N. and Ross, Neil J. and van de Wetering, John and Yeh, Lia",
    title = "{Qutrit Metaplectic Gates Are a Subset of Clifford+T}",
    eprint = "2202.09235",
    archivePrefix = "arXiv",
    primaryClass = "quant-ph",
    doi = "10.4230/LIPIcs.TQC.2022.12",
    journal = "Leibniz Int. Proc. Inf.",
    volume = "232",
    pages = "12:1--12:15",
    year = "2022"
}

@article{Bocharov:2018zfk,
    author = "Bocharov, Alex",
    title = "{A note on optimality of quantum circuits over metaplectic basis}",
    doi = "10.26421/QIC18.1-2-1",
    journal = "Quant. Inf. Comput.",
    volume = "18",
    number = "1-2",
    pages = "0001--0017",
    year = "2018"
}

@article{Shende:2004gqq,
    author = "Shende, Vivek V. and Markov, Igor L. and Bullock, Stephen S.",
    title = "{Minimal universal two-qubit controlled-NOT-based circuits}",
    eprint = "quant-ph/0308033",
    archivePrefix = "arXiv",
    doi = "10.1103/PhysRevA.69.062321",
    journal = "Phys. Rev. A",
    volume = "69",
    number = "6",
    pages = "062321",
    year = "2004"
}

@article{peruzzo2014variational,
  title={A variational eigenvalue solver on a photonic quantum processor},
  author={Peruzzo, Alberto and McClean, Jarrod and Shadbolt, Peter and Yung, Man-Hong and Zhou, Xiao-Qi and Love, Peter J and Aspuru-Guzik, Al{\'a}n and O’brien, Jeremy L},
  journal={Nature communications},
  volume={5},
  pages={4213},
  year={2014},
  publisher={Nature Publishing Group}
}

@article{Zache:2023cfj,
    author = "Zache, Torsten V. and Gonz\'alez-Cuadra, Daniel and Zoller, Peter",
    title = "{Fermion-qudit quantum processors for simulating lattice gauge theories with matter}",
    eprint = "2303.08683",
    archivePrefix = "arXiv",
    primaryClass = "quant-ph",
    doi = "10.22331/q-2023-10-16-1140",
    journal = "Quantum",
    volume = "7",
    pages = "1140",
    year = "2023"
}

@article{Zache:2023dko,
    author = "Zache, Torsten V. and Gonz\'alez-Cuadra, Daniel and Zoller, Peter",
    title = "{Quantum and Classical Spin-Network Algorithms for q-Deformed Kogut-Susskind Gauge Theories}",
    eprint = "2304.02527",
    archivePrefix = "arXiv",
    primaryClass = "quant-ph",
    doi = "10.1103/PhysRevLett.131.171902",
    journal = "Phys. Rev. Lett.",
    volume = "131",
    number = "17",
    pages = "171902",
    year = "2023"
}

@article{Bender:2018rdp,
      author         = "Bender, Julian and Zohar, Erez and Farace, Alessandro and
                        Cirac, J. Ignacio",
      title          = "{Digital quantum simulation of lattice gauge theories in
                        three spatial dimensions}",
      journal        = "New J. Phys.",
      volume         = "20",
      year           = "2018",
      number         = "9",
      pages          = "093001",
      doi            = "10.1088/1367-2630/aadb71",
      eprint         = "1804.02082",
      archivePrefix  = "arXiv",
      primaryClass   = "quant-ph",
      SLACcitation   = "%%CITATION = ARXIV:1804.02082;%%"
}

@misc{kitaev,
    author = "Kitaev, A. Yu.",
    title = "{Quantum measurements and the Abelian stabilizer problem}",
    eprint = "quant-ph/9511026",
    archivePrefix = "arXiv",
    month = "11",
    year = "1995",
    url={https://arxiv.org/abs/quant-ph/9511026}
}

@book{NC,
  title={Quantum computation and quantum information},
  author={Nielsen, Michael A and Chuang, Isaac},
  year={2002},
  publisher={Cambridge University Press}
}

@inproceedings{hales2000improved,
  title={An improved quantum Fourier transform algorithm and applications},
  author={Hales, Lisa and Hallgren, Sean},
  booktitle={Proceedings 41st Annual Symposium on Foundations of Computer Science},
  pages={515--525},
  year={2000},
  organization={IEEE}
}

@article{Gustafson:2023kvd,
    author = "Gustafson, Erik J. and Lamm, Henry and Lovelace, Felicity",
    title = "{Primitive quantum gates for an SU(2) discrete subgroup: Binary octahedral}",
    eprint = "2312.10285",
    archivePrefix = "arXiv",
    primaryClass = "hep-lat",
    reportNumber = "FERMILAB-PUB-23-753-SQMS-T",
    doi = "10.1103/PhysRevD.109.054503",
    journal = "Phys. Rev. D",
    volume = "109",
    number = "5",
    pages = "054503",
    year = "2024"
}

@article{Zohar:2013zla,
      author         = "Zohar, Erez and Cirac, J. Ignacio and Reznik, Benni",
      title          = "{Quantum simulations of gauge theories with ultracold
                        atoms: local gauge invariance from angular momentum
                        conservation}",
      journal        = "Phys. Rev.",
      volume         = "A88",
      year           = "2013",
      pages          = "023617",
      doi            = "10.1103/PhysRevA.88.023617",
      eprint         = "1303.5040",
      archivePrefix  = "arXiv",
      primaryClass   = "quant-ph",
      SLACcitation   = "%%CITATION = ARXIV:1303.5040;%%"
}

@article{Brower:1997ha,
	Archiveprefix = {arXiv},
	Author = {Brower, R. and Chandrasekharan, S. and Wiese, U. J.},
	Doi = {10.1103/PhysRevD.60.094502},
	Eprint = {hep-th/9704106},
	Journal = {Phys. Rev.},
	Pages = {094502},
	Primaryclass = {hep-th},
	Reportnumber = {MIT-CTP-2623},
	Slaccitation = {%%CITATION = HEP-TH/9704106;%%},
	Title = {{QCD as a quantum link model}},
	Volume = {D60},
	Year = {1999}}

@inproceedings{Pueschel:1998zzo,
  title={{Fast quantum Fourier transforms for a class of non-abelian groups}},
  author={P{\"u}schel, Markus and R{\"o}tteler, Martin and Beth, Thomas},
  booktitle={Applied Algebra, Algebraic Algorithms and Error-Correcting Codes: 13th International Symposium, AAECC-13 Honolulu, Hawaii, USA, November 15--19, 1999 Proceedings 13},
  pages={148--159},
  year={1999},
  organization={Springer},
  url={https://arxiv.org/abs/quant-ph/9807064}
}

@article{moore2003generic,
  title={{Generic quantum Fourier transforms}},
  author={Moore, Cristopher and Rockmore, Daniel and Russell, Alexander},
  journal={ACM Transactions on Algorithms (TALG)},
  volume={2},
  number={4},
  pages={707--723},
  year={2006},
  publisher={ACM New York, NY, USA},
  url={https://arxiv.org/abs/quant-ph/0304064}
}

@misc{Hoyer:1997qc,
    author = "Hoyer, Peter",
    title = "{Efficient quantum transforms}",
    eprint = "quant-ph/9702028",
    archivePrefix = "arXiv",
    month = "2",
    year = "1997"
}

@inproceedings{beals1997quantum,
  title={Quantum computation of Fourier transforms over symmetric groups},
  author={Beals, Robert},
  booktitle={Proceedings of the twenty-ninth annual ACM symposium on Theory of computing},
  pages={48--53},
  year={1997}
}

@article{Carena:2022hpz,
    author = "Carena, Marcela and Gustafson, Erik J. and Lamm, Henry and Li, Ying-Ying and Liu, Wanqiang",
    title = "{Gauge theory couplings on anisotropic lattices}",
    eprint = "2208.10417",
    archivePrefix = "arXiv",
    primaryClass = "hep-lat",
    reportNumber = "FERMILAB-PUB-21-702-T",
    doi = "10.1103/PhysRevD.106.114504",
    journal = "Phys. Rev. D",
    volume = "106",
    number = "11",
    pages = "114504",
    year = "2022"
}

@article{PhysRevA.52.3457,
  title = {Elementary gates for quantum computation},
  author = {Barenco, Adriano and Bennett, Charles H. and Cleve, Richard and DiVincenzo, David P. and Margolus, Norman and Shor, Peter and Sleator, Tycho and Smolin, John A. and Weinfurter, Harald},
  journal = {Phys. Rev. A},
  volume = {52},
  issue = {5},
  pages = {3457--3467},
  numpages = {0},
  year = {1995},
  month = {Nov},
  publisher = {American Physical Society},
  doi = {10.1103/PhysRevA.52.3457},
  url = {https://link.aps.org/doi/10.1103/PhysRevA.52.3457}
}

@misc{2019arXiv190401671B,
       author = {{Baker}, Jonathan M. and {Duckering}, Casey and {Hoover}, Alexander and {Chong}, Frederic T.},
        title = "{Decomposing Quantum Generalized Toffoli with an Arbitrary Number of Ancilla}",
         year = 2019,
        month = apr,
archivePrefix = "arXiv",
       eprint = "1904.01671",
 primaryClass = "quant-ph",
}

@article{Gattringer:2016kco,
    author = "Gattringer, Christof and Langfeld, Kurt",
    title = "{Approaches to the sign problem in lattice field theory}",
    eprint = "1603.09517",
    archivePrefix = "arXiv",
    primaryClass = "hep-lat",
    doi = "10.1142/S0217751X16430077",
    journal = "Int. J. Mod. Phys. A",
    volume = "31",
    number = "22",
    pages = "1643007",
    year = "2016"
}

@ARTICLE{1715533,  author={Ketkar, A. and Klappenecker, A. and Kumar, S. and Sarvepalli, P.K.},  journal={IEEE Transactions on Information Theory},   title={Nonbinary Stabilizer Codes Over Finite Fields},   year={2006},  volume={52},  number={11},  pages={4892-4914},  doi={10.1109/TIT.2006.883612}}

@inproceedings{Grassl:1999qrs,
    author = "Grassl, Markus and Geiselmann, Willi and Beth, Thomas",
    title = "{Quantum Reed-Solomon Codes}",
    booktitle = "{Applied Algebra, Algebraic Algorithms and Error-Correcting Codes (AAECC-13)}",
    series = "Lecture Notes in Computer Science",
    volume = "1719",
    pages = "231--244",
    publisher = "Springer",
    eprint = "quant-ph/9910059",
    archivePrefix = "arXiv",
    primaryClass = "quant-ph",
    doi = "10.1007/3-540-46796-3_23",
    year = "1999"
}

@article{Wang:2020qhd,
    author = "Wang, Yuchen and Hu, Zixuan and Sanders, Barry C. and Kais, Sabre",
    title = "{Qudits and High-Dimensional Quantum Computing}",
    journal = "Front. in Phys.",
    volume = "8",
    pages = "589504",
    year = "2020",
    eprint = "2008.00959",
    archivePrefix = "arXiv",
    primaryClass = "quant-ph",
    doi = "10.3389/fphy.2020.589504"
}

@article{Troyer:2004ge,
      author         = "Troyer, Matthias and Wiese, Uwe-Jens",
      title          = "{Computational complexity and fundamental limitations to
                        fermionic quantum Monte Carlo simulations}",
      journal        = "Phys. Rev. Lett.",
      volume         = "94",
      year           = "2005",
      pages          = "170201",
      doi            = "10.1103/PhysRevLett.94.170201",
      eprint         = "cond-mat/0408370",
      archivePrefix  = "arXiv",
      primaryClass   = "cond-mat",
      SLACcitation   = "%%CITATION = COND-MAT/0408370;%%"
}

@article{PhysRevD.99.114507,
  title = {Tensor renormalization group study of the non-Abelian Higgs model in two dimensions},
  author = {Bazavov, Alexei and Catterall, Simon and Jha, Raghav G. and Unmuth-Yockey, Judah},
  journal = {Phys. Rev. D},
  volume = {99},
  issue = {11},
  pages = {114507},
  numpages = {12},
  year = {2019},
  month = {Jun},
  publisher = {American Physical Society},
  doi = {10.1103/PhysRevD.99.114507},
  url = {https://link.aps.org/doi/10.1103/PhysRevD.99.114507}
}

@article{Buser:2020uzs,
    author = "Buser, Alexander J. and Bhattacharya, Tanmoy and Cincio, Lukasz and Gupta, Rajan",
    title = "{State preparation and measurement in a quantum simulation of the $O$(3) sigma model}",
    eprint = "2006.15746",
    archivePrefix = "arXiv",
    primaryClass = "quant-ph",
    reportNumber = "LA-UR-20-24538",
    doi = "10.1103/PhysRevD.102.114514",
    journal = "Phys. Rev. D",
    volume = "102",
    number = "11",
    pages = "114514",
    year = "2020"
}

@article{Bhattacharya:2020gpm,
    author = "Bhattacharya, Tanmoy and Buser, Alexander J. and Chandrasekharan, Shailesh and Gupta, Rajan and Singh, Hersh",
    title = "{Qubit regularization of asymptotic freedom}",
    eprint = "2012.02153",
    archivePrefix = "arXiv",
    primaryClass = "hep-lat",
    reportNumber = "LA-UR-20-29558",
    doi = "10.1103/PhysRevLett.126.172001",
    journal = "Phys. Rev. Lett.",
    volume = "126",
    number = "17",
    pages = "172001",
    year = "2021"
}

@article{Zohar:2012ay,
      author         = "Zohar, Erez and Cirac, J. Ignacio and Reznik, Benni",
      title          = "{Simulating Compact Quantum Electrodynamics with
                        ultracold atoms: Probing confinement and nonperturbative
                        effects}",
      journal        = "Phys. Rev. Lett.",
      volume         = "109",
      year           = "2012",
      pages          = "125302",
      doi            = "10.1103/PhysRevLett.109.125302",
      eprint         = "1204.6574",
      archivePrefix  = "arXiv",
      primaryClass   = "quant-ph",
      SLACcitation   = "%%CITATION = ARXIV:1204.6574;%%"
}

@article{Wiese:2014rla,
      author         = "Wiese, Uwe-Jens",
      title          = "{Towards Quantum Simulating QCD}",
      booktitle      = "{Proceedings, 24th International Conference on
                        Ultra-Relativistic Nucleus-Nucleus Collisions (Quark
                        Matter 2014): Darmstadt, Germany, May 19-24, 2014}",
      journal        = "Nucl. Phys.",
      volume         = "A931",
      year           = "2014",
      pages          = "246-256",
      doi            = "10.1016/j.nuclphysa.2014.09.102",
      eprint         = "1409.7414",
      archivePrefix  = "arXiv",
      primaryClass   = "hep-th",
      SLACcitation   = "%%CITATION = ARXIV:1409.7414;%%"
}

@article{Zohar:2016iic,
      author         = "Zohar, Erez and Farace, Alessandro and Reznik, Benni and
                        Cirac, J. Ignacio",
      title          = "{Digital lattice gauge theories}",
      journal        = "Phys. Rev.",
      volume         = "A95",
      year           = "2017",
      number         = "2",
      pages          = "023604",
      doi            = "10.1103/PhysRevA.95.023604",
      eprint         = "1607.08121",
      archivePrefix  = "arXiv",
      primaryClass   = "quant-ph",
      SLACcitation   = "%%CITATION = ARXIV:1607.08121;%%"
}

@article{Jordan:2017lea,
    author = "Jordan, Stephen P. and Krovi, Hari and Lee, Keith S.M. and Preskill, John",
    title = "{BQP-completeness of Scattering in Scalar Quantum Field Theory}",
    eprint = "1703.00454",
    archivePrefix = "arXiv",
    primaryClass = "quant-ph",
    doi = "10.22331/q-2018-01-08-44",
    journal = "Quantum",
    volume = "2",
    pages = "44",
    year = "2018"
}

@article{Zohar:2015hwa,
      author         = "Zohar, Erez and Cirac, J. Ignacio and Reznik, Benni",
      title          = "{Quantum Simulations of Lattice Gauge Theories using
                        Ultracold Atoms in Optical Lattices}",
      journal        = "Rept. Prog. Phys.",
      volume         = "79",
      year           = "2016",
      number         = "1",
      pages          = "014401",
      doi            = "10.1088/0034-4885/79/1/014401",
      eprint         = "1503.02312",
      archivePrefix  = "arXiv",
      primaryClass   = "quant-ph",
      SLACcitation   = "%%CITATION = ARXIV:1503.02312;%%"
}

@article{Kane:2023jdo,
    author = "Kane, Christopher F. and Gomes, Niladri and Kreshchuk, Michael",
    title = "{Nearly optimal state preparation for quantum simulations of lattice gauge theories}",
    eprint = "2310.13757",
    archivePrefix = "arXiv",
    primaryClass = "quant-ph",
    reportNumber = "Phys.Rev.A 110 (2024) 1, 012455",
    doi = "10.1103/PhysRevA.110.012455",
    journal = "Phys. Rev. A",
    volume = "110",
    number = "1",
    pages = "012455",
    year = "2024"
}

@article{Gustafson:2022xdt,
    author = "Gustafson, Erik J. and Lamm, Henry and Lovelace, Felicity and Musk, Damian",
    title = "{Primitive quantum gates for an SU(2) discrete subgroup: Binary tetrahedral}",
    eprint = "2208.12309",
    archivePrefix = "arXiv",
    primaryClass = "quant-ph",
    reportNumber = "FERMILAB-PUB-22-583-SQMS-T",
    doi = "10.1103/PhysRevD.106.114501",
    journal = "Phys. Rev. D",
    volume = "106",
    number = "11",
    pages = "114501",
    year = "2022"
}

@misc{Gustafson:2022hjf,
    author = "Gustafson, Erik J.",
    title = "{Stout Smearing on a Quantum Computer}",
    eprint = "2211.05607",
    archivePrefix = "arXiv",
    primaryClass = "hep-lat",
    reportNumber = "FERMILAB-PUB-22-838-T",
    month = "11",
    year = "2022"
}

@article{Xie:2022jgj,
	archiveprefix = {arXiv},
	author = {Xie, Xu-Dan and Guo, Xingyu and Xing, Hongxi and Xue, Zheng-Yuan and Zhang, Dan-Bo and Zhu, Shi-Liang},
	collaboration = {QuNu},
	doi = {10.1103/PhysRevD.106.054509},
	eprint = {2205.12767},
	journal = {Phys. Rev. D},
	number = {5},
	pages = {054509},
	primaryclass = {quant-ph},
	title = {{Variational thermal quantum simulation of the lattice Schwinger model}},
	volume = {106},
	year = {2022}
}

@article{motta2020determining,
  title={Determining eigenstates and thermal states on a quantum computer using quantum imaginary time evolution},
  author={Motta, Mario and Sun, Chong and Tan, Adrian TK and O’Rourke, Matthew J and Ye, Erika and Minnich, Austin J and Brandao, Fernando GSL and Chan, Garnet Kin-Lic},
  journal={Nature Physics},
  volume={16},
  number={2},
  pages={205--210},
  year={2020},
  publisher={Nature Publishing Group}
}

@article{PhysRevLett.108.080402,
  title = {Thermalization in Nature and on a Quantum Computer},
  author = {Riera, Arnau and Gogolin, Christian and Eisert, Jens},
  journal = {Phys. Rev. Lett.},
  volume = {108},
  issue = {8},
  pages = {080402},
  numpages = {5},
  year = {2012},
  month = {Feb},
  publisher = {American Physical Society},
  doi = {10.1103/PhysRevLett.108.080402},
  url = {https://link.aps.org/doi/10.1103/PhysRevLett.108.080402}
}

@article{Charles:2023zbl,
    author = "Charles, Clement and Gustafson, Erik J. and Hardt, Elizabeth and Herren, Florian and Hogan, Norman and Lamm, Henry and Starecheski, Sara and Van de Water, Ruth S. and Wagman, Michael L.",
    title = "{Simulating Z2 lattice gauge theory on a quantum computer}",
    eprint = "2305.02361",
    archivePrefix = "arXiv",
    primaryClass = "hep-lat",
    reportNumber = "FERMILAB-PUB-23-171-SQMS-T",
    doi = "10.1103/PhysRevE.109.015307",
    journal = "Phys. Rev. E",
    volume = "109",
    number = "1",
    pages = "015307",
    year = "2024"
}

@article{Clemente:2020lpr,
    author = "Clemente, Giuseppe and others",
    collaboration = "QuBiPF",
    title = "{Quantum computation of thermal averages in the presence of a sign problem}",
    eprint = "2001.05328",
    archivePrefix = "arXiv",
    primaryClass = "hep-lat",
    doi = "10.1103/PhysRevD.101.074510",
    journal = "Phys. Rev. D",
    volume = "101",
    number = "7",
    pages = "074510",
    year = "2020"
}

@article{PhysRevD.11.395,
  title = {Hamiltonian formulation of {W}ilson's lattice gauge theories},
  author = {Kogut, John and Susskind, Leonard},
  journal = {Phys. Rev. D},
  volume = {11},
  issue = {2},
  pages = {395--408},
  numpages = {0},
  year = {1975},
  month = {Jan},
  publisher = {American Physical Society},
  doi = {10.1103/PhysRevD.11.395},
  url = {https://link.aps.org/doi/10.1103/PhysRevD.11.395}
}

@article{Assi:2024pdn,
    author = "Assi, Beno{\^\i}t and Lamm, Henry",
    title = "{Digitization and subduction of SU(N) gauge theories}",
    eprint = "2405.12204",
    archivePrefix = "arXiv",
    primaryClass = "hep-lat",
    reportNumber = "FERMILAB-PUB-24-0177-T",
    doi = "10.1103/PhysRevD.110.074511",
    journal = "Phys. Rev. D",
    volume = "110",
    number = "7",
    pages = "074511",
    year = "2024"
}

@article{Johnson:1982yq,
    author = "Johnson, R. C.",
    title = "{Angular momentum on a lattice}",
    reportNumber = "CERN-TH-3245-REV, CERN-TH-3245",
    doi = "10.1016/0370-2693(82)90134-4",
    journal = "Phys. Lett. B",
    volume = "114",
    pages = "147--151",
    year = "1982"
}

@article{Moore:2005dw,
    author = "Moore, David C. and Fleming, George Tamminga",
    title = "{Angular momentum on the lattice: The Case of non-zero linear momentum}",
    eprint = "hep-lat/0507018",
    archivePrefix = "arXiv",
    doi = "10.1103/PhysRevD.73.014504",
    journal = "Phys. Rev. D",
    volume = "73",
    pages = "014504",
    year = "2006",
    note = "[Erratum: Phys.Rev.D 74, 079905 (2006)]"
}

@article{Lamm:2024jnl,
    author = "Lamm, Henry and Li, Ying-Ying and Shu, Jing and Wang, Yi-Lin and Xu, Bin",
    title = "{Block encodings of discrete subgroups on a quantum computer}",
    eprint = "2405.12890",
    archivePrefix = "arXiv",
    primaryClass = "hep-lat",
    reportNumber = "USTC-ICTS/PCFT-24-15, FERMILAB-PUB-24-0242-T",
    doi = "10.1103/PhysRevD.110.054505",
    journal = "Phys. Rev. D",
    volume = "110",
    number = "5",
    pages = "054505",
    year = "2024"
}

@article{Harlow:2018tng,
    author = "Harlow, Daniel and Ooguri, Hirosi",
    title = "{Symmetries in quantum field theory and quantum gravity}",
    eprint = "1810.05338",
    archivePrefix = "arXiv",
    primaryClass = "hep-th",
    doi = "10.1007/s00220-021-04040-y",
    journal = "Commun. Math. Phys.",
    volume = "383",
    number = "3",
    pages = "1669--1804",
    year = "2021"
}

@article{Creutz:1979zg,
	Author = {Creutz, Michael and Jacobs, Laurence and Rebbi, Claudio},
	Doi = {10.1103/PhysRevD.20.1915},
	Journal = {Phys. Rev.},
	Pages = {1915},
	Reportnumber = {BNL-26307},
	Slaccitation = {%%CITATION = PHRVA,D20,1915;%%},
	Title = {{Monte Carlo Study of Abelian Lattice Gauge Theories}},
	Volume = {D20},
	Year = {1979}}

@article{Zohar:2014qma,
	Archiveprefix = {arXiv},
	Author = {Zohar, Erez and Burrello, Michele},
	Doi = {10.1103/PhysRevD.91.054506},
	Eprint = {1409.3085},
	Journal = {Phys. Rev.},
	Number = {5},
	Pages = {054506},
	Primaryclass = {quant-ph},
	Slaccitation = {%%CITATION = ARXIV:1409.3085;%%},
	Title = {{Formulation of lattice gauge theories for quantum simulations}},
	Volume = {D91},
	Year = {2015}}

@article{Bhanot:1981xp,
	Author = {Bhanot, G. and Rebbi, C.},
	Doi = {10.1103/PhysRevD.24.3319},
	Journal = {Phys. Rev.},
	Pages = {3319},
	Reportnumber = {Print-81-0392 (BNL)},
	Slaccitation = {%%CITATION = PHRVA,D24,3319;%%},
	Title = {{Monte Carlo Simulations of Lattice Models With Finite Subgroups of SU(3) as Gauge Groups}},
	Volume = {D24},
	Year = {1981}}

@article{Halimeh:2020xfd,
    author = "Halimeh, Jad C. and Ott, Robert and McCulloch, Ian P. and Yang, Bing and Hauke, Philipp",
    title = "{Robustness of gauge-invariant dynamics against defects in ultracold-atom gauge theories}",
    eprint = "2005.10249",
    archivePrefix = "arXiv",
    primaryClass = "cond-mat.quant-gas",
    doi = "10.1103/PhysRevResearch.2.033361",
    journal = "Phys. Rev. Res.",
    volume = "2",
    number = "3",
    pages = "033361",
    year = "2020"
}

@article{Weingarten:1981jy,
      author         = "Weingarten, Don",
      title          = "{Monte Carlo Evaluation of Hadron Masses in Lattice Gauge
                        Theories with Fermions}",
      journal        = "Phys. Lett.",
      volume         = "109B",
      year           = "1982",
      pages          = "57",
      doi            = "10.1016/0370-2693(82)90463-4",
      note           = "[,631(1981)]",
      reportNumber   = "IUHET-69",
      SLACcitation   = "%%CITATION = PHLTA,109B,57;%%"
}

@article{Weingarten:1980hx,
      author         = "Weingarten, D. H. and Petcher, D. N.",
      title          = "{Monte Carlo Integration for Lattice Gauge Theories with
                        Fermions}",
      journal        = "Phys. Lett.",
      volume         = "99B",
      year           = "1981",
      pages          = "333-338",
      doi            = "10.1016/0370-2693(81)90112-X",
      reportNumber   = "IUHET-61",
      SLACcitation   = "%%CITATION = PHLTA,99B,333;%%"
}

@article{Fradkin:1978dv,
    author = "Fradkin, Eduardo H. and Shenker, Stephen H.",
    doi = "10.1103/PhysRevD.19.3682",
    journal = "Phys. Rev. D",
    pages = "3682--3697",
    reportNumber = "SLAC-PUB-2238",
    title = "{Phase Diagrams of Lattice Gauge Theories with Higgs Fields}",
    volume = "19",
    year = "1979"
}

@article{Petcher:1980cq,
	Author = {Petcher, D. and Weingarten, D. H.},
	Doi = {10.1103/PhysRevD.22.2465},
	Journal = {Phys. Rev.},
	Pages = {2465},
	Reportnumber = {Print-80-0404 (INDIANA), IUHET-53},
	Slaccitation = {%%CITATION = PHRVA,D22,2465;%%},
	Title = {{Monte Carlo Calculations and a Model of the Phase Structure for Gauge Theories on Discrete Subgroups of SU(2)}},
	Volume = {D22},
	Year = {1980}}

@article{Ale:2024uxf,
    author = "Ale, Victor and Bauer, Nora M. and Jha, Raghav G. and Ringer, Felix and Siopsis, George",
    title = "{Quantum computation of SU(2) lattice gauge theory with continuous variables}",
    eprint = "2410.14580",
    archivePrefix = "arXiv",
    primaryClass = "hep-lat",
    reportNumber = "JLAB-THY-24-4217",
    doi = "10.1007/JHEP06(2025)084",
    journal = "JHEP",
    volume = "06",
    pages = "084",
    year = "2025"
}

@article{Mueller:2024mmk,
    author = "Mueller, Niklas and Wang, Tianyi and Katz, Or and Davoudi, Zohreh and Cetina, Marko",
    title = "{Quantum computing universal thermalization dynamics in a (2{\,}+{\,}1)D Lattice Gauge Theory}",
    eprint = "2408.00069",
    archivePrefix = "arXiv",
    primaryClass = "quant-ph",
    reportNumber = "IQuS@UW-21-085, UMD-PP-024-08",
    doi = "10.1038/s41467-025-60177-7",
    journal = "Nature Commun.",
    volume = "16",
    number = "1",
    pages = "5492",
    year = "2025"
}

@article{Pardo:2024edu,
    author = "Pardo, Guy and Bender, Julian and Katz, Nadav and Zohar, Erez",
    title = "{Truncation-free quantum simulation of pure-gauge compact QED using Josephson arrays}",
    eprint = "2410.11413",
    archivePrefix = "arXiv",
    primaryClass = "hep-lat",
    reportNumber = "MIT-CTP/5817",
    doi = "10.1088/2058-9565/adce2a",
    journal = "Quantum Sci. Technol.",
    volume = "10",
    number = "3",
    pages = "035011",
    year = "2025"
}

@article{Raychowdhury:2018osk,
    author = "Raychowdhury, Indrakshi and Stryker, Jesse R.",
    title = "{Solving Gauss's Law on Digital Quantum Computers with Loop-String-Hadron Digitization}",
    eprint = "1812.07554",
    archivePrefix = "arXiv",
    primaryClass = "hep-lat",
    reportNumber = "INT-PUB-18-062",
    doi = "10.1103/PhysRevResearch.2.033039",
    journal = "Phys. Rev. Res.",
    volume = "2",
    number = "3",
    pages = "033039",
    year = "2020"
}

@article{Yamamoto:2020eqi,
    author = "Yamamoto, Arata",
    title = "{Real-time simulation of (2+1)-dimensional lattice gauge theory on qubits}",
    eprint = "2008.11395",
    archivePrefix = "arXiv",
    primaryClass = "hep-lat",
    doi = "10.1093/ptep/ptaa171",
    journal = "PTEP",
    volume = "2021",
    number = "1",
    pages = "013B06",
    year = "2021"
}

@article{Mathis:2020fuo,
    author = "Mathis, Simon V. and Mazzola, Guglielmo and Tavernelli, Ivano",
    title = "{Toward scalable simulations of Lattice Gauge Theories on quantum computers}",
    eprint = "2005.10271",
    archivePrefix = "arXiv",
    primaryClass = "quant-ph",
    doi = "10.1103/PhysRevD.102.094501",
    journal = "Phys. Rev. D",
    volume = "102",
    number = "9",
    pages = "094501",
    year = "2020"
}

@misc{Carena:2021ltu,
    author = "Carena, Marcela and Lamm, Henry and Li, Ying-Ying and Liu, Wanqiang",
    title = "{Lattice Renormalization of Quantum Simulations}",
    eprint = "2107.01166",
    archivePrefix = "arXiv",
    primaryClass = "hep-lat",
    reportNumber = "FERMILAB-PUB-21-222-T",
    month = "7",
    year = "2021"
}

@misc{Harmalkar:2020mpd,
    author = "Harmalkar, Siddhartha and Lamm, Henry and Lawrence, Scott",
    archivePrefix = "arXiv",
    collaboration = "NuQS",
    eprint = "2001.11490",
    month = "1",
    primaryClass = "hep-lat",
    reportNumber = "FERMILAB-PUB-20-045-T",
    title = "{Quantum Simulation of Field Theories Without State Preparation}",
    year = "2020"
}

@article{Hackett:2018cel,
    author = "Hackett, Daniel C. and Howe, Kiel and Hughes, Ciaran and Jay, William and Neil, Ethan T. and Simone, James N.",
    archivePrefix = "arXiv",
    doi = "10.1103/PhysRevA.99.062341",
    eprint = "1811.03629",
    journal = "Phys.\ Rev.\ A",
    number = "6",
    pages = "062341",
    primaryClass = "quant-ph",
    reportNumber = "FERMILAB-PUB-18-615-T",
    title = "{Digitizing Gauge Fields: Lattice Monte Carlo Results for Future Quantum Computers}",
    volume = "99",
    year = "2019"
}

@article{Bauer:2021gek,
    author = "Bauer, Christian W. and Grabowska, Dorota M.",
    title = "{Efficient representation for simulating U(1) gauge theories on digital quantum computers at all values of the coupling}",
    eprint = "2111.08015",
    archivePrefix = "arXiv",
    primaryClass = "hep-ph",
    reportNumber = "CERN-TH-2021-188",
    doi = "10.1103/PhysRevD.107.L031503",
    journal = "Phys. Rev. D",
    volume = "107",
    number = "3",
    pages = "L031503",
    year = "2023"
}

@article{Murairi:2024xpc,
    author = "Murairi, Edison M. and Sohaib Alam, M. and Lamm, Henry and Hadfield, Stuart and Gustafson, Erik",
    title = "{Highly-efficient quantum Fourier transformations for certain non-Abelian groups}",
    eprint = "2408.00075",
    archivePrefix = "arXiv",
    primaryClass = "quant-ph",
    reportNumber = "FERMILAB-PUB-24-0241-SQMS-T",
    doi = "10.1103/PhysRevD.110.074501",
    journal = "Phys. Rev. D",
    volume = "110",
    number = "7",
    pages = "074501",
    year = "2024"
}

@article{Gustafson:2024kym,
    author = "Gustafson, Erik J. and Ji, Yao and Lamm, Henry and Murairi, Edison M. and Perez, Sebastian Osorio and Zhu, Shuchen",
    title = "{Primitive quantum gates for an SU(3) discrete subgroup: \ensuremath{\Sigma}(36\texttimes{}3)}",
    eprint = "2405.05973",
    archivePrefix = "arXiv",
    primaryClass = "hep-lat",
    reportNumber = "FERMILAB-PUB-24-0132-SQMS-T, TUM-HEP-1506/24, FERMILAB-PUB-24-0132-SQMS-T; TUM-HEP-1506/24",
    doi = "10.1103/PhysRevD.110.034515",
    journal = "Phys. Rev. D",
    volume = "110",
    number = "3",
    pages = "034515",
    year = "2024"
}

@misc{Murairi:2022zdg,
    author = "Murairi, Edison M. and Cervia, Michael J. and Kumar, Hersh and Bedaque, Paulo F. and Alexandru, Andrei",
    title = "{How many quantum gates do gauge theories require?}",
    eprint = "2208.11789",
    archivePrefix = "arXiv",
    primaryClass = "hep-lat",
    month = "8",
    year = "2022"
}

@article{Hartung:2022hoz,
    author = "Hartung, Tobias and Jakobs, Timo and Jansen, Karl and Ostmeyer, Johann and Urbach, Carsten",
    title = "{Digitising SU(2) gauge fields and the freezing transition}",
    eprint = "2201.09625",
    archivePrefix = "arXiv",
    primaryClass = "hep-lat",
    doi = "10.1140/epjc/s10052-022-10192-5",
    journal = "Eur. Phys. J. C",
    volume = "82",
    number = "3",
    pages = "237",
    year = "2022"
}

@article{Farrell:2023fgd,
    author = "Farrell, Roland C. and Illa, Marc and Ciavarella, Anthony N. and Savage, Martin J.",
    title = "{Scalable Circuits for Preparing Ground States on Digital Quantum Computers: The Schwinger Model Vacuum on 100 Qubits}",
    eprint = "2308.04481",
    archivePrefix = "arXiv",
    primaryClass = "quant-ph",
    reportNumber = "IQuS@UW-21-060, NT@UW-23-13",
    doi = "10.1103/PRXQuantum.5.020315",
    journal = "PRX Quantum",
    volume = "5",
    number = "2",
    pages = "020315",
    year = "2024"
}

@article{Alexandru:2021jpm,
    author = "Alexandru, Andrei and Bedaque, Paulo F. and Brett, Ruair{\'\i} and Lamm, Henry",
    title = "{Spectrum of digitized QCD: Glueballs in a S(1080) gauge theory}",
    eprint = "2112.08482",
    archivePrefix = "arXiv",
    primaryClass = "hep-lat",
    reportNumber = "FERMILAB-PUB-21-683-T",
    doi = "10.1103/PhysRevD.105.114508",
    journal = "Phys. Rev. D",
    volume = "105",
    number = "11",
    pages = "114508",
    year = "2022"
}

@article{Horn:1979fy,
    author = "Horn, D. and Weinstein, M. and Yankielowicz, S.",
    title = "{Hamiltonian Approach to Z(N) Lattice Gauge Theories}",
    reportNumber = "TAUP-723-79",
    doi = "10.1103/PhysRevD.19.3715",
    journal = "Phys. Rev. D",
    volume = "19",
    pages = "3715",
    year = "1979"
}

@article{Ji:2020kjk,
    author = "Ji, Yao and Lamm, Henry and Zhu, Shuchen",
    collaboration = "NuQS",
    title = "{Gluon Field Digitization via Group Space Decimation for Quantum Computers}",
    eprint = "2005.14221",
    archivePrefix = "arXiv",
    primaryClass = "hep-lat",
    reportNumber = "FERMILAB-PUB-20-198-T",
    doi = "10.1103/PhysRevD.102.114513",
    journal = "Phys. Rev. D",
    volume = "102",
    number = "11",
    pages = "114513",
    year = "2020"
}

@article{Fromm:2022vaj,
    author = "Fromm, Michael and Philipsen, Owe and Winterowd, Christopher",
    title = "{Dihedral lattice gauge theories on a quantum annealer}",
    eprint = "2206.14679",
    archivePrefix = "arXiv",
    primaryClass = "hep-lat",
    doi = "10.1140/epjqt/s40507-023-00188-9",
    journal = "EPJ Quant. Technol.",
    volume = "10",
    number = "1",
    pages = "31",
    year = "2023"
}

@article{Carena:2024dzu,
    author = "Carena, Marcela and Lamm, Henry and Li, Ying-Ying and Liu, Wanqiang",
    title = "{Quantum error thresholds for gauge-redundant digitizations of lattice field theories}",
    eprint = "2402.16780",
    archivePrefix = "arXiv",
    primaryClass = "hep-lat",
    reportNumber = "USTC-ICTS/PCFT-24-06, FERMILAB-PUB-23-570-T",
    doi = "10.1103/PhysRevD.110.054516",
    journal = "Phys. Rev. D",
    volume = "110",
    number = "5",
    pages = "054516",
    year = "2024"
}

@article{Mathew:2022nep,
    author = "Mathew, Emil and Raychowdhury, Indrakshi",
    title = "{Protecting local and global symmetries in simulating (1+1)D non-Abelian gauge theories}",
    eprint = "2206.07444",
    archivePrefix = "arXiv",
    primaryClass = "hep-lat",
    doi = "10.1103/PhysRevD.106.054510",
    journal = "Phys. Rev. D",
    volume = "106",
    number = "5",
    pages = "054510",
    year = "2022"
}

@article{Irmejs:2022gwv,
    author = "Irmejs, Reinis and Banuls, Mari Carmen and Cirac, J. Ignacio",
    title = "{Quantum simulation of Z2 lattice gauge theory with minimal resources}",
    doi = "10.1103/PhysRevD.108.074503",
    journal = "Phys. Rev. D",
    volume = "108",
    number = "7",
    pages = "074503",
    year = "2023"
}

@article{Popov:2023xft,
    author = "Popov, Pavel P. and Meth, Michael and Lewenstein, Maciej and Hauke, Philipp and Ringbauer, Martin and Zohar, Erez and Kasper, Valentin",
    title = "{Variational quantum simulation of U(1) lattice gauge theories with qudit systems}",
    doi = "10.1103/PhysRevResearch.6.013202",
    journal = "Phys. Rev. Res.",
    volume = "6",
    number = "1",
    pages = "013202",
    year = "2024"
}

@misc{Ciavarella:2024fzw,
    author = "Ciavarella, Anthony N. and Bauer, Christian W.",
    title = "{Quantum Simulation of SU(3) Lattice Yang Mills Theory at Leading Order in Large N}",
    eprint = "2402.10265",
    archivePrefix = "arXiv",
    primaryClass = "hep-ph",
    month = "2",
    year = "2024"
}

@article{Kavaki:2024ijd,
    author = "Kavaki, Ali H. Z. and Lewis, Randy",
    title = "{From square plaquettes to triamond lattices for SU(2) gauge theory}",
    eprint = "2401.14570",
    archivePrefix = "arXiv",
    primaryClass = "hep-lat",
    doi = "10.1038/s42005-024-01697-4",
    journal = "Commun. Phys.",
    volume = "7",
    number = "1",
    pages = "208",
    year = "2024"
}

@misc{budde2024quantum,
    author = "Budde, Thea and Krsti{\'c} Marinkovi{\'c}, Marina and Barros, Joao C. Pinto",
    title = "{Quantum many-body scars for arbitrary integer spin in 2+1D Abelian gauge theories}",
    eprint = "2403.08892",
    archivePrefix = "arXiv",
    primaryClass = "hep-lat",
    doi = "10.1103/PhysRevD.110.094506",
    journal = "Phys. Rev. D",
    volume = "110",
    number = "9",
    pages = "094506",
    year = "2024"
}

@article{Alexandru:2020wrj,
    author = "Alexandru, Andrei and Basar, Gokce and Bedaque, Paulo F. and Warrington, Neill C.",
    title = "{Complex paths around the sign problem}",
    eprint = "2007.05436",
    archivePrefix = "arXiv",
    primaryClass = "hep-lat",
    doi = "10.1103/RevModPhys.94.015006",
    journal = "Rev. Mod. Phys.",
    volume = "94",
    number = "1",
    pages = "015006",
    year = "2022"
}

@misc{Wills:2026kom,
    author = "Wills, Adam",
    title = "{A Review of Galois Qudits}",
    eprint = "2605.18981",
    archivePrefix = "arXiv",
    primaryClass = "quant-ph",
    month = "5",
    year = "2026"
}

@article{Alam:2021uuq,
    author = "Alam, M. Sohaib and Hadfield, Stuart and Lamm, Henry and Li, Andy C. Y.",
    collaboration = "SQMS",
    title = "{Primitive quantum gates for dihedral gauge theories}",
    eprint = "2108.13305",
    archivePrefix = "arXiv",
    primaryClass = "quant-ph",
    reportNumber = "FERMILAB-PUB-21-364-QIS-T",
    doi = "10.1103/PhysRevD.105.114501",
    journal = "Phys. Rev. D",
    volume = "105",
    number = "11",
    pages = "114501",
    year = "2022"
}

@misc{Grabowska:2022uos,
    author = "Grabowska, Dorota M. and Kane, Christopher and Nachman, Benjamin and Bauer, Christian W.",
    title = "{Overcoming exponential scaling with system size in Trotter-Suzuki implementations of constrained Hamiltonians: 2+1 U(1) lattice gauge theories}",
    eprint = "2208.03333",
    archivePrefix = "arXiv",
    primaryClass = "quant-ph",
    reportNumber = "CERN-TH-2022-133",
    month = "8",
    year = "2022"
}

@phdthesis{romers2007discrete,
  title={Discrete gauge theories in two spatial dimensions},
  author={Romers, JC},
  year={2007},
  school={Master’s thesis, Universiteit van Amsterdam}
}

@article{Kogut:1980qb,
    author = "Kogut, John B.",
    title = "{1/n Expansions and the Phase Diagram of Discrete Lattice Gauge Theories With Matter Fields}",
    reportNumber = "ILL-TH-80-01",
    doi = "10.1103/PhysRevD.21.2316",
    journal = "Phys. Rev. D",
    volume = "21",
    pages = "2316",
    year = "1980"
}

@misc{Bauer:2022hpo,
    author = "Bauer, Christian W. and others",
    title = "{Quantum Simulation for High Energy Physics}",
    eprint = "2204.03381",
    archivePrefix = "arXiv",
    primaryClass = "quant-ph",
    reportNumber = "UMD-PP-022-04, LA-UR-22-22100, RIKEN-iTHEMS-Report-22, IQuS@UW-21-027, MITRE-21-03848-2, FERMILAB-PUB-22-249-SQMS-T",
    month = "4",
    year = "2022"
}

@article{Ji:2022qvr,
    author = "Ji, Yao and Lamm, Henry and Zhu, Shuchen",
    collaboration = "NuQS",
    title = "{Gluon digitization via character expansion for quantum computers}",
    eprint = "2203.02330",
    archivePrefix = "arXiv",
    primaryClass = "hep-lat",
    reportNumber = "FERMILAB-PUB-21-496-T",
    doi = "10.1103/PhysRevD.107.114503",
    journal = "Phys. Rev. D",
    volume = "107",
    number = "11",
    pages = "114503",
    year = "2023"
}

@article{Gonzalez-Cuadra:2022hxt,
    author = "Gonz{\'a}lez-Cuadra, Daniel and Zache, Torsten V. and Carrasco, Jose and Kraus, Barbara and Zoller, Peter",
    title = "{Hardware Efficient Quantum Simulation of Non-Abelian Gauge Theories with Qudits on Rydberg Platforms}",
    eprint = "2203.15541",
    archivePrefix = "arXiv",
    primaryClass = "quant-ph",
    doi = "10.1103/PhysRevLett.129.160501",
    journal = "Phys. Rev. Lett.",
    volume = "129",
    number = "16",
    pages = "160501",
    year = "2022"
}

@article{Carena:2022kpg,
    author = "Carena, Marcela and Lamm, Henry and Li, Ying-Ying and Liu, Wanqiang",
    title = "{Improved Hamiltonians for Quantum Simulations of Gauge Theories}",
    eprint = "2203.02823",
    archivePrefix = "arXiv",
    primaryClass = "hep-lat",
    reportNumber = "FERMILAB-PUB-21-674-T",
    doi = "10.1103/PhysRevLett.129.051601",
    journal = "Phys. Rev. Lett.",
    volume = "129",
    number = "5",
    pages = "051601",
    year = "2022"
}

@article{Meth:2023wzd,
    author = "Meth, Michael and others",
    title = "{Simulating two-dimensional lattice gauge theories on a qudit quantum computer}",
    eprint = "2310.12110",
    archivePrefix = "arXiv",
    primaryClass = "quant-ph",
    doi = "10.1038/s41567-025-02797-w",
    journal = "Nature Phys.",
    volume = "21",
    number = "4",
    pages = "570--576",
    year = "2025"
}

@misc{Balaji:2025afl,
    author = "Balaji, Praveen and Conefrey-Shinozaki, Cian{\'a}n and Draper, Patrick and Elhaderi, Jason K. and Gupta, Drishti and Hidalgo, Luis and Lytle, Andrew and Rinaldi, Enrico",
    title = "{Quantum circuits for SU(3) lattice gauge theory}",
    eprint = "2503.08866",
    archivePrefix = "arXiv",
    primaryClass = "hep-lat",
    doi = "10.1103/k8f6-yft8",
    journal = "Phys. Rev. D",
    volume = "112",
    number = "5",
    pages = "054511",
    year = "2025"
}

@article{Bocharov:2014awg,
    author = "Bocharov, Alex and Roetteler, Martin and Svore, Krysta M.",
    title = "{Efficient Synthesis of Universal Repeat-Until-Success Quantum Circuits}",
    eprint = "1404.5320",
    archivePrefix = "arXiv",
    primaryClass = "quant-ph",
    doi = "10.1103/PhysRevLett.114.080502",
    journal = "Phys. Rev. Lett.",
    volume = "114",
    number = "8",
    pages = "080502",
    year = "2015"
}

@article{Gustafson:2025buk,
    author = "Gustafson, Erik J. and Lamm, Henry and Liu, Diyi and Murairi, Edison M. and Zhu, Shuchen",
    title = "{Synthesis of single-qutrit circuits from Clifford+R gates}",
    eprint = "2503.20203",
    archivePrefix = "arXiv",
    primaryClass = "quant-ph",
    reportNumber = "FERMILAB-PUB-25-0002-SQMS-T",
    doi = "10.1103/98q1-3yv8",
    journal = "Phys. Rev. A",
    volume = "112",
    number = "6",
    pages = "062414",
    year = "2025"
}

@misc{Balaji:2025yua,
    author = "Balaji, Praveen and Conefrey-Shinozaki, Cianan and Draper, Patrick and Elhaderi, Jason K. and Gupta, Drishti and Hidalgo, Luis and Lytle, Andrew",
    title = "{Perturbation theory, irrep truncations, and state preparation methods for quantum simulations of SU(3) lattice gauge theory}",
    eprint = "2509.25865",
    archivePrefix = "arXiv",
    primaryClass = "hep-lat",
    month = "9",
    year = "2025"
}

@misc{Crane:2024tlj,
    author = "Crane, Eleanor and others",
    title = "{Hybrid Oscillator-Qubit Quantum Processors: Simulating Fermions, Bosons, and Gauge Fields}",
    eprint = "2409.03747",
    archivePrefix = "arXiv",
    primaryClass = "quant-ph",
    month = "9",
    year = "2024"
}

@inproceedings{Shor:1996qc,
    author = "Shor, Peter W.",
    title = "{Fault-tolerant quantum computation}",
    eprint = "quant-ph/9605011",
    archivePrefix = "arXiv",
    doi = "10.1109/SFCS.1996.548464",
    month = "5",
    year = "1996"
}

@article{Brock:2024vkc,
    author = "Brock, Benjamin L. and Singh, Shraddha and Eickbusch, Alec and Sivak, Volodymyr V. and Ding, Andy Z. and Frunzio, Luigi and Girvin, Steven M. and Devoret, Michel H.",
    title = "{Quantum error correction of qudits beyond break-even}",
    eprint = "2409.15065",
    archivePrefix = "arXiv",
    primaryClass = "quant-ph",
    doi = "10.1038/s41586-025-08899-y",
    journal = "Nature",
    volume = "641",
    number = "8063",
    pages = "612--618",
    year = "2025"
}

@misc{Ryan-Anderson:2022tdl,
    author = "Ryan-Anderson, C. and others",
    title = "{Implementing Fault-tolerant Entangling Gates on the Five-qubit Code and the Color Code}",
    eprint = "2208.01863",
    archivePrefix = "arXiv",
    primaryClass = "quant-ph",
    month = "8",
    year = "2022"
}

@misc{Paetznick:2024ztu,
    author = "Paetznick, A. and others",
    title = "{Demonstration of logical qubits and repeated error correction with better-than-physical error rates}",
    eprint = "2404.02280",
    archivePrefix = "arXiv",
    primaryClass = "quant-ph",
    month = "4",
    year = "2024"
}

@article{Ryan-Anderson:2024zyh,
    author = "Ryan-Anderson, C. and others",
    title = "{High-fidelity teleportation of a logical qubit using transversal gates and lattice surgery}",
    eprint = "2404.16728",
    archivePrefix = "arXiv",
    primaryClass = "quant-ph",
    doi = "10.1126/science.adp6016",
    journal = "Science",
    volume = "385",
    number = "6715",
    pages = "adp6016",
    year = "2024"
}

@article{Li:2023cla,
    author = "Li, Zhiyuan and others",
    title = "{Error per single-qubit gate below 10$^{-4}$ in a superconducting qubit}",
    eprint = "2302.08690",
    archivePrefix = "arXiv",
    primaryClass = "quant-ph",
    doi = "10.1038/s41534-023-00781-x",
    journal = "npj Quantum Inf.",
    volume = "9",
    number = "1",
    pages = "111",
    year = "2023"
}

@article{Jiang:2025ufg,
    author = "Jiang, Jacky and Klco, Natalie and Di Matteo, Olivia",
    title = "{Non-Abelian dynamics on a cube: Improving quantum compilation through qudit-based simulations}",
    eprint = "2506.10945",
    archivePrefix = "arXiv",
    primaryClass = "quant-ph",
    doi = "10.1103/rrmd-fqx6",
    journal = "Phys. Rev. D",
    volume = "112",
    number = "7",
    pages = "074512",
    year = "2025"
}

@article{GoogleQuantumAIandCollaborators:2024efv,
    author = "Acharya, Rajeev and others",
    collaboration = "Google Quantum AI and Collaborators",
    title = "{Quantum error correction below the surface code threshold}",
    eprint = "2408.13687",
    archivePrefix = "arXiv",
    primaryClass = "quant-ph",
    doi = "10.1038/s41586-024-08449-y",
    journal = "Nature",
    volume = "638",
    number = "8052",
    pages = "920--926",
    year = "2025"
}

@article{Barends:2014zey,
    author = "Barends, R. and others",
    title = "{Superconducting quantum circuits at the surface code threshold for fault tolerance}",
    eprint = "1402.4848",
    archivePrefix = "arXiv",
    primaryClass = "quant-ph",
    doi = "10.1038/nature13171",
    journal = "Nature",
    volume = "508",
    pages = "500--503",
    year = "2014"
}

@misc{Marxer:2025ufr,
    author = "Marxer, Fabian and others",
    title = "{Above 99.9{\%} Fidelity Single-Qubit Gates, Two-Qubit Gates, and Readout in a Single Superconducting Quantum Device}",
    eprint = "2508.16437",
    archivePrefix = "arXiv",
    primaryClass = "quant-ph",
    month = "8",
    year = "2025"
}

@misc{Regent:2025cos,
    author = "R{\'e}gent, Fran{\c{c}}ois-Marie Le",
    title = "{Awesome Quantum Computing Experiments: Benchmarking Experimental Progress Towards Fault-Tolerant Quantum Computation}",
    eprint = "2507.03678",
    archivePrefix = "arXiv",
    primaryClass = "quant-ph",
    month = "7",
    year = "2025"
}

@misc{Eisert:2025ytq,
    author = "Eisert, Jens and Preskill, John",
    title = "{Mind the gaps: The fraught road to quantum advantage}",
    eprint = "2510.19928",
    archivePrefix = "arXiv",
    primaryClass = "quant-ph",
    month = "10",
    year = "2025"
}

@misc{Kurkcuoglu:2025gik,
    author = {K{\"u}rk{\c{c}}{\"u}oglu, Doga Murat and Lamm, Henry and Ogunkoya, Oluwadara and Pierattelli, Leonardo},
    title = "{The 2T-quoctit: a two-mode bosonic qudit for high energy physics}",
    eprint = "2509.21941",
    archivePrefix = "arXiv",
    primaryClass = "quant-ph",
    reportNumber = "FERMILAB-PUB-25-0143-SQMS-STUDENT-T",
    month = "9",
    year = "2025"
}

@article{Gidney:2018add,
    author = "Gidney, Craig",
    title = "{Halving the cost of quantum addition}",
    journal = "Quantum",
    volume = "2",
    pages = "74",
    year = "2018",
    doi = "10.22331/q-2018-06-18-74",
    eprint = "1709.06648",
    archivePrefix = "arXiv",
    primaryClass = "quant-ph"
}

@misc{Froland:2026rzt,
    author = "Froland, Henry and Grabowska, Dorota M. and Grieninger, Sebastian and Hartse, Jeremy and Lashbrook, Anne L. and Li, Zhiyao and Li, Ziyuan and Powell, Sarah J. M. and Savage, Martin J. and Yao, Xiaojun and Zemlevskiy, Nikita A.",
    title = "{Realizing Error Suppression in Partially Fault-Tolerant Quantum Simulations with IBM Quantum Computers}",
    eprint = "2607.24947",
    archivePrefix = "arXiv",
    primaryClass = "quant-ph",
    reportNumber = "IQuS@UW-21-130",
    month = "7",
    year = "2026"
}

@misc{Chen:2026sul,
    author = "Chen, Haobin and Cheng, Lin and Guo, Xingyu",
    title = "{Digital Quantum Simulation of Nonequilibrium Dynamics in the Schwinger Model under a Strong External Electric Field}",
    eprint = "2607.02894",
    archivePrefix = "arXiv",
    primaryClass = "hep-lat",
    month = "7",
    year = "2026"
}

@misc{Lee:2026qao,
    author = "Lee, Sangjin and Choi, Sangkook",
    title = "{Mitigating Trotter Errors via Post-Processed Symmetry Restoration}",
    eprint = "2606.20242",
    archivePrefix = "arXiv",
    primaryClass = "quant-ph",
    month = "6",
    year = "2026"
}

@misc{Lamm:2026xqk,
    author = "Lamm, Henry and Roggero, Alessandro and Singh, Hersh and Spagnoli, Luca",
    title = "{Exact chiral symmetry with quantum signal processing}",
    eprint = "2607.28524",
    archivePrefix = "arXiv",
    primaryClass = "hep-lat",
    month = "7",
    year = "2026"
}

@misc{Rouxinol:2026vjl,
    author = "Rouxinol, Gabriel and Bender, Julian and Emonts, Patrick and Grossi, Michele and Halimeh, Jad C.",
    title = "{Quantum Phase Diagram of the $2+1$D Untruncated SU$(2)$ Lattice Gauge Theory with Dynamical Fermions}",
    eprint = "2607.26132",
    archivePrefix = "arXiv",
    primaryClass = "hep-lat",
    month = "7",
    year = "2026"
}

@misc{Sukeno:2026vjm,
    author = "Sukeno, Hiroki and Rinaldi, Enrico and Okuda, Takuya",
    title = "{Measurement-based simulation of lattice gauge theory dynamics with adaptive quantum circuits on a trapped-ion processor}",
    eprint = "2608.04290",
    archivePrefix = "arXiv",
    primaryClass = "quant-ph",
    month = "8",
    year = "2026"
}

@misc{Froland:2026aff,
    author = "Froland, Henry and Grabowska, Dorota M.",
    title = "{Measuring Non-Stabilizerness in an SU(2) Lattice Gauge Theory}",
    eprint = "2606.14842",
    archivePrefix = "arXiv",
    primaryClass = "quant-ph",
    month = "6",
    year = "2026"
}

@misc{Jha:2026ror,
    author = "Jha, Raghav G. and Toga, Goksu C. and Taher, Jaber I. and Bakalov, Bojko N. and Kemper, Alexander F.",
    title = "{Magic and entanglement in 1+1-dimensional SU(2) lattice gauge theory}",
    eprint = "2606.09971",
    archivePrefix = "arXiv",
    primaryClass = "quant-ph",
    month = "6",
    year = "2026"
}

@article{Alexandru:2022son,
    author = "Alexandru, Andrei and Bedaque, Paulo F. and Carosso, Andrea and Cervia, Michael J. and Sheng, Andy",
    title = "{Qubitization strategies for bosonic field theories}",
    eprint = "2209.00098",
    archivePrefix = "arXiv",
    primaryClass = "hep-lat",
    doi = "10.1103/PhysRevD.107.034503",
    journal = "Phys. Rev. D",
    volume = "107",
    number = "3",
    pages = "034503",
    year = "2023"
}

@misc{Bedaque:2025eon,
    author = "Bedaque, Paulo F. and Murairi, Edison M. and Rupak, Gautam and Simonyan, Valery S.",
    title = "{Quantum computation of mass gap in an asymptotically free theory}",
    eprint = "2512.21282",
    archivePrefix = "arXiv",
    primaryClass = "quant-ph",
    reportNumber = "FERMILAB-PUB-25-0849-T",
    month = "12",
    year = "2025"
}

@article{Buser:2020cvn,
    author = "Buser, Alexander J. and Gharibyan, Hrant and Hanada, Masanori and Honda, Masazumi and Liu, Junyu",
    title = "{Quantum simulation of gauge theory via orbifold lattice}",
    eprint = "2011.06576",
    archivePrefix = "arXiv",
    primaryClass = "hep-th",
    reportNumber = "CALT-TH-2020-024, DMUS-MP-20/11, YITP-20-121",
    doi = "10.1007/JHEP09(2021)034",
    journal = "JHEP",
    volume = "09",
    pages = "034",
    year = "2021"
}

@article{Bergner:2024qjl,
    author = "Bergner, Georg and Hanada, Masanori and Rinaldi, Enrico and Schafer, Andreas",
    title = "{Toward QCD on quantum computer: orbifold lattice approach}",
    eprint = "2401.12045",
    archivePrefix = "arXiv",
    primaryClass = "hep-th",
    reportNumber = "RIKEN-iTHEMS-Report-24",
    doi = "10.1007/JHEP05(2024)234",
    journal = "JHEP",
    volume = "05",
    pages = "234",
    year = "2024"
}

@article{Halimeh:2024bth,
    author = {Halimeh, Jad C. and Hanada, Masanori and Matsuura, Shunji and Nori, Franco and Rinaldi, Enrico and Sch{\"a}fer, Andreas},
    title = "{A universal framework for the quantum simulation of Yang{\textendash}Mills theory}",
    eprint = "2411.13161",
    archivePrefix = "arXiv",
    primaryClass = "quant-ph",
    reportNumber = "RIKEN-iTHEMS-Report-24",
    doi = "10.1038/s42005-025-02421-6",
    journal = "Commun. Phys.",
    volume = "9",
    number = "1",
    pages = "67",
    year = "2026"
}

@misc{Bergner:2025zkj,
    author = "Bergner, Georg and Hanada, Masanori and Mendicelli, Emanuele",
    title = "{Exponential speedup in quantum simulation of Kogut-Susskind Hamiltonian via orbifold lattice}",
    eprint = "2506.00755",
    archivePrefix = "arXiv",
    primaryClass = "quant-ph",
    month = "5",
    year = "2025"
}

@misc{Halimeh:2025ivn,
    author = "Halimeh, Jad C. and Hanada, Masanori and Matsuura, Shunji",
    title = "{Universal framework with exponential speedup for the quantum simulation of quantum field theories including QCD}",
    eprint = "2506.18966",
    archivePrefix = "arXiv",
    primaryClass = "quant-ph",
    year = "2025"
}

@misc{Hanada:2025goy,
    author = "Hanada, Masanori and Matsuura, Shunji and Sch{\"a}fer, Andreas and Sun, Jinzhao",
    title = "{Gauge Symmetry in Quantum Simulation}",
    eprint = "2512.22932",
    archivePrefix = "arXiv",
    primaryClass = "quant-ph",
    year = "2025"
}

@misc{Webb-Mack:2026bkg,
    author = "Webb-Mack, Zo{\"e} and Klco, Natalie",
    title = "{Deforming the Trail: Baseline Quantum Circuitry for SU(2)$_k$ Lattice Gauge Theory}",
    eprint = "2605.15076",
    archivePrefix = "arXiv",
    primaryClass = "quant-ph",
    year = "2026"
}

@article{DAndrea:2023qnr,
    author = "D'Andrea, Irian and Bauer, Christian W. and Grabowska, Dorota M. and Freytsis, Marat",
    title = "{A new basis for Hamiltonian SU(2) simulations}",
    eprint = "2307.11829",
    archivePrefix = "arXiv",
    primaryClass = "hep-ph",
    doi = "10.1103/PhysRevD.109.074501",
    journal = "Phys. Rev. D",
    volume = "109",
    number = "7",
    pages = "074501",
    year = "2024"
}

@article{Romiti:2023hbd,
    author = "Romiti, Simone and Urbach, Carsten",
    title = "{Digitizing lattice gauge theories in the magnetic basis: reducing the breaking of the fundamental commutation relations}",
    eprint = "2311.11928",
    archivePrefix = "arXiv",
    primaryClass = "hep-lat",
    doi = "10.1140/epjc/s10052-024-13037-5",
    journal = "Eur. Phys. J. C",
    volume = "84",
    pages = "708",
    year = "2024"
}

@article{Kadam:2025trs,
    author = "Kadam, Saurabh V. and Naskar, Aahiri and Raychowdhury, Indrakshi and Stryker, Jesse R.",
    title = "{Loop-string-hadron approach to SU(3) lattice Yang-Mills theory, II: Operator representation for the trivalent vertex}",
    eprint = "2512.11796",
    archivePrefix = "arXiv",
    primaryClass = "hep-lat",
    doi = "10.1103/PhysRevD.111.074516",
    journal = "Phys. Rev. D",
    volume = "111",
    number = "7",
    pages = "074516",
    year = "2025"
}

@misc{Lamm:2026zzl,
      author        = "Lamm, Henry",
      title         = "{Ether of Orbifolds}",
      eprint        = "2603.29091",
      archivePrefix = "arXiv",
      primaryClass  = "hep-lat",
      reportNumber  = "FERMILAB-PUB-26-0210-T",
      month         = "3",
      year          = "2026"
}

@misc{Hidalgo:2026zsz,
      author        = "Hidalgo, Luis and Draper, Patrick",
      title         = "{Lattice Quantum Chromodynamics for Quantum Simulations}",
      eprint        = "2607.26445",
      archivePrefix = "arXiv",
      primaryClass  = "quant-ph",
      month         = "7",
      year          = "2026"
}

@misc{Rainaldi:2026wpt,
      author        = "Rainaldi, Tommaso and Montgomery, Jake and Yale, Christopher G. and McFarland, Brian K. and Revelle, Melissa C. and Lobser, Daniel and Tortorici, Edward C. and Clark, Susan and Siopsis, George and Grau, Matt and Ringer, Felix",
      title         = "{Benchmarking trigonometric continuous-variable gate primitives with trapped ions}",
      eprint        = "2607.14085",
      archivePrefix = "arXiv",
      primaryClass  = "quant-ph",
      reportNumber  = "SAND2026-23624O",
      month         = "7",
      year          = "2026"
}

@misc{Turco:2026cte,
      author        = "Turco, Matteo and Spagnoli, Luca and Roggero, Alessandro",
      title         = "{Binary Gauss Stabilizers for Abelian Lattice Gauge Theories}",
      eprint        = "2607.14861",
      archivePrefix = "arXiv",
      primaryClass  = "quant-ph",
      month         = "7",
      year          = "2026"
}

@misc{Andrade:2026kbz,
      author        = "Andrade, Barbara and Millar, Declan and Anderson, Lewis and Pascuzzi, Vincent R. and Lewenstein, Maciej and Tavernelli, Ivano and Halimeh, Jad C. and Grass, Tobias",
      title         = "{Dynamic Induction of Lattice Gauge Theories on a Quantum Computer}",
      eprint        = "2608.02756",
      archivePrefix = "arXiv",
      primaryClass  = "quant-ph",
      month         = "8",
      year          = "2026"
}

@misc{Gandon:2026das,
      author        = "Gandon, Anthony and Mariani, Alessandro and Banerjee, Debasish and Huffman, Emilie and Kanwar, Gurtej and Tacchino, Francesco and Wiese, Uwe-Jens and Tavernelli, Ivano",
      title         = "{String dynamics of a (2+1)D U(1) quantum link model on a digital quantum computer}",
      eprint        = "2606.19601",
      archivePrefix = "arXiv",
      primaryClass  = "quant-ph",
      month         = "6",
      year          = "2026"
}

@misc{Froland:2026tfx,
      author        = "Froland, Henry and Grabowska, Dorota M. and Grieninger, Sebastian and Hartse, Jeremy and Lashbrook, Anne L. and Li, Zhiyao and Li, Ziyuan and Powell, Sarah J.M. and Savage, Martin J. and Yao, Xiaojun and Zemlevskiy, Nikita A.",
      title         = "{The Utility of Sparse Error Detection in Quantum Simulations}",
      eprint        = "2608.02944",
      archivePrefix = "arXiv",
      primaryClass  = "quant-ph",
      reportNumber  = "IQuS@UW-21-132, NTG@UW-26-19",
      month         = "8",
      year          = "2026"
}

@article{Kalra:2025synthesis,
    author = {Kalra, Amolak Ratan and Mosca, Michele and Valluri, Dinesh},
    title = {{Synthesis and Arithmetic of Single Qutrit Circuits}},
    journal = {Quantum},
    volume = {9},
    pages = {1647},
    year = {2025},
    doi = {10.22331/q-2025-02-26-1647},
    eprint = {2311.08696},
    archivePrefix = {arXiv},
    primaryClass = {quant-ph}
}

@misc{Evra:2024arithmeticity,
    author = {Evra, Shai and Parzanchevski, Ori},
    title = {{Arithmeticity, thinness and efficiency of qutrit Clifford+T gates}},
    year = {2024},
    eprint = {2401.16120},
    archivePrefix = {arXiv},
    primaryClass = {quant-ph}
}

@misc{Davoudi:2026pcu,
    author = "Davoudi, Zohreh and Stryker, Jesse R.",
    title = "{Exponential-in-$N_c^2$ cost reduction of product-formula-based quantum simulations of quantum chromodynamics}",
    eprint = "2608.21258",
    archivePrefix = "arXiv",
    primaryClass = "hep-lat",
    reportNumber = "UMD-PP-026-06",
    month = "8",
    year = "2026"
}

@misc{Pasztor:2026pte,
    author = "Pasztor, Attila and Pesznyak, David",
    title = "{Gauge field digitization in the Hamiltonian limit}",
    eprint = "2609.07886",
    archivePrefix = "arXiv",
    primaryClass = "hep-lat",
    month = "9",
    year = "2026"
}

@misc{Saha:2026blc,
    author = "Saha, Amit and Arzani, Francesco",
    title = "{Efficient Synthesis of Multi-Controlled Toffoli Gates with Ternary Clifford$+P_9$ Gates}",
    eprint = "2609.29884",
    archivePrefix = "arXiv",
    primaryClass = "quant-ph",
    month = "9",
    year = "2026"
}

@misc{Bruinsma:2026fke,
    author = "Bruinsma, Carli and Posta, Pietro M. and Stokvis, Joppe and Grinko, Dmitry and Ozols, Maris",
    title = "{Quantum Fourier transform toolbox}",
    eprint = "2608.28573",
    archivePrefix = "arXiv",
    primaryClass = "quant-ph",
    month = "8",
    year = "2026"
}

@misc{Draper:2026bcj,
    author = "Draper, Patrick",
    title = "{Block Encoding Non-Abelian Lattice Gauge Theory}",
    eprint = "2608.17115",
    archivePrefix = "arXiv",
    primaryClass = "quant-ph",
    month = "8",
    year = "2026"
}

@misc{Li:2026ppp,
    author = "Li, Tianyin and Li, Ying-Ying and Wang, Xiaoyang and Xing, Hongxi",
    title = "{Efficient Quantum Simulations of Yang-Mills theory with Maximal-tree Gauge}",
    eprint = "2608.27267",
    archivePrefix = "arXiv",
    primaryClass = "hep-lat",
    reportNumber = "RIKEN-iTHEMS-Report-26",
    month = "8",
    year = "2026"
}

@misc{Lamm:2026cfq,
    author = "Lamm, Henry",
    title = "{MORE Thermal Gauge Theories at Finite $\theta$ and $\mu$ from Real-Time Quantum Simulation}",
    eprint = "2609.38569",
    archivePrefix = "arXiv",
    primaryClass = "hep-lat",
    reportNumber = "FERMILAB-PUB-26-0658-T",
    month = "9",
    year = "2026"
}

@misc{Warman:2026flz,
    author = "Warman, Alison and Tantivasadakarn, Nathanan and Schafer-Nameki, Sakura",
    title = "{Exact Recovery for Non-Abelian Surface Codes}",
    eprint = "2610.03677",
    archivePrefix = "arXiv",
    primaryClass = "quant-ph",
    month = "10",
    year = "2026"
}

@misc{Ale:2026hqe,
    author = "Ale, Victor and Rainaldi, Tommaso and Rico, Enrique and Ringer, Felix and Siopsis, George",
    title = "{Encoding Compact U(1) Gauge Fields in Bosonic Modes with GKP Stabilization}",
    eprint = "2609.00167",
    archivePrefix = "arXiv",
    primaryClass = "quant-ph",
    month = "8",
    year = "2026"
}

@misc{Garofalo:2026pwd,
    author = "Garofalo, Marco and Hartung, Tobias and Jakobs, Timo and Ludwig, Paul and Ostmeyer, Johann and Urbach, Carsten",
    title = "{No-go theorems for dual Hamiltonians of Non-Abelian Lattice Gauge Theories}",
    eprint = "2609.39780",
    archivePrefix = "arXiv",
    primaryClass = "hep-lat",
    month = "9",
    year = "2026"
}
